\documentclass[usenatbib]{mn2e}

\usepackage[pdftex]{graphicx}

\usepackage[latin1]{inputenc}
\usepackage[T1]{fontenc}
\usepackage[english]{babel}
\usepackage[pdftex]{graphicx}
\usepackage{graphicx}
\usepackage{appendix}
\usepackage{setspace}
\usepackage{color}

\usepackage{subfigure}
\usepackage{wrapfig}
\usepackage[babel]{csquotes}
\usepackage{natbib}
\usepackage{wasysym}
\usepackage{amssymb}
\usepackage{tabularx}
\usepackage{aas_macros}
\usepackage{fancyhdr}										
\usepackage{colortbl}										
\usepackage{longtable}										
\usepackage{nomencl}
\usepackage{natbib}
\usepackage{rotating}
\usepackage[section]{placeins}
\usepackage{vmargin}
\usepackage{hyperref}
\usepackage{array}
\title[{\itshape RedGOLD} Cluster detection in the CFHT-LS W1]{The {\itshape RedGOLD} Cluster Detection Algorithm and its Cluster Candidate Catalogue for the  CFHT-LS W1}
\author[Licitra et al. 2015]{Rossella Licitra$^{1,2}, $\thanks{E-mail:
rossella.licitra@obspm.fr} 
Simona Mei$^{1,2,3}$, 
Anand Raichoor $^{1,4}$,
Thomas Erben $^{5}$,
\newauthor
 Hendrik Hildebrandt $^{5}$
\\
\\
$^{1}$GEPI, Observatoire de Paris, PSL Research University,  CNRS, University of Paris Diderot,\\
 61, Avenue de l'Observatoire 75014, Paris France \\
$^{2}$University of Paris Denis Diderot, University of Paris Sorbonne Cit\'e (PSC), 75205 Paris Cedex
  13, France\\
$^{3}$California Institute of Technology, Pasadena, CA 91125, USA\\
$^{4}$CEA, Centre de Saclay, IRFU/SPP, F-91191 Gif-sur-Yvette, France\\
$^{5}$Argelander-Institut fur Astronomie, University of Bonn, Auf dem Hugel 71, D-53121 Bonn, Germany}
\begin{document}

\date{}

\pagerange{\pageref{firstpage}--\pageref{lastpage}} \pubyear{2002}

\maketitle

\label{firstpage}

\begin{abstract}

We present {\itshape RedGOLD}, a new optical/NIR galaxy cluster detection algorithm, and apply it to the CFHT-LS W1 field. 
{\itshape RedGOLD} searches for red-sequence galaxy overdensities while minimising contamination from dusty star-forming galaxies. It imposes an NFW profile and calculates cluster detection significance and richness. We optimise these latter two parameters using both simulations and X-ray detected cluster catalogues, and obtain a catalogue $\sim 80\%$ pure up to $z \sim 1$, and  $\sim 100\%$ ($\sim 70\%$) complete at $z\le 0.6$ ( $z\lesssim1$) for galaxy clusters with $M \gtrsim 10^{14}\ {\rm M_{\odot}}$ at the CFHT-LS Wide depth. In the CFHT-LS W1, we detect 11 cluster candidates per $\rm deg^2$ out to $z\sim1.1$. When we optimise both completeness and purity,  {\itshape RedGOLD} obtains a cluster catalogue with higher completeness and purity than other public catalogues, obtained using CFHT-LS W1 observations, for $M \gtrsim 10^{14}\ {\rm M_{\odot}}$. We use X--ray detected cluster samples to extend the study of the X--ray temperature--optical richness relation to a lower mass threshold, and find a mass scatter at fixed richness of $\sigma_{lnM|\lambda}=0.39\pm0.07$ and  $\sigma_{lnM|\lambda}=0.30\pm0.13$ for the \citet{Gozaliasl2014} and \citet{Mehrtens2012} samples. When considering similar mass ranges as previous work, we recover a smaller scatter in mass at fixed richness.
We recover $93\%$ of the redMaPPer detections, and find that its richness estimates is on average $\sim 40-50\%$ larger than ours at $z>0.3$.  {\itshape RedGOLD} recovers X-ray cluster spectroscopic redshifts  at better than $5\%$ up to $z\sim1$, and the centres within a few tens of arcseconds.

\end{abstract}


\section{Introduction}

Galaxy clusters are powerful probes of our cosmological models and the evolution of galaxies in dense environments.
Galaxy clusters can be detected in different ways, tracing different cluster components.
Using X--ray and submillimeter observations, it is possible to trace their gas, by its
 bremsstrahlung radiation  \citep[e.g.,][]{Voit2005}, and through the \textit{Sunyaev-Zel'dovich effect} \citep{Sunyaev1970}.
It is also possible to consider the stellar component of cluster galaxies and study their radiation 
in the optical or in the infrared. The analysis of multi-wavelength data permits us to detect clusters
searching for  early-type galaxy (ETG) overdensities, which dominate the inner regions of galaxy clusters, 
in agreement with the morphology-density relation \citep{Dressler1980}. 

So far, many different methods have been developed to detect galaxy clusters using optical data: 
some works are based on the search of spatial overdensities through friends-of-friends algorithms  \citep[e.g.,][]{Wen2012}, adaptive kernel
techniques \citep[e.g.,][]{Mazure2007,Adami2010} or Voronoi tessellations \citep[e.g.,][]{Kim2002}. 
Other methods are based on the detection of galaxies that lie on the red-sequence \citep{Gladders2000,Thanjavur2009, Rykoff2014}. In some cases,  also the existence of a brightest central galaxy \citep{Koester2007} is required. 
A widely used technique is the \textit{Matched filter cluster detection} \citep{Postman1996} that relies on detecting galaxies in one passband and
searches galaxy clusters analysing  the galaxy distribution,  with the assumption of model profiles that fit the data (for example a  characteristic galaxy cluster luminosity and a radial profile \citep[e.g.,][]{Olsen2007, Grove2009}.
\citet{Milkeraitis2010} proposed a revised method, the so-called \textit{3D-Matched-Filter}, which uses a finding algorithm based on the luminosity and radial profile of galaxy clusters, and photometric redshifts.  

The existence of several detection algorithms depends on the fact that the ideal method to detect galaxy clusters would produce a catalogue that includes all the real clusters (i.e. complete) and is not contaminated by false detections (i.e. pure).
However, beside the noise and systematics in the observations, all detection techniques are affected by biases and selection effects, 
because of their basic assumptions on the nature or the morphology of galaxy clusters. 
As a consequence, the resulting cluster catalogue will reflect these assumptions, missing structures that do not fit  the adopted cluster properties:
for example, X--ray selected cluster catalogues are incomplete against gas--poor clusters, while 
optically detected cluster catalogues are contaminated by galaxy projections and may be incomplete against fossil groups, because of the fainter magnitudes of the companion galaxies
with respect to the central one \citep{Jones2003, Proctor2011}.

This implies that the different detection techniques are complementary and, while the ideal method to detect all galaxy clusters does not exist, each method can be optimised for a certain class of clusters and groups. 

In the next decade, large scale deep surveys in the optical and the near-infrared have been planned for understanding the nature of dark energy, such as the  Large Synoptic Survey Telescope \citep[LSST,][]{lsstsb2012}, the European Space Agency's Euclid
mission \citep{laureijs2011}, and the U.S. National
Aeronautics and Space Administrations WFIRST Mission \footnote{http://wfirst.gsfc.nasa.gov}. These surveys will use galaxy clusters as cosmological probes, and, to do so, will need accurate estimates of the cluster mass. Their cluster samples will be detected by the analysis of multi-wavelength optical and infrared data, and cluster mass estimates will be derived from galaxy counts in these wavelengths (e.g., the optical richness). For this reason, there is a large effort in the cluster community to improve existing cluster detection algorithms in the optical and the near-infrared, and to optimise their performance.

In this paper we present {\itshape RedGOLD},  our cluster detection algorithm based on a revised red--sequence technique, and apply it to the CFHT-LS \citep[Canada-France-Hawaii Telescope
Legacy Survey;][] {Gwyn2012} Wide~1 (W1) field.
To validate and optimise our detection technique, we present a direct comparison with X--ray detected cluster catalogues and previous public catalogues based on different detection techniques in the optical.

The paper is organised as follows: in section \ref{sec:Observations} we describe the observations
and the survey properties. We briefly present the photometric redshift estimates in section \ref{sec:photoz}. In section \ref{sec:technique} we present our detection technique and the optical richness provided by our algorithm. 
Section \ref{sec:compl} is focused on the estimate of the completeness and the purity of our algorithm
using both simulations and observations.
In section \ref{sec:cfht} we discuss our  results obtained applying the algorithm to the  Canada-France-Hawaii Lensing Survey \citep[hereafter referred to as CFHTLenS;][]{Heymans2012} optical data and
the comparison with existing publicly available cluster catalogues.
 Finally, in section \ref{sec:conclusions} we present our conclusions.

We assume a standard cosmological model with $\Omega_m=0.3, \Omega_{\Lambda}=0.7$ and 
$H_0=70\ {\rm km\ s^{-1}\ Mpc^{-1}}$. If not differently specified, magnitudes are given in the AB system \citep{Oke1983, Sirianni2005}.  

\section{Observations and data description} \label{sec:Observations}

We apply our algorithm to the CFHT-LS W1 field, using the CFHTLenS reduction. Here we briefly summarise the CFHTLenS data that we use, and we refer the reader to \citet{Erben2013} 
for a detailed description of the survey properties. 
The CFHT-LS covers $154\ {\rm deg^2}$ in 5 optical bands, $u^*, g, r, i, z$, observed with the MegaCam instrument \citep{Boulade2003}. The CFHTLenS survey analysis combined weak lensing data processing with THELI \citep{Erben2013}, shear measurement with lensfit \citep{Miller2013}, and photometric redshift measurement with PSF-matched photometry \citep{Hildebrandt2012}. A full systematic error analysis of the shear measurements in combination with the photometric redshifts is presented in \citet{Heymans2012}, with additional error analyses of the photometric redshift measurements presented in \citet{Benjamin2013}. 
The depth of the CFHT-LS deep and wide fields is $i\sim27.4$~mag and $i\sim25.7$~mag, respectively.

Among the four wide fields, we  use images from the CFHT-LS W1 field, centred on  
the position RA=02:18:00  and DEC=-07:00:00, processed as described in \citet{Raichoor2014}.
\citet{Raichoor2014} used a modified version of the THELI pipeline \citep{Erben2013} to reprocess the  CFHT-LS W1 fields:
 they calibrated the zero points on the Sloan Digital Sky Survey \citep[SDSS;][]{York2000} and, for that  reason, 
they analysed 62 out of the 72 pointings of the CFHT-LS W1, because the remaining 10 fields are not covered 
by the SDSS. They adopted the \textit{global} PSF homogenisation method, described in \citet{Hildebrandt2012}, because it significantly increases the  accuracy of the colour measurements, and, as a consequence, the accuracy of the photometric redshifts. The final CFHT--LS W1 area that we use is then $\sim 60\ {\rm deg^2}$.

We calibrate our cluster detection algorithm using the X--ray catalogue provided by \citet{Gozaliasl2014}.
It covers 3~$\rm deg^2$ in the CFHT-LS W1 field and includes 135 X--ray groups and clusters up to redshift 1.1 with masses between $9.5\times 10^{12}<M_{200}<3.8\times10^{14}\ M_{\odot}$\footnote{ $M_{200}$ is the cluster total mass in a sphere with mean density equal to 200 times the critical density of the Universe $\rho_c$. The radius of this sphere is defined as $R_{200}$. }.
The median mass is $M_{200}=5.9\times10^{13}\  M_{\odot}$.

We use both the  \citet{Gozaliasl2014} catalogue and the XMM Cluster Survey (XCS) catalogue \citep{Mehrtens2012} to study the X--ray temperature--optical richness relation. The XCS  serendipitously searches for galaxy clusters, using the whole available data set in the XMM--Newton Science Archive. This catalogue includes 503 X--ray detected clusters  up to $z\sim 1.5$, with 401 of them with an X--ray temperature measurement of $0.4<T_X<14.7$ keV. Of those, 27 detections are in the CFHT-LS W1 field, and we restrict our subsample to 20 objects with a temperature measurement of $0.6<T_X<7.5$ keV.

\section{The photo-z catalogue} \label{sec:photoz}

The photometric redshift estimates have been obtained as explained in \citet{Raichoor2014}: they used the bayesian codes {\itshape LePhare} \citep{Arnouts1999,Arnouts2002,Ilbert2006} and {\itshape BPZ} \citep{Benitez2000,Benitez2004,Coe2006}, and a set of 60 templates \citep{Capak2004}, obtained interpolating 
four empirical models \citep[Ell, Sbc, Scd, Im; ][]{Coleman1980}
and two starburst spectra \citep{Kinney1996}. For the {\itshape LePhare} photometric redshift estimates, they
included the reddening as a free parameter ($0 < E(B -V )< 0.25$), 
considering the Small Magellanic Cloud (SMC) extinction law for late type galaxies \citep{Prevot1984}.
They also introduced a new prior for the brightest objects.

To estimate the photometric redshift accuracy,  \citet{Raichoor2014} used spectroscopic redshift measurements from different surveys:   the VIMOS Public Extragalactic Redshift Survey \citep[VIPERS;][]{Guzzo2014}, and the F02 and F22 fields of the VIMOS VLT Deep Survey \citep[VVDS;][]{Lefevre2005,Lefevre2013}. 

As shown in \citet{Raichoor2014}, the photometric redshift quality decreases with increasing magnitude and with increasing redshift, with 
 $\sigma_{photoz}\sim 0.03 \times (1+z)$  at $i<23.5$~mag and a fraction of outliers \footnote{Following, \citet{Raichoor2014}, outliers are defined  as galaxies with $|{\frac{z_{phot}-z_{spec}}{1+z_{spec}}}|>0.15$} of  less than $ 9\%$.  Similarly, the bias, defined as the median of  $\Delta z = \frac{z_{\rm phot}-z_{\rm spec}}{1+z_{\rm spec}}$, is around zero for bright and low--redshift objects while it becomes larger for faint ($i>23.5$~mag) and high--redshift ($z>0.8$) sources.

\section{The Cluster Detection Algorithm {\it RedGOLD}} \label{sec:technique}

Our algorithm, which we name {\it RedGOLD}  (Red-sequence Galaxy Overdensity cLuster Detector), is based on the detection of red-sequence galaxy overdensities. 
It relies on the observational  evidence that galaxy clusters host a large population of passive (red)
and luminous (L>0.2 $\times L^*$) ETGs, mostly concentrated in their cores and tightly  distributed on a red-sequence on the colour-magnitude
diagram \citep{Bower1992}. This assumption is true for clusters of galaxies up to $z\sim1.5$ \citep[e.g.,][]{Mei2009,Snyder2012,Brodwin2013, Mei2015}. 

The method consists of two main steps, described in the following sections: (1) the detection of spatial overdensities of red early-type galaxies; (2)  the confirmation of a tight red-sequence in the colour-magnitude relation.

\subsection{Red galaxy overdensity detection} \label{sec:overdens}

In order to detect spatial overdensities of red early--type galaxies, we eliminate all saturated objects and consider only galaxies with $i < 23.5$~mag, to have uncertainties on photometric redshifts $\sigma_{photoz} \lesssim 0.03 \times (1+z)$. This applies to all our procedure from now on, except to the cluster candidate richness estimate.

To identify stars, we remove objects with the SExtractor $CLASS\_STAR>0.95$ and $i<22.5$~mag, following \citet{Raichoor2014}. 

We divide the entire galaxy sample in redshift slices in the range $0.1<z<1.2$ 
with a step of $\delta z=0.2$ and overlapping by $3 \times \sigma_{photoz}$. 
 To take into account the errors on the galaxy photometric redshifts, 
we also select all galaxies with a photometric redshift within 
 one $\sigma_{photoz}$ from a given redshift bin, where $\sigma_{photoz}$ is the error on the individual galaxy photometric redshift from \citet{Raichoor2014}.

In each redshift bin, early-type galaxies have well defined red-sequence colours (i.e., colours of typical old stellar populations) which can be predicted with stellar population models  \citep[e.g.,][]{Mei2009}.

We convert our observed magnitudes into absolute rest-frame magnitude, following \citet{Mei2009},  using the corresponding galaxy photometric redshift and taking into account the filter corrections as in \citet{Mei2009}. We adopt both a {\it k}-correction and an evolution correction as in \citet{Mei2009}. 
To have the lowest possible contamination, we first select passive galaxies in two colours simultaneously. We choose two pairs of filters at each redshift
bin, corresponding to the (U-B) and (B-V) rest-frame colours: doing so, the colour that corresponds to the (U-B) rest-frame colour straddles the  4000\ \AA \ break, and the joint colour cut, which corresponds to the (B-V) rest--frame, allows to separate red galaxies with ongoing or recent star-formation events from red passive galaxies  \citep{Larson1978}.

To compute predicted colours in each redshift bin, we use single burst \citet{Bruzual2003} (BC03) stellar population models. We assume a passive evolution, a galaxy formation redshift $z_{form}=3$  and a solar metallicity, $Z= 0.02$. In this work we adopt a Salpeter initial mass function \citep{Salpeter1955}.

In addition to our colour selection, we require that red galaxies are also defined as ETGs according to the classification provided by the Spectral Energy Distribution (SED) fitting used to estimate photometric redshifts, i.e. objects which show ETG spectral characteristics. In fact, in the redshift range that we consider, the galaxy morphological classification based on galaxy shapes and structural parameters  is possible only for the brightest galaxies (and the lowest redshifts) 
with ground--based observations. Typically, the magnitude and redshift limits for ground--based observations morphological classification are $r<22$~mag and $z<0.5$ for ETGs \citep{Povic2013}.

To identify galaxy overdensities, in each MegaCam field, and for each redshift slice:
\begin{itemize}

\item  We divide the coordinates space in overlapping circular cells of fixed comoving radius $r_{grid}=500$~kpc, and with centres separated by $500$~kpc;

\item We count the number of red ETG galaxies $N_{gal}$ in each cell, and we build the galaxy count distribution in different redshift bins. We consider the background contribution $N_{bkg}$ as the mode of this distribution, and calculate its standard deviation $\sigma_{bkg}$, in each redshift bin;  

\item We estimate the detection significance $\sigma_{det}=\frac{N_{gal}-N_{bkg}}{\sigma_{bkg}}$ in each cell.
\end{itemize}
 
Since clusters are structures denser in red ETGs than the average red ETG background,  we find our preliminary overdensity-based detections as systems characterised by red ETG densities larger than $N_{bkg}+\sigma_{det} \times \sigma_{bkg}$.

\begin{figure}
    \begin{center}
           \includegraphics[width=\columnwidth]{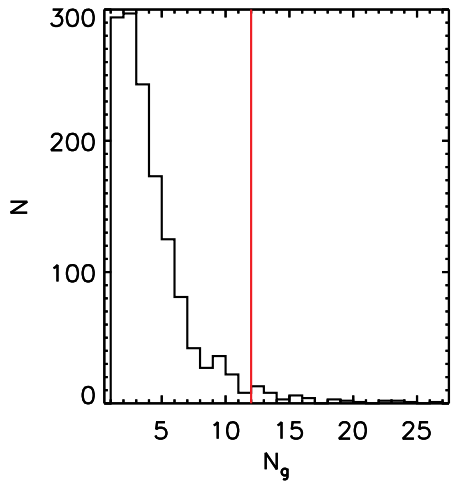}   
      \end{center}
    \caption{Example of our galaxy count distribution $N_g$. The red vertical line represents the $N_{bkg}+3 \times \sigma_{bkg}$ limit at z=0.5 for one MegaCam pointing in the CFHT-LS W1.}
    \label{fig:backdistr}
  \end{figure}

The choice of $\sigma_{det}$ changes the cluster catalogue completeness and purity (see section \ref{sec:compl}, in which we discuss our choice of $\sigma_{det}$). 
 In Figure \ref {fig:backdistr}, we show an example of the galaxy count distribution for 
one CFHT-LS W1 MegaCam pointing at $z=0.5$: the red vertical line represents the detection limit $N_{bkg}+3 \times \sigma_{bkg}$, implying that all structures lying  on the right side of the chosen $N_{bkg}+\sigma_{det} \times \sigma_{bkg}$ are cluster candidate detections with $\sigma_{det}\ge3$. 

A preliminary cluster redshift is assigned as the central value of the redshift bin.

In the CFHTLenS data, for each science image, a mask flags regions with less accurate photometry \citep[e.g. because of star haloes; ][]{Erben2013}. \citet{Rykoff2014} pointed out that 
masks have to be taken into account  not to underestimate the cluster richness and proposed a technique to extrapolate the richness measurement in regions with missing photometry
 (e.g. empty regions/holes).

In our case, we choose not to use an extrapolation technique. The way we take into account the presence of masks for the stars and
other saturated objects is by selecting only objects with an error in
photometry within the average distribution. In fact, the area over which the CFHTLenS catalogue is empty is very small ($\sim10\%$) and the main difference in the photometry of galaxies in masked areas is the larger photometry uncertainties. We build a photometry uncertainty distribution in magnitude
bins using \citet{Raichoor2014} photometry and photometric errors. We discard all objects that
have uncertainties more than 3-$\sigma$ the average uncertainty
distribution in the red overdensity calculation.

{\it A posteriori}, we verify that our procedure does not affect our detection efficiency and does not discard a significant number of real cluster members, leading to an underestimation of the cluster richness. We describe this procedure in more detail in section~\ref{sec:compl} and section~\ref{sec:cfht}. 

\subsubsection{Cluster centring}

A key step of all detection algorithms is the cluster centring. The estimate of the centre
is very important, as a miscentring can lead to an increasing in the scatter and in the slope of the colour-magnitude   relation. The miscentring can also lead
to a bias in the weak-lensing mass measurements \citep[up to $\sim 30\%$;] []{George2012}, and in the cluster richness estimates \citep[e.g.,][]{Johnston2007,Rozo2011}.

The very simple  idea underlying our centring technique is that, since red ETGs are mostly concentrated in the inner regions of galaxy clusters, we can centre our preliminary detections using local galaxy densities. Since \citet{George2011} has shown that centroids trace overdensity centres less efficiently than the brightest galaxies, we search for the brightest galaxy in the most overdense region.
We consider all red ETGs brighter than $0.2 \times {\rm L^*}$, and select as cluster centre the galaxy with the highest number of red ETG galaxies within a calibrated fraction of its cell, weighted on luminosity. For our application to the CFHT-LS W1, we calibrate the radius of this region with the available X--ray cluster centres from \citet{Gozaliasl2014}, minimising the distance between our centres and the X--ray centres. 
If two or more galaxies correspond to the highest local density value,
we simply centre our detection on the brightest one. 

\subsection{Colour-magnitude relation and red-sequence}

To confirm our cluster candidates and refine our photometric redshift estimate, we analyse their colour-magnitude diagram. 

Following \citet{Mei2009}, we convert our observed colours at a given redshift into $(U-B)_{z=0}$ rest-frame colours  \footnote{we use the Johnson U and B sensitivity curve respectively from \citet{Bessell1990} and \citet{MaizApellaniz2006}, following \citet{Mei2009}} using the relation :

\begin{equation}
(U-B)_{z=0}=Z_{point|(U-B)}+S_{(U-B)} \times col_{obs} \ ,
\end{equation}
where $col_{obs}$ is the observed colour, and $Z_{point|(U-B)}$ and $S_{(U-B)}$ are the fit zero point and slope, respectively.
We use BC03 single burst stellar population models, assuming a passive evolution, a formation redshift $1<z_{form}<7$ and three different values of metallicities (Z=0.008, 0.02 and 0.05). 
The rest-frame magnitudes are computed in the Vega system while observed magnitudes are in the AB system, following \citet{Mei2009}.

We convert observed magnitudes in absolute B rest-frame magnitudes, $M_B$, by fitting the relation:
\begin{equation}
M_{B, z=0}= mag_{obs}+Z_{point|M_B}+S_{M_B} \times col_{obs} \ ,
\end{equation}

 where $mag_{obs}$ and $col_{obs}$ are the observed magnitude and colours for which we perform this conversion, and $Z_{point|M_B}$ and $S_{M_B}$ are the zero point and the slope of the linear fit, respectively. 

In the age and metallicity range that we consider (consistent with the ETG old populations), there is a linear relation between the colours that straddle the 4000\ \AA~break and the $(U-B)_{z=0}$ rest-frame colour \citep{Mei2009}. Errors on the relation zero point and scatter, due to the sampling that we are using, are estimated through jackknife. Errors on the relation zero point and
scatter, due to the sampling that we are using,
are estimated through jackknife applied to galaxies.

The observed magnitudes 
are chosen to be the closest to the rest-frame U and B magnitudes. 

From the estimated $(U-B)$ and $M_B$, we perform a robust linear fit on the $(U-B)$ vs $M_B$ relation, using the Tukey's biweight \citep{Press1992}.
\citet{Mei2009} did not find a significant evolution in the colour-magnitude parameters as a function of redshift \citep[confirmed by] []{Snyder2012}: the average slope and intrinsic scatter for early-type galaxies (E+S0) are $\frac{\Delta{(U-B)_{z=0}}}{\Delta_{M_{B,z=0}}}= -0.046\pm 0.023$
and $ \sigma_{(U-B)_{z=0}}=0.061\pm 0.015$~mag, respectively. 
For this reason, we impose that our detections have a red-sequence scatter and slope within $3\sigma$  of
the expected value estimated from \citet{Mei2009} (mean value of the red-sequence parameters plus three times the scatter, adding in quadrature the photometric errors).

To refine the cluster candidate redshift, we use the ETG median photometric redshift. For each cluster candidate, we assign cluster membership to galaxies corresponding to the spatial overdensity if the difference between the galaxy photometric redshift and the cluster candidate redshift is within 3 times the uncertainty on photometric redshifts, $\sigma_{photoz}$.
In our samples, the uncertainty on photometric redshifts is larger than the intrinsic redshift dispersion due to the cluster galaxy dispersion velocities.

\subsection{Multiple detections} \label{sec:multdet}

A very common problem for every cluster detection algorithm is that of multiple detections, i.e. the fact
that the same structure is detected multiple times (e.g., in different redshift bins or different spatial regions). 

Since we centre our preliminary detections on the red ETG with
the highest local density, two preliminary detections that are spatially close can converge in similar centre positions.

For this reason, we develop an algorithm to clean the final cluster catalogue, in order to minimise the contamination
due to multiple detections: in particular, we iteratively filter our catalogue, 
checking for detections characterised by at least half of members in common and with a final cluster candidate redshift difference
of $\Delta z \le 0.1$.

When a multiple detection is found, we retain as its centre the centre of the greatest red overdensity, characterised by the highest signal-to-noise ratio i.e. $\sigma_{det}$, weighted on luminosity.
Applying these corrections, we are able to remove overlapping detections.

\subsection{Optical richness}\label{sec:rich}

The estimate of the cluster richness is a key point when studying galaxy clusters, because the cluster richness is a proxy for the cluster mass, which is not directly measurable. 

Several mass proxies have been adopted in the literature.
For example, the X--ray luminosity $L_X$ is commonly adopted as a mass $M$ proxy. 
\citet{Vikhlinin2009} found that the scatter in mass at fixed $L_X$ is $\sigma_{ln M|L_X}\sim0.25$ and \citet{Mantz2010}  $\sigma_{ln M|L_X}\sim0.32$. When using weak--lensing analysis to obtain cluster mass measurements, the scatter in the halo mass at fixed weak--lensing mass is of the same order of magnitude, $\sigma_{ln M|WL}\sim0.3$ \citep{Becker2011,VonDerLinden2014}. 
Similarly, a key measurement of optical detected clusters is the cluster richness, adopted as cluster mass proxy. 
For the MaxBCG catalogue \citep{Koester2007}, using X--ray and weak lensing mass estimates, \citet{Rozo2009a} found a relatively high scatter for the mass--richness relation at fixed richness $N_{200}$, $\sigma_{lnM|N_{200}}\sim0.45$.
Later,  \citet{Rozo2009b} adopted an optimised cluster richness estimator $\lambda$, estimated considering 
the radial cluster density profile, the cluster luminosity function and the cluster galaxy colour distribution. 
When assuming the optimised $\lambda$ richness estimator, \citet{Rozo2009b} found $\sigma_{ln L_X|\lambda}=0.69$, representing a significant improvement with respect to the X--luminosity--$N_{200}$ relation, $\sigma_{ln L_X|N_{200}}=0.86$. 
\citet{Rykoff2012} analysed different possible sources of increased scatter in the mass--richness relation and optimised the $\lambda$ estimator, finding a smaller value for the scatter $\sigma_{ln L_X|\lambda}\sim0.63$, which corresponds to a scatter in mass at fixed richness $\sigma_{ln M|\lambda}\sim 0.3$.

With the goal of minimising the scatter in the mass--richness relation and to find the best optical mass proxy, several optical richness definitions have been adopted in the literature. These definitions  can be divided in two main groups. For the first group, the richness estimate is based on galaxy counts within a given magnitude range in a given spatial region \citep[e.g.][]{Koester2007,Andreon2010}. For the second group,  the richness is measured from the galaxy spatial distribution, assuming  cluster profile models as  the Navarro--Frenk--White \citep[NFW; ][]{Navarro1996} and the galaxy luminosity function, such as the Schechter \citep{Schechter1976} luminosity function  \citep[e.g.,][]{Postman1996,High2010,Rozo2009b,Ascaso2012,Rykoff2014}.

We define our richness estimate in the following way: 
\begin{itemize}
\item We count for each redshift bin, the number of red ETGs (as defined above) in a given scaling radius $R_{scale}$ and brighter than $0.2\times {\rm L^*}$; 
\item We set an initial cluster candidate comoving size $R_{scale}=1.0\ {\rm Mpc}$ and we estimate the corresponding richness;
\item  We iteratively scale $R_{scale}$ according to the relation $R_{scale}=(\lambda/100)^{\beta}$, with $\beta=0.2$, until the difference in richness for two successive iterations is less than $N_{bkg}$.
\end{itemize}

Following \citet{Rykoff2014}, we adopt $\beta=0.2$. {\it A posteriori}, we test different values of $\beta$ and we find that $\beta=0.2$ minimise the number of {\itshape RedGOLD} detections without an X--ray counterpart in the Gozaliasl's catalogue.
Typical values of  $R_{scale}$ are between 0.5 and 1.0~Mpc.

At each iteration, we subtract the background contribution that corresponds to the same area and to the redshift bin in which the galaxy counts are computed.

\subsection{Concentration parameter}\label{sec:conc}

Galaxy clusters are characterised by
similar radial profiles and the galaxy density in the cluster centre can be an order of magnitude (or
more) higher than that in the peripheral regions (e.g. following the dark matter NFW profile). To take this into account,  we impose an additional constraint on the radial distribution of the red--sequence galaxies \citep[see also][]{Rykoff2014}.  Doing this, we are assuming that red galaxies follow the same profile distribution as dark matter. Our assumption is justified both theoretically by the fact that galaxies are collissionless as dark matter, and observationally \citep{Lin2004, Collister2005, Holland2015}. Because of the singularity of the NFW profile at $R=0$, we adopt a core radius $r_{core} = 0.1\ h^{-1}$ Mpc and
we assume that the surface density profile is constant for $r \lesssim r_{core}$, following \citet{Rykoff2012}.

We estimate a typical cluster NFW surface density profile $\Sigma(r)$ following \citet{Bartelmann1996},  in four radii corresponding to $R_{025}=0.25 \times {R_{200}},\ R_{050}=0.5\times {R_{200}},\ R_{075}=0.75\times {R_{200}}$,  and $R_{1}=1\times{R_{200}}, ${\footnote{In this work, we estimate $R_{200}$ fitting the relation $R_{200}-R_{scale}$ used to estimate the richness $\lambda$, for the {\it RedGOLD} detections with a counterpart in the X--ray catalogue by \citet{Gozaliasl2014} (see section~\ref{sec:compl})}} and we compare the ratios of the observed surface density profile estimated at different radii  with the value predicted by the NFW profile.

We adopt the mass--redshift--concentration relation from \citet{Duffy2008}:
\begin{equation}
c=A \left (\frac{M}{M_{piv}} \right)^B(1+z)^C
\end{equation}
with $M_{piv}=2\times 10^{12}\ {\rm h^{-1}M_{\odot}}$. \citet{Duffy2008} estimated  the best--fit parameters A, B and C for {\it relaxed} systems ($A=6.71\pm0.12$, $B=-0.091\pm0.009$, $C=-0.44\pm0.05$), and for the {\it full} cluster sample ($A=5.71\pm0.12$, $B=-0.084\pm0.006$, $C=-0.47\pm0.04$), for the dark matter halo mass range $10^{11}-10^{15}\ {\rm M_{\odot}}$ and the redshift range $0<z<2$. These values are in agreement with recent work
in the literature \citep[e.g.][]{Dutton2014, Klypin2014}. 

We identify cluster candidates allowing the {\it full} cluster sample concentration to vary  within $3\sigma_c$, with $\sigma_c$ being the uncertainty on the concentration $c$ from propagation of the uncertainties on $A$, $B$, and $C$ given above. 
In particular, we compare the ratios between the observed surface density profile at the four different radii ($\Sigma(R_{025})/\Sigma(R_{050})$, $\Sigma(R_{050})/\Sigma(R_{075})$ and $\Sigma(R_{075})/\Sigma(R_{1})$) with the theoretical values obtained above, and retain only the {\it RedGOLD} detections with observed profiles consistent with the NFW profile for all four radii, in the NWF profile range within $3\sigma_c$.

None of our cluster candidates are over--concentrated, i.e. show a concentration larger than $c+3\sigma_c$.
Imposing our limit on the galaxy radial profile, we discard ~6\% of our candidates, all with a shallower (with respect to \citet{Duffy2008}) galaxy distribution. We visually check these
excluded detections and we find that they are poorer and smaller systems.

\section{Completeness and purity of our algorithm} \label{sec:compl}

\subsection{Completeness versus Purity} \label{sec:complpur}
In the literature, there are different methods to detect galaxy clusters. However,  
any technique suffers for some incompleteness and contamination effects: 
for each algorithm, it is necessary to find a good compromise between {\itshape completeness} and {\itshape purity}. 
The {\itshape Completeness}  is defined as the ratio of detected structures which correspond to a true cluster $N^{det}_{true}$ to the total number of true clusters $N_{true}$:

\begin{equation}
Completeness = \frac{N^{det}_{true}}{N_{true}}\ .
\end{equation}

The {\itshape Purity} is the total number of detection $N_{det}$ minus the fraction of false detections  $N_{false}$ to the number of detected objects.

\begin{equation}
Purity = \frac{N_{det}-N_{false}}{N_{det}}
\end{equation}

The completeness quantifies how well a detection algorithm is able to find true clusters (i.e. the probability that a detection algorithm will detect true clusters), while the purity estimates 
the percentage of true clusters (as opposite to false detections) detected by the algorithm (i.e. the probability that a detection corresponds to a true cluster).
These are two key quantities  to determine the goodness of any cluster catalogue and  the ideal algorithm is characterised by simultaneously high values of completeness 
and purity.

In practice, it is very difficult to maximise both quantities at the same time and it is common to find instead a good compromise. 
To find the best compromise between completeness and purity,
 we first test {\itshape RedGOLD} on semi-analytic simulations and then on already known X--ray detected clusters.

In both definitions of completeness and purity, it is important to define a true and a false cluster.  Following the literature  \citep[e.g, ][]{Finoguenov2003, Lin2004, Evrard2008, Finoguenov2009, McGee2009, Mead2010, George2011, ChiangOverzier2013, Gillis2013, Shankar2013}, we define a true cluster  as a dark matter halo more massive than $10^{14}\ \rm M_{\odot} $, since numerical simulations show that 90$\%$  of the dark matter haloes more massive than $10^{14}\ M_{\odot}$ are a very regular virialised cluster population up to redshift $z\sim1.5$  \citep[e.g.,][]{Evrard2008, ChiangOverzier2013}. We define true galaxy groups, dark matter haloes with mass  $10^{13}\ M_{\odot} < M < 10^{14}\ M_{\odot}$. Within our definitions, dark matter haloes with lower masses are not considered as a group or cluster detection, but as field galaxies. 

Since we want to optimise {\it RedGOLD} to detect galaxy clusters, we estimate its completeness with respect to dark matter haloes more massive than $10^{14}\ \rm M_{\odot}$. However, because of the scatter $\sigma_{mass}$ in the scaling relations between cluster dark matter halo mass and measured mass proxies \citep{Rozo2014b,Rozo2014c}, we cannot consider as false detections the cluster candidates with mass within $\sim 3 \times \ \sigma_{mass}$ from a typical scaling relation.  As discussed in section~\ref{sec:rich} the typical scatter in the observed mass scaling relations is $\sigma_{ln M|M_{proxy}}\sim 0.3$, where $M$ is the mass estimate for the dark matter halo, and $M_{proxy}$ is the used mass proxy (e.g. $L_X$, $N_{200}$, $\lambda$, etc). For this reason, we estimate the purity of our algorithm with respect to dark matter haloes more massive than $10^{13}\ M_{\odot}$. We have also tested a lower limit in mass, and when estimating the purity with respect to dark matter haloes more massive than $10^{12.5}\ \rm M_{\odot}$, its estimated value changes  by only $9\times10^{-4}$.

\subsection{Completeness and purity of our algorithm from Millennium Simulation}\label{sec:MS}
 
We apply our detection algorithm to the Millennium Simulation \citep{Springel2005}: 
among the different realisations of mock galaxy catalogues based on semi-analytic models, we use  the lightcones by \citet{Henriques2012},
which consist of 24 independent beams, and have been built from the model by \citet{Guo2011}. In fact, the \citet{Guo2011} semi-analytic model matches  the local 
SDSS luminosity and stellar mass function, obtaining a good agreement with the observations. Before using them to test {\itshape RedGOLD}, we have taken into consideration some properties of the model that can introduce systematics in our detection procedure.
In fact, although many improvements have been made with respect to previous simulations (e.g. the stellar mass function), 
the Guo model still shows some discrepancies with the observations: in particular, galaxy colours are difficult to reproduce 
in an accurate way, since they depend on different parameters, as metallicity, star-formation history and dust.

\citet{Guo2011} showed that at z=0 there is a discrepancy between the colours predicted in their models and the SDSS observations, over-predicting the fraction of red dwarf galaxies ($M<10^{9.5}\ {\rm M_{\odot}}$), with colours redder than observed. On the other hand, at $M>10^{10.5}\ {\rm M_{\odot}}$, the colours are bluer with respect to the observations.

Since \citet{Guo2011} simulated ETG spectral energy distributions for galaxies with $B/T\ge 0.7$ (see also \citet{Shankar2014}), we select galaxies with $B/T\ge 0.7$ as ETGs. We find that the ETG abundance 
in galaxy clusters is not well reproduced, and the ETG fraction is systematically underestimated.
Clearly, this deeply affects the results obtained with our algorithm, since it relies on the search of 
red--sequence ETGs. 

This discrepancy affecting semi-analytic models has been already noted in previous works:
for example, \citet{Cohn2007} compared the red-sequence of simulated galaxies in the Millennium Simulation \citep{Springel2005,Croton2006, Kitzbichler2007}  with the observations, finding that the simulated red-sequence
has a larger scatter and a positive slope while the observed slope is negative. Also \citet{Hilbert2010} investigated the effect of this discrepancy between
models and observations, finding that it is crucial to correct for it when using optical cluster finding algorithms with simulations. In particular, they measured 
the mean colour of red--sequence galaxies for mock galaxies as a function of redshift and they compared it with the same relation obtained from SDSS galaxies,
finding that the mean red--sequence galaxy colour obtained from semi-analytic models is quite close to the SDSS ones at very low redshift,
but the discrepancy between the two become significant at higher redshifts. They explicitly noted that, without any adjustment in colours,
they would have not found almost any clusters at $z > 0.25$.

As our detection technique relies on the search of the red--sequence ETGs, we have to take into account these effects. 
As a consequence, to have a reliable estimate of completeness and purity, we correct the mock catalogues in order to 
obtain a realistic galaxy type distribution in simulated clusters, and accurate colours. 

For the lightcones by \citet{Henriques2012}, up to the galaxy luminosities that we are considering
in this work, we find that:

\begin{itemize} 
\item all clusters have a negligible fraction of  bulge-dominated red galaxies; $\sim 70\%$ of haloes more massive than $10^{14}\ {\rm M_{\odot}}$ and at $z \le 1.1$ have less than 5 bulge-dominated 
members with colours matching the BC03 predictions for passive galaxies  \citep[see also][]{Ascaso2015}; on the other hand, observed clusters up to $z\sim1.5$ show ETG fractions of $70-80\%$ \citep[e.g.,][]{Postman2005, Desai2007, Mei2009, Mei2012, Brodwin2013}; 
\item $\sim 10\%$ of the simulated clusters show positive slope (while observations show negative slopes) and/or wider scatter of red-sequence galaxies with respect to observations;
\item at a given redshift, on the red-sequence there is a shift between simulated ETG colours from the semi-analytic model and ETG colours predicted from the BC03 stellar population models, with the lightcone colours being bluer.
\end{itemize}

\begin{figure}
    \begin{center}
        \includegraphics[width=0.8\columnwidth]{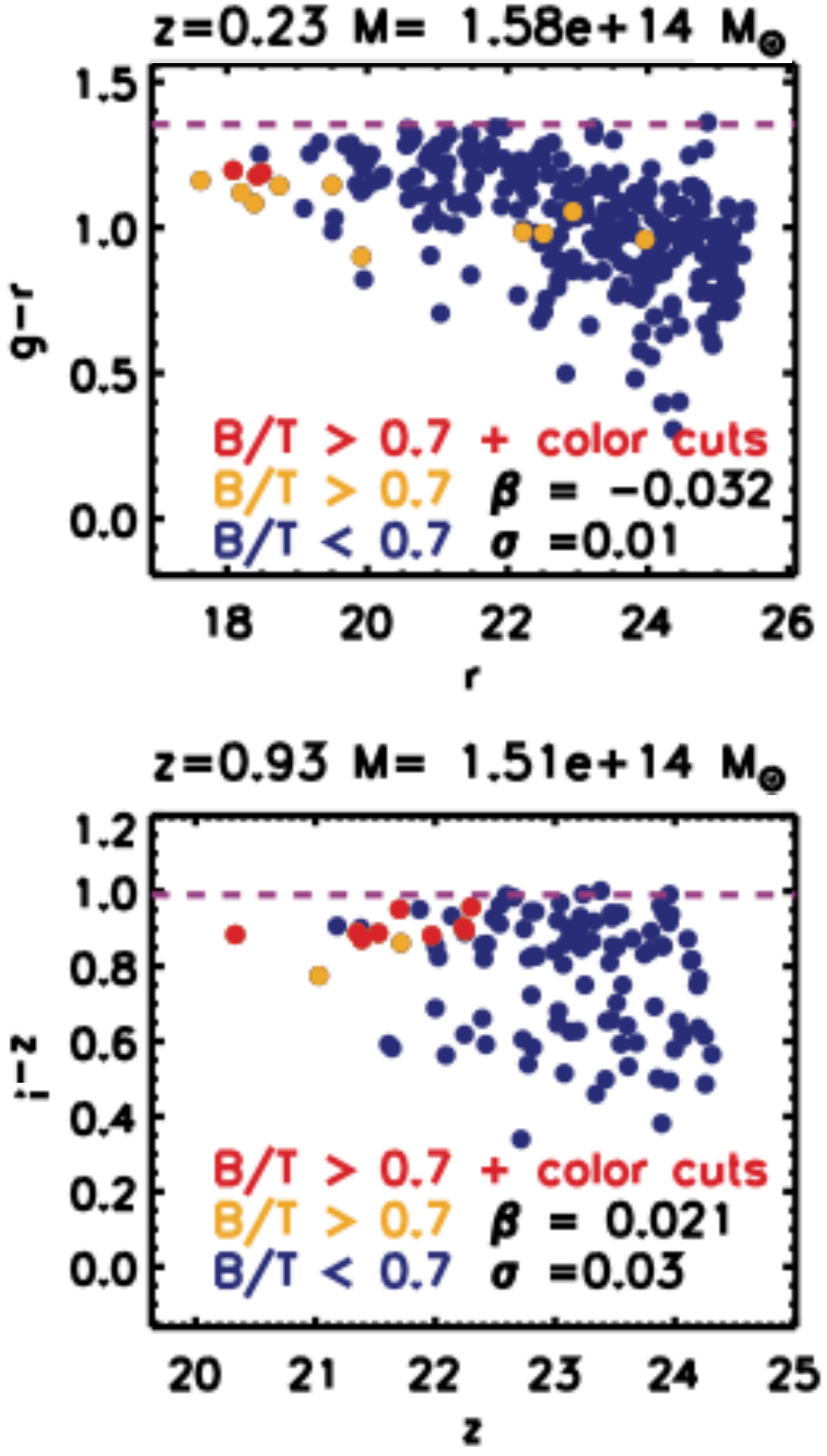}    
      \end{center}
    \caption{Original colour-magnitude relation of two clusters in the \protect\citet{Henriques2012} lightcones at $z=0.23$ (top panel) and $z=0.93$ (bottom panel). Blue circles are cluster members, orange symbols represent members with $B/T \ge 0.7$ (i.e. ETGs) and red circles are ETGs characterised by colours in agreement with predictions
     from BC03 models for passive galaxies at the same redshift. The purple dashed line shows the mean colour predicted by the BC03 models.  The $\sigma$ and $\beta$ values refer to the red--sequence scatter and slope.}
    \label{fig:ExHenOr}
  \end{figure}

Figure \ref {fig:ExHenOr}  shows two examples of the colour-magnitude relation for two massive clusters in the lightcone catalogues,
at redshift $z=0.23$ and $z=0.93$: blue circles are cluster members, 
orange circles represent ETGs, i.e. members with $B/T \ge 0.7$, and red circles are ETGs with colours
in agreement with those predicted by BC03 models for passive galaxies at the same redshifts. The purple dashed line shows the ETG colour predictions by single burst BC03 models, assuming passive evolution, $z_{form}=3$  and solar metallicity. Both problems are clearly visible: the total number of ETGs is negligible and only a small fraction of  ETGs matches the BC03 predicted colours, leading, in one of these two cases, to a positive red-sequence slope.

These results imply that we have to correct for both the colour mismatch of red ETGs and the low fraction of early--type galaxies.
In the following section, we will describe how we implement these corrections and give a final estimate for purity and completeness.

\subsubsection{Mock catalogue corrections}

The first modification that we need to apply to run {\itshape RedGOLD} on the \citet{Henriques2012}  lightcones is to obtain red-sequence colours from the simulations to identify red overdensities (instead of using predictions from the BC03 models, that are inconsistent with the red-sequence in the simulations). 

Since we should modify colours for both early and late type galaxies in all environments,
we do not change the colours in the mock catalogues to avoid to introduce biases in the galaxy properties and their large-scale distribution.
For this reason,  instead of changing the colours in the lightcones, we estimate the expected red-sequence ETG colours used by {\itshape RedGOLD} to match the  \citet{Henriques2012} red-sequence colours. 

We consider all ETGs (objects with $B/T\ge 0.7$) brighter than $0.2\times{\rm L^*}$ from the lightcone cluster catalogues, in narrow redshift slices of 0.05, and we build the histogram of galaxy colours for each redshift bin. 
In each redshift bin, we fit this distribution with a Gaussian and we obtain its mean $\bar{c}$ and its standard deviation $\sigma_{col}$ as a function of redshift.
These are the expected red-sequence ETG colour and its intrinsic scatter, which we use for our {\it RedGOLD}  red overdensity detections.

The second discrepancy in the  \citet{Henriques2012} simulation is the low fraction of ETGs on cluster red-sequences (i.e. galaxies with $B/T \ge 0.7$): in fact, since {\itshape RedGOLD} detects red early-type galaxy overdensities, 
it is necessary that the clusters in the simulation have realistic ETG fractions. 

Although Guo models are able to reproduce the galaxy distribution for different 
morphological types in the local Universe \citep[see Figure 4 in][]{Guo2011}, 
there is a lack of early-type galaxies in clusters at $z \apprge 0.1$. 
In observed clusters, groups and the field up to $z\sim1.5$, the ETG fraction is of $\sim 70\% \pm 10\%$, $\sim 50\% \pm 10\%$, and $\sim 30\% \pm 10\%$, respectively, up to the magnitude limits considered in this work 
\citep[e.g.,][]{Treu2003, Desai2007, Postman2005, Smith2005, Mei2009, George2011, Mei2012}. 
To correct for this discrepancy, we modify \citet{Guo2011} galaxy morphologies in the cluster, group and field red-sequence, to reproduce these observed fractions.
Since our detection code searches for red ETG overdensities, if we modify the ETG fractions only in clusters,
we would obtain optimistic values for the completeness and purity, as groups and field ETGs are not enhanced.
For this reason, to have a coherent scenario, we also modify the ETG fraction in groups and in the field.

We distribute the cluster, group and field ETGs around the mean red-sequence colour, following a Gaussian distribution with standard deviation equal to the intrinsic red-sequence scatters that we have derived above for the lightcones (i.e. $68\%$ of the ETGs will be distributed in 1~$\sigma_{col}$). Since we modify the percentages in the same way at all luminosities, we do not expect to change in a significant way the shape of the ETG luminosity function in the luminosity range considered for the cluster detection with {\itshape RedGOLD}.

\begin{figure}
    \begin{center}
        \includegraphics[width=0.8\columnwidth]{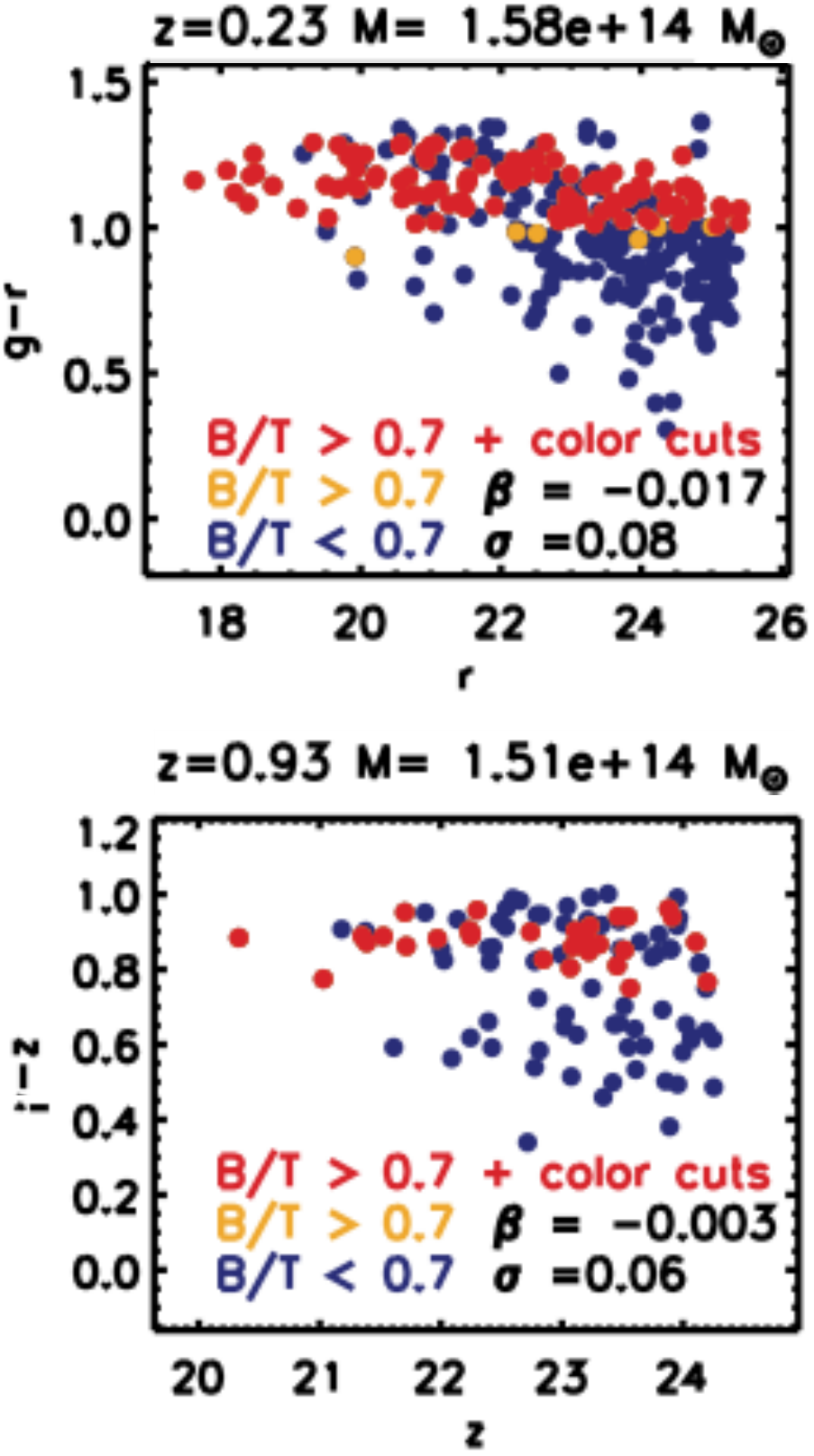}    
      \end{center}
    \caption{Modified colour-magnitude relation of the two clusters in the \protect\citet{Henriques2012} lightcones at $z=0.23$ (top panel) and $z=0.93$ (bottom panel) shown in  Fig.~\ref{fig:ExHenOr} after our correction procedure. Symbols are the same as Fig.~\ref{fig:ExHenOr}.
     Within  {\itshape RedGOLD}, we use the average red-sequence colours at each redshift from the \protect\citet{Henriques2012}, and modify the ETG 
    fractions to be consistent with the observations. The $\sigma$ and $\beta$ values refer to the red--sequence scatter and slope.} 
    \label{fig:ExHenMod}
  \end{figure}

In Figure \ref {fig:ExHenMod}, we show the corrected colour-magnitude-relation for the two clusters shown in Fig. \ref{fig:ExHenOr}, after applying this procedure. For the {\itshape RedGOLD} detection procedure, we use the average red-sequence colours at each redshift from  \citet{Henriques2012}  colours, and modify ETG fractions to be consistent with the observations. 
When these corrections are applied, we find that only $\sim 5\%$ of clusters have less than 5 ETGs or wrong values
for the red-sequence scatter and/or slope, and in all cases they are massive structures lying  on the edges of the lightcones.
 When applying these corrections, the red--sequence is well reproduced: both the red--sequence scatter $\sigma$ and slope $\beta$ are in agreement with the observations.

\subsubsection{Magnitude and colour uncertainties}

Since we want to have galaxy simulations with a photometric accuracy that is representative of the CFHTLenS data, we modify simulated galaxy magnitudes from the \citet{Henriques2012} lightcones, to reproduce the CFHTLenS photometric errors.

We convert the SDSS magnitudes in the \citet{Henriques2012} catalogues in CFHT/MegaCam magnitudes, $u^*, g, r, i, z$, following \citet{Ferrarese2012}. For each bandpass, we then compute the mean photometric error $\bar{\epsilon}$ and the corresponding uncertainty $\sigma_{\epsilon}$ 
(in magnitude bins of 0.1 mag) in the CFHTLenS data, and we use them to correct magnitudes and errors in  \citet{Henriques2012} catalogues.
 We add the mean error to each simulated magnitude, following a Gaussian distribution, and randomly assign magnitude uncertainties as a function of magnitude.

For the same reason, we modify the redshifts in the lightcones to reproduce the same accuracy of the CFHTLenS redshift estimates: in particular, we change the photometric redshifts randomly extracting values from a Gaussian centred on the true redshift value and with a standard deviation $\sigma_{photoz}$, following the values reported in \citet{Raichoor2014} as a function of magnitude. 
Following the same procedure used to add uncertainties to magnitudes, we add uncertainties in photometric redshifts. To reproduce the outlier fraction as observed in the CFHTLenS, we assign to a percentage of objects, that corresponds to the observed percentage of outliers, a random photometric redshift, which differs from the original of $\Delta z >0.15$ (according to the definition of outliers).

\subsubsection{CFHTLenS masked regions}

To test the effect of the masks on our detections, we build a second series of simulation to take into account the CFHTLenS masked regions from \citet{Erben2013},  which include both masked regions without any source detections (e.g. empty regions/holes), and masked regions with higher photometry uncertainties. 
Firstly, we build an empirical size distribution of the holes and of the regions with photometric uncertainties higher than the average (i.e. the observed masked regions) from the CFHTLenS.
Then, we add to our modified Millennium Simulations random masked circular regions extracted from this distribution. We assign to the galaxies in the regions with photometric uncertainties higher than  the average, a random distribution of uncertainties derived from the one observed in the CFHTLenS corresponding masked regions. To do so, we build an uncertainty distribution for each magnitude bin, and calculate its mean and standard deviation.

We call these simulations, the masked modified Millennium. We run  {\itshape RedGOLD}  on both the modified Millennium Simulation (i.e. without masks) and the masked modified Millennium. As explained in section~\ref{sec:overdens}, in both cases we select only objects with an error in
photometry within the average distribution.

\subsubsection{Results}

Our main goal is to test {\itshape RedGOLD}  as red ETG overdensity cluster detector (steps described in the first three subsections of section~\ref{sec:technique}), applying it to the simulations. We run {\itshape RedGOLD} on the modified \citet{Henriques2012} galaxy catalogues, and obtain a cluster candidate catalogue. For each detection, we obtain position, redshift and detection significance, and 
estimate purity and completeness as a function of significance, redshift, and halo mass. 

We do not impose a cluster profile and do not estimate richness in this simple test. In fact, since the mock catalogues show a lack of bright red--sequence galaxies, richness measurements are biased towards lower values, and are not correlated with dark matter halo mass in the same way as in the observations. Moreover, since we use the scaling radius $R_{scale}$ (estimated from the cluster richness) to derive $R_{200}$, we cannot impose any limit on the cluster profile.
As a consequence, the results obtained using the Millennium Simulations might represent a pessimistic scenario.

\begin{figure}
    \begin{center}
        \includegraphics[width=\columnwidth]{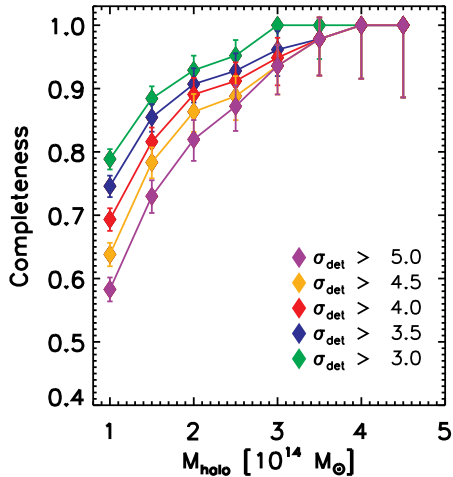}    
      \end{center}
    \caption{Completeness as a function of the halo mass  in the entire redshift range $0<z<1.1$ obtained using the lightcones by \protect\citet{Henriques2012}. Green, blue, red, orange and purple symbols refer to $\sigma_{det}\ge3, 3.5, 4, 4.5, 5$, respectively. Our completeness is always $>80\%$ for the most massive clusters 
($M>2.5 \times 10^{14}\ {\rm M_{\odot}}$), and does not change significantly for different values of $\sigma_{det}$.  In the mass range $10^{14} <M<2.5\times 10^{14} \ {\rm M_{\odot}}$, the completeness changes significantly
when considering different detection significance thresholds.}
    \label{fig:ComplHaloMMill}
  \end{figure}

\begin{figure}
    \begin{center}
        \includegraphics[width=\columnwidth]{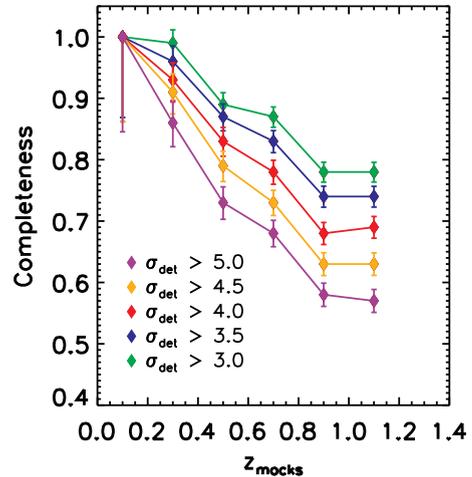}    
      \end{center}
    \caption{Completeness as a function of the redshift for haloes more massive than $10^{14}\ M_{\odot}$, obtained using the lightcones by \protect\citet{Henriques2012}. Symbols are as in Fig. \protect\ref{fig:ComplHaloMMill}. At low redshift ($z\lesssim 0.4$) {\itshape RedGOLD} is always $>80\%$ complete for all the considered $\sigma_{det}$ values. At higher redshift, though, increasing the detection significance corresponds to higher difference in the completeness as a function of $\sigma_{det}$.}
    \label{fig:ComplZMill}
  \end{figure}

To match the {\it RedGOLD} cluster candidates with the simulated dark matter haloes, we adopt a maximum projected distance between the centres corresponding to $R_{200}$ and $\Delta z=|z_{sim}-z_{RedGOLD}|\le3\times \sigma_{photoz}=3\times 0.03\times(1+z)$, where $z_{sim}$ is the cluster redshift in the simulations and $z_{RedGOLD}$ is the cluster redshift
estimated by our algorithm.

In Fig. \ref{fig:ComplHaloMMill} and  Fig. \ref{fig:ComplZMill}, we show the cluster completeness as a function of the dark matter halo mass in the entire redshift range $0<z<1.1$ and as a
function of the redshift  for haloes more massive than $10^{14}\ M_{\odot}$, respectively, for different values of the detection significance, $\sigma_{det}$. Green, blue, red, orange and purple symbols refer to $\sigma_{det}\ge3, 3.5, 4, 4.5, 5$, respectively.

The error bars represent the uncertainties estimated following \citet{Gehrels1986}. These approximations provide the lower and upper limit of a binomial distribution within the $84\%$ confidence limit (i.e. $1\sigma$) and hold even when the
completeness and the purity are estimated from small numbers (e.g. at high mass or low redshift). Using this conservative approach, our uncertainties are slightly overestimated \citep{Cameron2011}.

We define as clusters all dark matter haloes with mass $M_{halo}\ge10^{14}\ {\rm M_{\odot}}$ (see section~\ref{sec:compl}).

When we consider the entire redshift range $0<z<1.1$, our completeness is always $>80\%$ for the most massive clusters 
($M_{halo}>2.5 \times 10^{14}\ {\rm M_{\odot}}$), and does not change significantly for different values of $\sigma_{det}$. On the other hand, in the mass range $10^{14} <M_{halo}<2.5\times 10^{14} \ {\rm M_{\odot}}$, the completeness changes significantly
when considering different detection significance thresholds: at $\sigma_{det}\ge5$, {\itshape RedGOLD} misses $\sim40\%$ of  the less massive clusters ($M_{halo}\sim 10^{14}\ {\rm M_{\odot}}$).  When we consider all masses ($M_{halo}>10^{14}\ {\rm M_{\odot}}$), at low redshift ($z\lesssim 0.4$) {\itshape RedGOLD} is always $>80\%$ complete from $\sigma_{det}\ge3$ to $\sigma_{det}\ge5$. At higher redshift, though, increasing the detection significance corresponds to larger 
 differences in the completeness as a function of $\sigma_{det}$ (Fig.~\ref{fig:ComplZMill}).

\begin{figure}
    \begin{center}
        \includegraphics[width=\columnwidth]{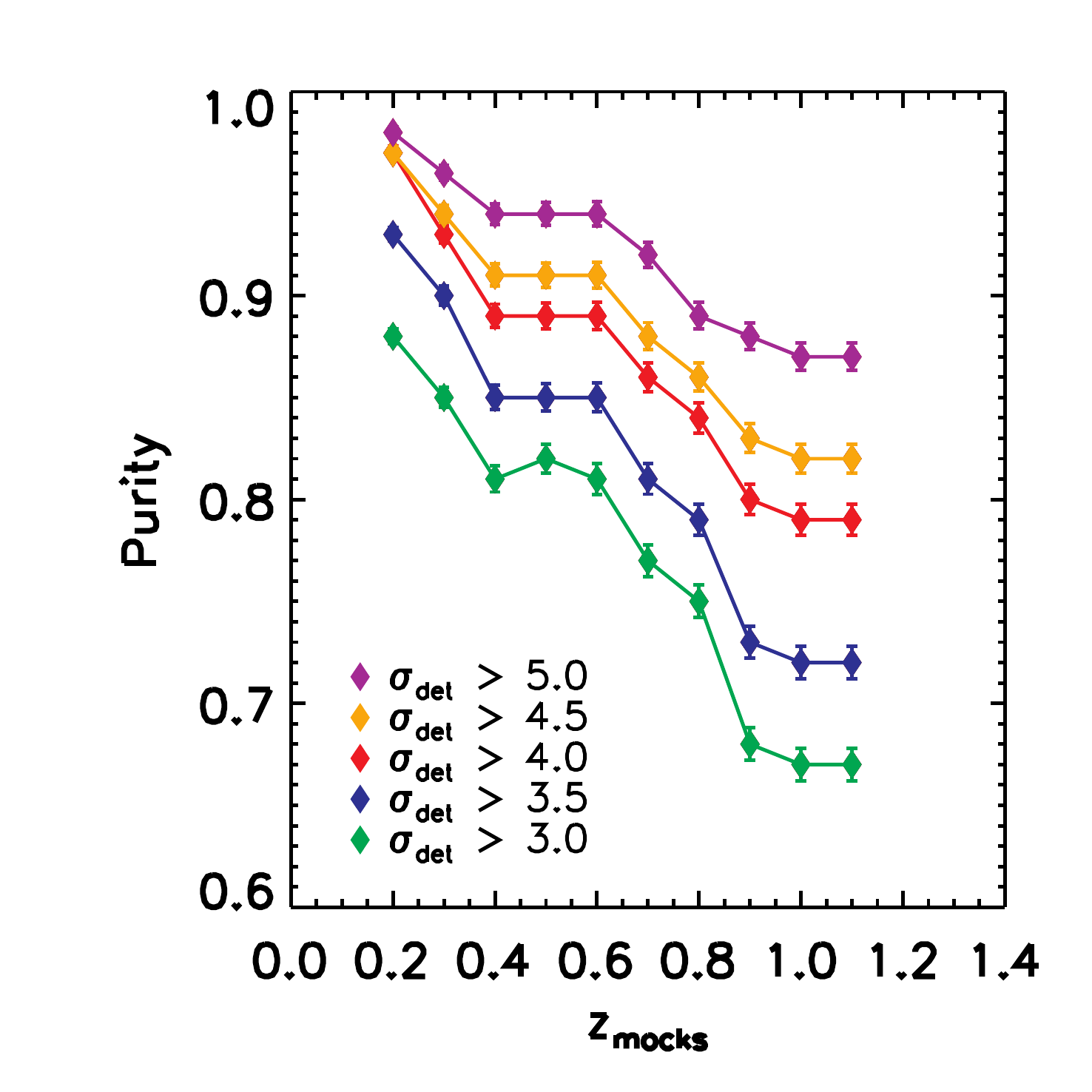}    
      \end{center}
    \caption{Purity as a function of the redshift, obtained using the lightcones by \protect\citet{Henriques2012}. Symbols are as in Fig. \protect\ref{fig:ComplHaloMMill}. The purity is  $\gtrsim 80\%$ up to redshift $z\sim1.2$ and $\sigma_{det} > 4$.} 
    \label{fig:PurZMill}
  \end{figure}
  
In Fig.~\ref{fig:PurZMill}, we plot the purity as a function of the redshift. To estimate the purity, we consider all detected haloes with more than five members  and more massive than $M_{halo} =10^{13}\ {\rm M_{\odot}}$ (see section~\ref{sec:complpur}). Similar choices have been adopted in previous work \citep{Milkeraitis2010, Soares-Santos2011}. 

As in Fig.~\ref{fig:ComplZMill}, we show our results as a function of the redshift and the detection significance. 
The purity as a function of redshift reaches higher values for higher $\sigma_{det}$ thresholds, as expected. For $\sigma_{det}\ge5$,
{\itshape RedGOLD} is pure at $>90\%$ at all redshifts, but, as shown in Fig.  \ref{fig:ComplHaloMMill} and \ref{fig:ComplZMill}, 
the completeness is significantly lower than for other thresholds. In all cases, the purity is  $\gtrsim 80\%$ up to redshift $z\sim1.2$ and $\sigma_{det} > 4$. This means that even if we reach a relatively low completeness ($\sim65\%$) in detecting clusters at $1<z<1.2$, we can still obtain a very high purity at this significance. 

At $\sigma_{det}\ge4, 4.5$ the purity is comparable with that reached considering  $\sigma_{det}\ge5$, 
being $>80-85\%$ in the whole redshift range. 
At $\sigma_{det}\ge 3, 3.5$, the purity starts to be significantly lower, especially at $z\gtrsim 0.6$. 
To keep a purity $>80\%$ up to $z\sim1$, our results show that we require a $\sigma_{det}\ge4$.

\begin{figure}
    \begin{center}
        \includegraphics[width=\columnwidth]{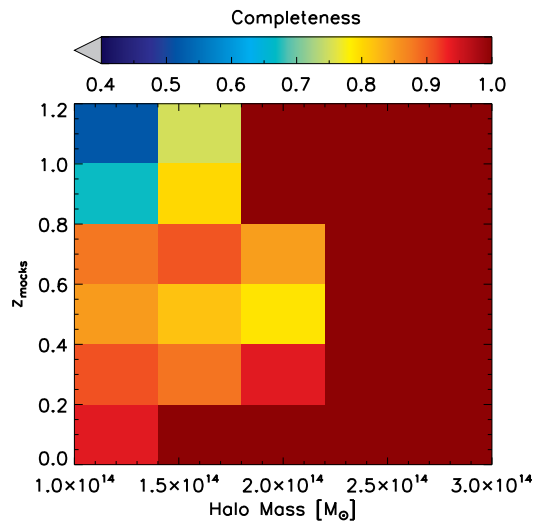}    
      \end{center}
    \caption{Completeness as a function of the halo mass and the redshift, obtained using the lightcones by \protect\citet{Henriques2012}, and assuming $\sigma_{det}\ge4$. The  completeness is $>80\%$ for $M_{halo} \gtrsim 2 \times 10^{14}\ {\rm M_{\odot}}$ and $z<1.1$. For  $ 10^{14}\ {\rm M_{\odot}} <M_{halo} \lesssim 2 \times 10^{14}\ {\rm M_{\odot}}$, it decreases at $65-70\%$ at $z>0.8$, and significantly depends on the halo mass. }
    \label{fig:ColMap}
  \end{figure}

Fig.~\ref{fig:ColMap} shows the completeness as a function of the halo mass and the redshift, 
assuming $\sigma_{det}\ge4$.
 {\it RedGOLD} always reaches a  completeness $>80\%$ for $M_{halo} \gtrsim 2 \times 10^{14}\ {\rm M_{\odot}}$ and $z<1.1$. For  
$ 10^{14}\ {\rm M_{\odot}} <M_{halo} \lesssim 2 \times 10^{14}\ {\rm M_{\odot}}$, the completeness decreases at $\sim65-70\%$ at $z>0.8$, and significantly depends on the halo mass.

When running {\itshape RedGOLD}  on the masked modified Millennium, the recovered purity and completeness levels do not differ from those obtained without considering the masked regions.

\subsection{Completeness and purity of our algorithm from X--ray detected clusters}\label{sec:complpurX}

To optimise the values of  {\itshape RedGOLD} $\lambda$ and $\sigma_{det}$ using observations, we run the algorithm on the CFHTLenS data, and compare the obtained cluster catalogues with the X--ray confirmed galaxy clusters from \citet{Gozaliasl2014}. 

\begin{figure}
    \begin{center}
        \includegraphics[width=\columnwidth]{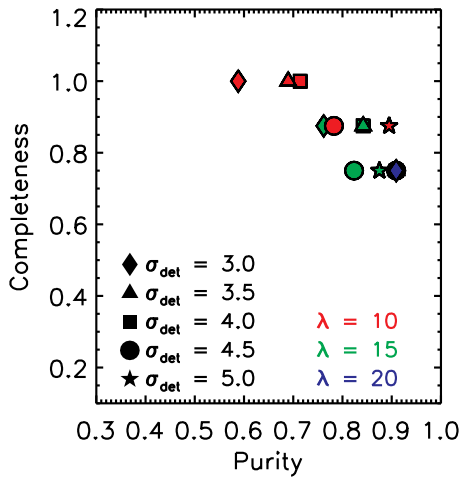}    
      \end{center}
    \caption{Completeness as a function of the purity for different thresholds of $\lambda$ and $\sigma_{det}$ up to $z\sim0.6$ for the full \citet{Gozaliasl2014} sample. Red, green and blue symbols represent $\lambda\ge 10, 15, 20$ while diamonds, triangles, squares, circles and stars indicate $\sigma_{det}\ge 3, 3.5, 4, 4.5, 5$, respectively.}
    \label{fig:ComplPur_lam_var_zlim_All}
  \end{figure}

In this case, the completeness is estimated with respect to the X--ray detected catalogue from \citet{Gozaliasl2014} as the ratio between the number of X--ray detected clusters with $M_{200}\ge10^{14}\ {\rm M_{\odot}}$ recovered by {\itshape RedGOLD} to the total number of X--ray detections with $M_{200}\ge10^{14}\ {\rm M_{\odot}}$.
 Similarly, the purity is the ratio between  the number of detections found by {\itshape RedGOLD} with an X--ray counterpart in the Gozaliasl's catalogue to the total number of the {\itshape RedGOLD} detections. Our estimated purity is a lower limit, because the Gozaliasl's catalogue purity and completeness are not published, and, as we show below, their catalogue is not complete at their mass limit. 
 To optimise these two quantities, we test different values of each parameters and we retain  those that maximise both completeness and purity. 

To match the {\it RedGOLD} cluster candidates with the X-ray detected catalogue by \citet{Gozaliasl2014}, we
adopt a maximum projected distance between the centres corresponding to $R_{200}+\sigma_{R200}$, where
$\sigma_{R200}$ is the estimated error on the $R_{200}$ measurement.
Moreover, we require that the maximum
redshift difference is  $\Delta z=|z_{Goz}-z_{RedGOLD}|\le3\times \sigma_{photoz}=3\times 0.03\times(1+z)$, where $z_{Goz}$ is the cluster redshift in the \citet{Gozaliasl2014} catalogue and $z_{RedGOLD}$ is the cluster redshift estimated by {\it RedGOLD}.

\begin{figure}
    \begin{center}
        \includegraphics[width=\columnwidth]{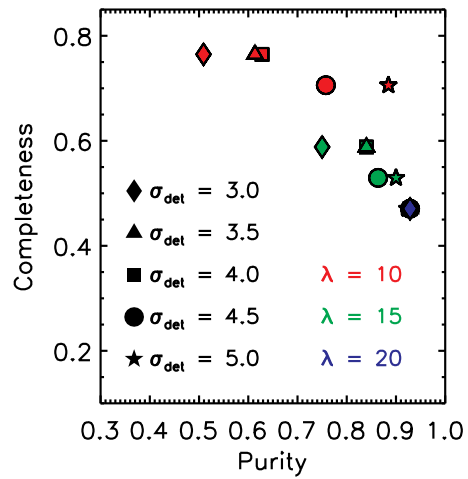}    
      \end{center}
        \caption{Completeness as a function of the purity for different thresholds of $\lambda$ and $\sigma_{det}$ in the whole redshift range for the full \citet{Gozaliasl2014} sample. Symbols are as in Fig.~\ref{fig:ComplPur_lam_var_zlim_All}.}
 \label{fig:ComplPur_lam_var_All}
  \end{figure}

Fig.~\ref{fig:ComplPur_lam_var_zlim_All} shows the estimated completeness as a function of the purity 
up to $z\sim0.6$, while Fig.~\ref{fig:ComplPur_lam_var_All} shows the results estimated in the whole redshift range, for different limits on  $\lambda$ and  $\sigma_{det}$.
Red, green and blue colours refer to $\lambda \ge 10, 15, 20$, respectively, while  diamonds, triangles, squares, circles and stars refer to $\sigma_{det}\ge 3, 3.5, 4, 4.5, 5$, respectively. 
For low values of $\lambda$ and $\sigma_{det}$, the completeness is higher  but the purity reaches lower values. For $z\lesssim0.6$, the optimal values of $\lambda \ge10$ and $\sigma_{det}\ge4$ keep the completeness at
$\sim100\%$ and the purity at $>70\%$.

When considering the entire redshift range,
$\lambda\ge10$ and $\sigma_{det}\ge4.5$ are the best values to obtain a completeness of $\sim70\%$ and a purity of $\sim80\%$, and the estimated completeness is lower than that estimated at $z \le 0.6$. This is expected since  half of the X--ray detections in the Gozaliasl's catalogue with $M_{200}\ge 10^{14}\ {\rm M_{\odot}}$ is at $z\gtrsim 0.6$ and {\itshape RedGOLD} is expected to have a lower completeness at high redshift at the CFHTLenS depth, as shown in the previous section. The Gozaliasl's  sample does not include clusters at redshift $0.6<z<0.8$, for masses $M_{200}\ge 10^{14}\ {\rm M_{\odot}}$. For this reason, our lower redshift analysis stops at $z\sim0.6$.
Since we do not know the completeness of the Gozaliasl's catalogue, our estimated purity is a lower limit. As an example, one {\itshape RedGOLD}  detection without an X--ray counterpart in the Gozaliasl's catalogue is a spectroscopically confirmed structure at $z=0.33$ \citep{Andreon2004}.
Taking into account this detection, we recover a lower limit for the purity of $\sim80\%$ at $z\le0.6$.

This analysis shows that our {\itshape RedGOLD} detections are optimised in both completeness and purity for $\lambda \ge10$ and $\sigma_{det}\ge4$ at $z\le 0.6$, $\lambda \ge10$ and $\sigma_{det}\ge4.5$ for the higher redshifts.  For this parameter choice, our {\itshape RedGOLD} catalogue is expected to be $100\%$ and $70\%$ complete, at  $z\le 0.6$ and $0<z<1.1$, respectively, and $\sim 80\%$ pure, for $M_{200}\gtrsim 1 \times 10^{14}\ {\rm M_{\odot}}$.  These results are consistent with the limits in $\sigma_{det}$ that we obtain from the Millennium Simulations for clusters with masses $M_{halo}\gtrsim 1 \times 10^{14}\ {\rm M_{\odot}}$. We also note that our threshold $\lambda_{min}=10$, to obtain at least 10 bright galaxies within the scale radius, is in agreement with the literature \citep[e.g.,][]{Eisenhardt2008}.

We build our cluster catalogue considering $\lambda \ge10$ and $\sigma_{det}\ge4$ at $z\le 0.6$, $\lambda \ge10$ and $\sigma_{det}\ge4.5$ for the higher redshifts. If the reader is interested in different values of completeness and purity, we advice to change the cuts in $\lambda$ and $\sigma_{det}$.

\section{ {\itshape RedGOLD}  cluster candidate detections in the CFHT-LS W1 area}\label{sec:cfht}

\subsection{{\itshape RedGOLD} detections}

\begin{figure}
    \begin{center}
        \includegraphics[width=\columnwidth]{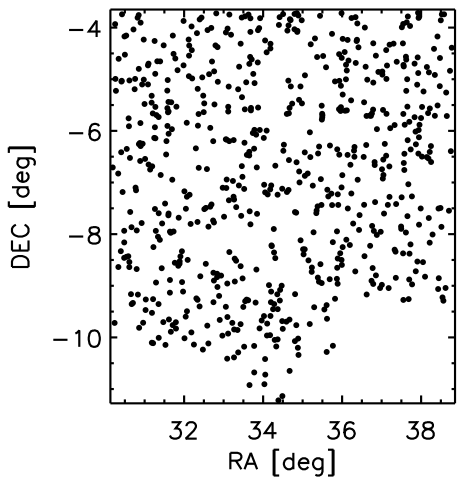}    
      \end{center}
    \caption{Spatial distribution of the {\itshape RedGOLD} CFHT-LS W1 cluster candidate detections in the $\sim 60\ {\rm deg^2}$. }
    \label{fig:spatDistrCFHTLS}
  \end{figure}
  
\begin{figure*}
    \begin{center}
   \includegraphics[width=12cm]{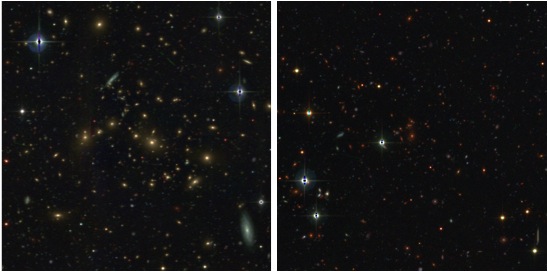}   
      \end{center}
    \caption{Optical images of two cluster candidates detected by {\it RedGOLD} at redshift
$z_{cluster}=0.19$ (left panel) and $z_{cluster}=0.44$ (right panel). Their detection significance and richness are of $\sigma_{det}=8.6$ and $\lambda=80.5$, and  $\sigma_{det}=11.1$ and $\lambda=54.1$, respectively.}
    \label{fig:cutouts}
  \end{figure*}

After applying our detection algorithm  with $\lambda \ge10$ and $\sigma_{det}\ge4$ at $z\le 0.6$, $\lambda \ge10$ and $\sigma_{det}\ge4.5$ at $z>0.6$, {\itshape RedGOLD} finds 652 
detections with $\lambda\ge 10$ up to $z\sim 1.1$ in the $\sim 60\ {\rm deg^2}$ of the CFHT-LS W1 field, i.e.
$\sim 11$ detections per $\rm deg^2$, of the same order of magnitude of theoretical predictions \citep{Weinberg2013}.  Fig. \ref{fig:spatDistrCFHTLS} shows the spatial distribution of the CFHT-LS W1 detections up to redshift $z\sim 1$. Fig.~\ref{fig:cutouts} shows two of our richest cluster candidates at $z_{cluster}=0.19$ ($\lambda=80.5$) and at $z_{cluster}=0.44$ ($\lambda=54.1$).

In $\sim18\ {\rm deg^2}$ of the area analysed in this work, published spectroscopy is available  from the SDSS, VVDS and VIPERS surveys. We find that $\sim58\%$ 
of the cluster candidates found in the same area,
imposing these lower limits on the cluster richness and the detection level, have at least one spectroscopic member in less than $1.5'$ with $|z_{spec}-z_{cluster}|<0.1$. 

 For each detection, we estimate its richness as described in section~\ref{sec:rich}. The presence of saturated objects (stars and bright galaxies) leads to larger uncertainties on galaxy photometry, and as a consequence, on photometric redshifts. To take this into account, we use the photometric error distribution in each magnitude bin from \citet{Raichoor2014}, and we exclude from the richness calculation galaxies with photometric errors larger than the average uncertainty plus three times its standard deviation (in each magnitude bin). 
 
 To test that this procedure does not significantly underestimate our richness, for each detected cluster candidate, we estimate the richness $\lambda_{mask}$, including also sources that are not included in our richness estimate because have large photometric errors in the \citet{Raichoor2014} CFHTLenS photometric catalogue.
 
Less than $7\%$ of the {\it RedGOLD} cluster candidates (obtained without imposing our lower limits on $\lambda$, $\sigma_{det}$ and the radial galaxy distribution) have a fraction of masked bright potential cluster members $>10\%$. These cluster candidates are very small systems with a mean redshift $\bar z_{cluster}=0.7$ and a mean richness $\bar \lambda_{mask}\sim8$.
If we consider only the {\it RedGOLD} detections obtained imposing our lower limits, we find that   $\sim 2$\%  have a fraction of masked bright potential cluster members $>10\%$. These detections are also small structures at high redshift, with a mean richness $\bar \lambda_{mask}=12$ and mean redshift $\bar z_{cluster}=0.7$. This means that our richness estimate is not significantly affected by the presence of the CFHTLenS masks for at least $\sim 98$\% of the cluster candidates in our final catalogue, and the fraction of masked members impacts our richness measurements only at low richness and high redshift.

Our catalogue \footnote{Our catalogue will be published with the paper} includes: RA and DEC, the cluster redshift, the detection significance $\sigma_{det}$, the cluster richness $\lambda$ and the corresponding uncertainty  $\lambda_{err}$.

In the next sections, we compare our detections with already published cluster catalogues.

\subsection{Comparison with X--ray detected cluster catalogues}

 X--ray detected cluster catalogues in the same area include: (1) the X--ray group catalogue provided by 
\citet{Gozaliasl2014} in a subarea in the CFHT-LS W1 field; (2) the X--ray catalogue provided by \citet{Mehrtens2012}, 
and (3)  a sample of 33 spectroscopically confirmed  X--ray detected 
clusters.

\subsubsection{Comparison with the X--ray catalogue by \protect\citet{Gozaliasl2014}}\label{sec:compGoz}

We have already shown the performance of {\it RedGOLD} in terms of purity and completeness with respect to the \citet{Gozaliasl2014} sample in section~\ref{sec:complpurX}.  The
\citet{Gozaliasl2014} catalogue includes 135 X--ray clusters and groups in 3~deg$^2$ in the CFHT-LS W1 area.
 In the area covered by the \citet{Gozaliasl2014} catalogue, {\it RedGOLD}  detects 38 cluster candidates, using the parameters optimised for the best simultaneous completeness and purity ($\lambda \ge10$ and $\sigma_{det}\ge4$ at $z\le 0.6$, $\lambda \ge10$ and $\sigma_{det}\ge4.5$ at $z>0.6$), and imposing an NFW profile. Of those, 28 clusters are in the  \citet{Gozaliasl2014} catalogue.

We cannot exclude that our additional ten detections without any X--ray counterpart are real galaxy groups, undetected in the X--rays. In fact, from visual inspection, they appear to be smaller systems  and could have an X--ray emission below the X--ray detection limit or without X--ray emission, if they are not-relaxed systems. As pointed out in section~\ref{sec:complpurX}, this is the case of a spectroscopically confirmed structure at $z\sim0.3$ \citep{Andreon2004}, which is
the richest {\it RedGOLD} detection without an X--ray counterpart. 
The 9 remaining detections have $\lambda<20$.

\begin{figure}
    \begin{center}
        \includegraphics[width=\columnwidth]{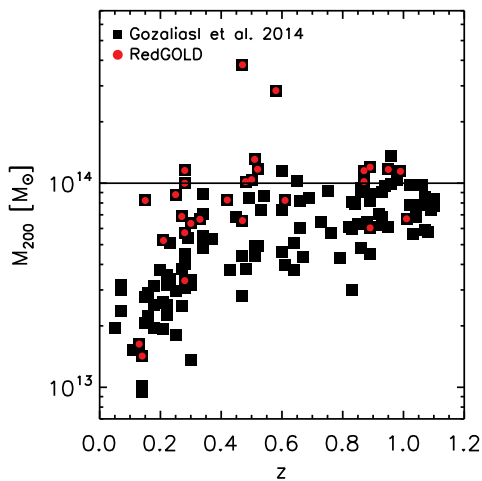}    
      \end{center}
    \caption{Cluster mass as a function of the redshift for the X--ray detected clusters from \protect\citet{Gozaliasl2014} (black squares) and for the {\itshape RedGOLD} detections with an associate X--ray counterpart (red circles). The black solid line shows the $10^{14}\ {\rm M_{\odot}}$ mass limit.}
    \label{fig:zMass_Goz}
  \end{figure}

Fig. \ref{fig:zMass_Goz} shows the cluster mass distribution as a function of the redshift for the X--ray detections from \citet{Gozaliasl2014} (black squares) and the clusters recovered by {\itshape RedGOLD} (red circles). The black solid line indicates the mass limit $M_{200}\ge 10^{14}\ {\rm M_{\odot}}$. 
 {\it RedGOLD} detects 13 of the 17 X--ray detections with $M_{200}\ge10^{14}\ {\rm M_{\odot}}$, in the entire redshift range, and all clusters with $z < 0.6$ (completeness of $\sim100\%$) in this mass range.  As already discussed in section~\ref{sec:complpurX}, this corresponds to a purity of $\sim80\%$, and a completeness of $100\%$ and $70\%$, at  $z\le 0.6$ and $0.6<z<1.1$, respectively, for $M_{200}\gtrsim 10^{14}\ {\rm M_{\odot}}$.

We examined the four unrecovered structures with $M_{200}\ge10^{14}\ {\rm M_{\odot}}$. 
All of them are at $z\ge 0.6$.
Two of the unrecovered X--ray detections at $z=0.65$ and $z=0.6$ appear to be optical poor systems. 
The other two are at higher redshift,  at z=0.96 and z=0.98, with masses $M_{200}=1.4 \pm 0.2 \times 10^{14}\ {\rm M_{\odot}}$ and $M_{200}=1.0 \pm 0.2 \times 10^{14}\ {\rm M_{\odot}}$,
respectively, where we expect our algorithm to be $\sim 65\%$ complete for our choice of parameters (see section~\ref{sec:complpurX}).

\begin{table*} 
\begin{center}
\begin{tabular}{c c c c c c}
\hline
Redshift & $N_{Gozaliasl}$& Cluster Mass & \% All Matched & \% With lower limits on $\lambda$ and $\sigma_{det}$\\
\hline
$z<0.6$  & 8 & $\ge 10^{14}\ {\rm M_{\odot}}$ & $100\%$ & $100\%$ \\
                  &16 & $\ge 7\times10^{13}\ {\rm M_{\odot}}$ & $75\%$ & $69\%$  \\ 
                  &60 & $< 7\times10^{13}\ {\rm M_{\odot}}$ & $20\%$ & $15\%$  \\ 
$z\ge 0.6$ &9& $\ge 10^{14}\ {\rm M_{\odot}}$ & $56\%$ & $56\%$  \\ 
                 &33  & $\ge 7\times10^{13}\ {\rm M_{\odot}}$ & $24\%$ & $18\%$  \\ 
                 &26  & $<7\times10^{13}\ {\rm M_{\odot}}$ & $15\%$ & $8\%$  \\ 
\hline
\end{tabular}
\caption{Comparison of our detections with the X--ray catalogue by \protect\citet{Gozaliasl2014}.}
\label{tab:Goz}
\end{center}
\end{table*}

Table~\ref{tab:Goz} summarises our results, listing the {\itshape RedGOLD} detections in the two different redshift bins for the different mass limits, without imposing any constraints on $\lambda$ and $\sigma_{det}$ in the fourth column, and considering the optimal values for the {\it RedGOLD} parameters in the last column (see section \ref{sec:complpurX}). 

\subsubsection{Comparison with the X--ray catalogue by \protect\citet{Mehrtens2012}}

We compare our detections also with the X--ray cluster catalogue by \citet{Mehrtens2012}. 

There are 27 X--ray cluster detections from \citet{Mehrtens2012} 
 in the region that we have analysed, 20 have a temperature measurement. We will consider these 20 for our analysis. 

As for the \citet{Gozaliasl2014} catalogue, to match the {\it RedGOLD} cluster candidates with the X-ray detected catalogue by \citet{Mehrtens2012}, we
adopt a maximum projected distance between the centres corresponding to $R_{200}+\sigma_{R200}$ and a maximum
redshift difference of  $\Delta z=|z_{Meh}-z_{RedGOLD}|\le3\times \sigma_{photoz}=3\times 0.03\times(1+z)$, where $z_{Meh}$ is the cluster redshift in the \citet{Mehrtens2012} catalogue.

 {\it RedGOLD} recovers 16 of the 20 \citet{Mehrtens2012} clusters, and their temperature ranges over $1<T_X<7.5$~keV (their median temperature is $T_X=4.1$~keV),
 without applying any constraints on $\lambda$, $\sigma_{det}$ and the radial galaxy distribution.
We discard two detections adopting the optimal values of the cluster richness and the sigma detection level, for a final recovery of $70\%$($80\%$) of their detections with (without) limits.

\begin{figure}
    \begin{center}
        \includegraphics[width=\columnwidth]{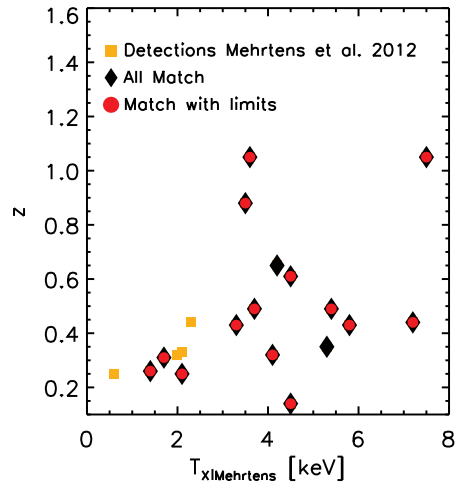}    
      \end{center}
    \caption{X--ray temperature $T_X$ as a function of the redshift $z$ for the clusters detected by {\itshape RedGOLD} with an X--ray counterpart in the \protect\citet{Mehrtens2012} catalogue with a temperature estimate. The red circles and black diamonds represent the detections when considering or not the lower limits on the cluster richness, the detection significance and the radial galaxy distribution, respectively. The orange squares represent  the four detections in the \protect\citet{Mehrtens2012} catalogue that we do not recover with {\itshape RedGOLD}. The performance of {\itshape RedGOLD} are very encouraging, with only four unmatched detections of the \protect\citet{Mehrtens2012} catalogue, all with $T_X\le 2.3$~keV.  }
    \label{fig:Txz_Mehr}
  \end{figure}

Fig.~\ref{fig:Txz_Mehr} shows the redshift--$T_X$ distribution of the clusters in the \citet{Mehrtens2012} catalogue (orange squares), our recovered detections with and without imposing our lower limits in red and black, respectively. 
The four undetected clusters have low temperatures ($T_X=0.6 - 2.3$~keV) as shown in Fig.~\ref{fig:TxLam_Mehr}, i.e. are poor clusters or groups. We recover 11(13) of the 13 clusters with $T_X > 2.5$~keV, i.e. the $85 (100)\%$ of the X--ray detected clusters by \citet{Mehrtens2012} with (without) considering the {\it RedGOLD} lower limits.

\subsubsection{The temperature--richness relation}

In this section, we discuss the scaling relation between optical and X--ray mass proxies, i.e. between the optical richness obtained with  {\itshape RedGOLD}  and the cluster X-ray temperature.

As already pointed out by \citet{Vikhlinin2006}  and \citet{Rasia2006}, the $\beta$--model does not accurately describe the cluster gas profile. This implies that the cluster masses  estimated
assuming a $\beta-$profile might be systematically underestimated up to a factor of $\sim40\%$ both when considering the isothermal and polytropic laws for the cluster temperatures \citep{Rasia2006}.

For this reason, we do not use the mass measurements to study scaling relations, but we study directly the optical richness--$T_X$ relation.  We use our recovered cluster detections up to $z=0.6$ in the \citet{Gozaliasl2014} catalogue to study the temperature--richness relation and compare our results with \citet{Rozo2014b}.

We fit the $T_X$--$\lambda$ relation in the following way:

\begin{equation}
\ln(T_X)=A+\alpha\ln(\lambda/\lambda_{pivot})\ , \label{TxLam}
\end{equation}

where $\lambda_{pivot}=median(\lambda)$, following \citep{Rozo2014b}.

\begin{figure}
    \begin{center}
        \includegraphics[width=\columnwidth]{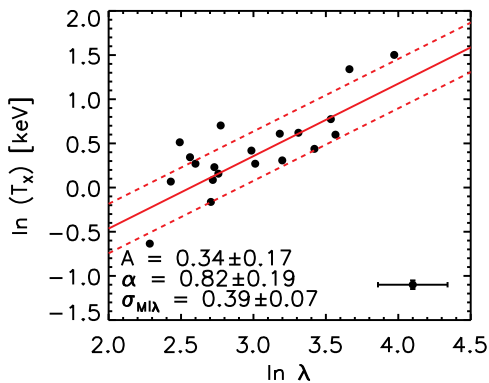}    
      \end{center}
    \caption{X--ray temperature $T_X$ as a function of the richness $\lambda$ (i.e. the $T_X-\lambda$ relation) for the 20 galaxy clusters detected by {\itshape RedGOLD} in the CFHT-LS W1 in common with the X--ray catalogue by \protect\citet{Gozaliasl2014} up to $z=0.6$.  We show the mean errors on the richness and the temperature in the bottom right corner.}
    \label{fig:TxLam_Goz}
  \end{figure}

Fig. \ref{fig:TxLam_Goz} shows the temperature--richness relation for the 20 galaxy clusters detected by {\itshape RedGOLD} in the CFHT-LS W1 field with a temperature measurement from \citet{Gozaliasl2014} up to $z=0.6$. Following  Eq. \ref{TxLam}, we perform a weighted fit on the errors and we obtain $A=0.34\pm0.17$, $\alpha=0.82\pm 0.19$, $\lambda_{pivot}=20.47$  and the scatter $\sigma=0.28\pm 0.04$.

 Assuming Eq. 2 from \citet{Rozo2014b} to estimate the scatter of the mass at fixed $\lambda$, we find a scatter of  $\sigma_{M|\lambda}=0.39\pm 0.07$. 
The values of the amplitude, slope and mass scatter at fixed richness inferred by the fit are shown in the plot. We show the mean errors on the richness and temperature in the bottom right corner.

\begin{figure}
    \begin{center}
        \includegraphics[width=\columnwidth]{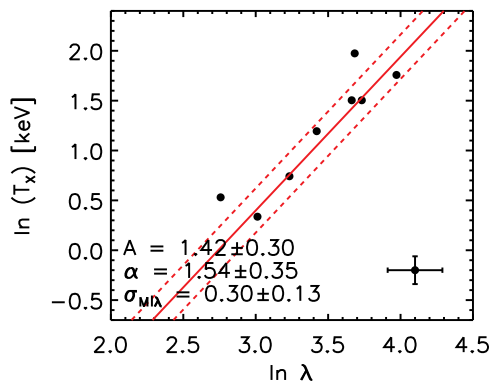}
      \end{center}
    \caption{X--ray temperature $T_X$ as a function of the richness $\lambda$ (i.e. the $T_X-\lambda$ relation) for the 8 galaxy clusters detected by {\itshape RedGOLD} in the CFHT-LS W1  in common with the X--ray catalogue by \protect\citet{Mehrtens2012} up to $z=0.6$. We show the mean errors on the richness and temperature in the bottom right corner.}
    \label{fig:TxLam_Mehr}
  \end{figure}

We conduct the same analysis using the temperature measurements provided by \citet{Mehrtens2012}. Fig. \ref{fig:TxLam_Mehr} shows the temperature--richness relation for the 8 galaxy clusters detected by {\itshape RedGOLD}  in the CFHT-LS W1 field with a temperature measurement from \citet{Mehrtens2012} and a temperature error less than 30\% up to $z=0.6$, following \citet{Rozo2014b}. Performing a weighted fit on the errors to study the temperature--richness relation, we find that $A=1.42\pm0.30$, $\alpha=1.54\pm 0.35$, $\lambda_{pivot}=35.38$  and a scatter $\sigma=0.22\pm 0.08$. Assuming Eq. 2 from \citet{Rozo2014b} to estimate the scatter of the mass at fixed $\lambda$, we find $\sigma_{M|\lambda}=0.30\pm 0.13$. As in Fig. \ref{fig:TxLam_Goz}, the values of the amplitude, slope and mass scatter at fixed richness inferred from  the fit are shown in the plot.

 Using the SDSS data and limiting the analysis to the redshift range $0.1<z<0.5$, \citet{Rozo2014b} found $A=1.206\pm0.044$, $\alpha=0.57\pm 0.10$, $\sigma=0.225\pm0.042$ and $\sigma_{M|\lambda}=0.30\pm 0.07$. 
The values of the slope  for our fit to the \citet{Gozaliasl2014} temperatures are consistent with  \citet{Rozo2014b} within $\approx 1 \sigma$ while for the slope estimated using the  \citet{Mehrtens2012} catalogue, our estimate is consistent with \citet{Rozo2014b} within 2$\sigma$. The scatter in mass at fixed richness obtained with the \citet{Gozaliasl2014} catalogue is comparable with \citet{Rozo2014b} but slightly higher while for the  \citet{Mehrtens2012} catalogue we obtain the same value $\sigma_{M|\lambda}=0.3$. 
 
The amplitude $A$ is significantly different when using the \citet{Gozaliasl2014} and \citet{Mehrtens2012} catalogues.
 For the fit to the \citet{Mehrtens2012} temperatures, the $A$ is consistent with \citet{Rozo2014b}, while the $A$ for \citet{Gozaliasl2014} is significantly lower than both the \citet{Rozo2014b} and our fit to  \citet{Mehrtens2012}. 
 The difference in the recovered amplitude of the temperature--richness relation is in part due to the different $\lambda_{pivot}$ for the two catalogues and to the different X--ray temperature definitions \citep[e.g. see][]{Rozo2014b}. While \citet{Gozaliasl2014} used core-excised temperatures, \citet{Mehrtens2012}  did not. Using our scaling relations and Gozaliasl's $M_{200}$, a temperature of $T_X \sim 1.8$~keV in the Gozaliasl's catalogue corresponds to $M_{200}\sim10^{14}\ \rm M_{\odot}$, and to a $\lambda \sim 30$. At this $\lambda$, the corresponding temperature in the \citet{Mehrtens2012} catalogue is $T_X \sim 2.7$~keV. 
 
  We are not able to investigate any evolution of the temperature--richness relation as a function of the redshift because of the small number of X--ray objects in the area.

These results are very promising because we are considering  a lower richness threshold (i.e. lower cluster mass) with respect to the \citet{Rozo2014b} cluster sample (see section ~\ref{sec:compRedMapper}) and are obtaining similar scatters. If, instead of using all the X--ray clusters in our area, we consider only higher richness thresholds, corresponding to $M_{200}\sim7\times10^{13}\ \rm M_{\odot}$ ($M_{200}\sim10^{14}\ \rm M_{\odot}$), we obtain a scatter in mass at fixed richness $\sigma_{M|\lambda}=0.27\pm0.08 \ (0.27\pm0.16)$ and $\sigma_{M|\lambda}=0.24\pm0.12\ (0.24\pm0.24)$ estimated from the \citet{Gozaliasl2014} and the \citep{Mehrtens2012} catalogue, respectively.
However, when considering these higher richness thresholds we only have between five and ten clusters to perform the fit and, for this reason, we will need to analyse a larger cluster sample to confirm these results.

\subsubsection{Spectroscopically confirmed X--ray clusters}

We also compare our results to a subsample of spectroscopically confirmed X--ray groups and clusters \citep{Pierre2006, Valtchanov2004, Andreon2004, Andreon2005, Pacaud2007, Miyazaki2007, Olsen2007, Berge2008}. In the CFHT-LS W1 area there are 33 spectroscopically confirmed groups/clusters between $0.1<z<1.1$. 
In Table~\ref{tab:cluspec}, we show the cluster ID, RA, DEC, redshift, and the corresponding reference for the spectroscopically confirmed clusters in the field.

To match the {\it RedGOLD} detections with the spectroscopically confirmed clusters, we adopt the same matching algorithm described for the X--ray detected catalogue, with
 a maximum projected distance between the centres corresponding to $R_{200}+\sigma_{R200}$ and a maximum
redshift difference of  $\Delta z=|z_{spec}-z_{RedGOLD}|\le3\times \sigma_{photoz}=3\times 0.03\times(1+z)$, where $z_{spec}$ is the cluster spectroscopic redshift.

\begin{table} 
\begin{center}
\resizebox{!}{4.8 cm}{
\begin{tabular}{c c c c c} 
\hline
Cluster ID  &  RA & DEC & z &  Reference  \\
\hline
XXLSSC 001 & 36.23792 & -3.81472 & 0.614 & (1)  \\  
XXLSSC 002 & 36.38542 & -3.91944 & 0.772 & (1) \\ 
XXLSSC 004 & 36.36833 & -5.11583 & 0.88 & (1) \\
XXLSSC 005 & 36.79042 & -4.30139 & 1.0   &  (1)   \\  
XXLSSC 006 & 35.44083 & -3.76889 & 0.429 & (2)   \\
XXLSSC 008 & 36.33417 & -3.80833 & 0.297 &  (2)   \\
RzCS 001      & 36.01792 & -5.28944 & 0.494 &  (2)    \\  
XXLSSC 012 & 37.11417 & -4.4300 & 0.433 &  (2)    \\   
XXLSSC 013 & 36.85792 & -4.5375 & 0.307 &  (2)    \\  
XXLSSC 014 & 36.64375 & -4.06528 & 0.344 &  (2)    \\  
XXLSSC 016 & 37.11750 & -4.99611 & 0.332 & (2)    \\  
XXLSSC 017 & 36.61417 & -4.99861 & 0.381 &  (2)   \\  
XXLSSC 018 & 36.00667 & -5.09028 & 0.322 & (2)    \\  
XXLSSC 019 & 36.04917 & -5.37972 & 0.494 & (2)    \\  
XXLSSC 020 & 36.63667 & -5.00889 & 0.494 & (2)    \\  
XXLSSC 022 & 36.91667 & -4.85806 & 0.29   & (4)    \\  
XXLSSC 025 & 36.35292 & -4.67861 & 0.26   & (4)    \\  
XXLSSC 027 & 37.01417 & -4.85083 & 0.29   & (6)    \\  
XXLSSC 029 & 36.01625 & -4.22444 & 1.05   & (3)    \\  
VVDS Cluster                    & 36.28917 & -4.54833 & 0.77    & (8)  \\
XXLSSC 038 & 36.85417 & -4.18972 & 0.58  & (4)     \\  
XXLSSC 044 & 36.13958 & -4.23472 & 0.26 & (4)    \\  
XXLSSC 049 & 35.98917 & -4.58806 & 0.49   & (6)    \\  
XXLSSC 053 & 36.12167 & -4.82333 & 0.49   & (5)   \\  
XXLSSC 007 & 36.03750 & -3.91917 & 0.557 & (2)    \\  
XXLSSC 040 & 35.52292 & -4.54639 & 0.32 & (6)    \\  
XXLSSC 041 & 36.37833 & -4.23972 & 0.14 & (4)   \\  
a & 36.34583 & -4.44444 & 0.46   & (4)    \\  
b & 36.37333 & -4.42972 &  0.92   & (4)    \\  
c & 36.54125 & -4.52222 &  0.82   & (4)    \\ 
d & 36.71625 & -4.16583 &  0.34   & (4)    \\  
XLSSCJ022534.2-042535 & 36.3925 & -4.42639 &  0.92 & (3)  \\  
XXLSSC 005b & 36.8 & -4.23056 &1.0 & (3)  \\  
\hline
\end{tabular}
}
\caption{List of the confirmed galaxy clusters in the CFHT-LS W1.  (1) \protect\citet{Valtchanov2004}, (2) \protect\citet{Andreon2004}, 
(3) \protect\citet{Andreon2005}, (4)\protect\citet{Pierre2006},  (5) \protect\citet{Miyazaki2007}, (6) \protect\citet{Pacaud2007},  (7) \protect\citet{Berge2008}, (8) \protect\citet{Lefevre2013}. }
\label{tab:cluspec}
\end{center}
\end{table}

 {\itshape RedGOLD} recovers 24 out of the 33 spectroscopically confirmed clusters without considering any lower limit on $\lambda$, $\sigma_{det}$ and the radial galaxy distribution. When adopting the lower limits on $\lambda$, $\sigma_{det}$ and assuming the radial galaxy distribution, we discard 5 detections because of the imposed constraints on $\lambda$ (they all have $\lambda<10$). We check the nine missing detections: four detections are C2 and C3 objects from
\citet{Pierre2006}. This class includes faint and poor galaxy structures and their detection implies higher contamination rate.

A cluster at z=1 (ID=XLSS005b) 
is undetected by {\itshape RedGOLD} because it is blended with XLSSC005 at approximately  the same redshift. 
An X--ray detected cluster at $z=0.92$ unrecovered by {\itshape RedGOLD} is an extremely poor system, undetected in $(R-z')$, but appearing as a galaxy overdensity in the K--band, as found by \citet{Andreon2005}.  
Finally, we are not able to recover three clusters at $z=0.322$, $z=0.381$ and $z=0.557$: the first one has a  central BCG, but there is no a clear red overdensity, the second one is detected also in the catalogue by \citet{Gozaliasl2014} and
has $M=8.5 \pm 0.7 \times10^{13}\ {\rm M_{\odot}}$, and the last one is an optically poor system.

\begin{figure*}
\centering
\begin{tabular}{@{}cccc@{}}
        \includegraphics[width=.5\textwidth]{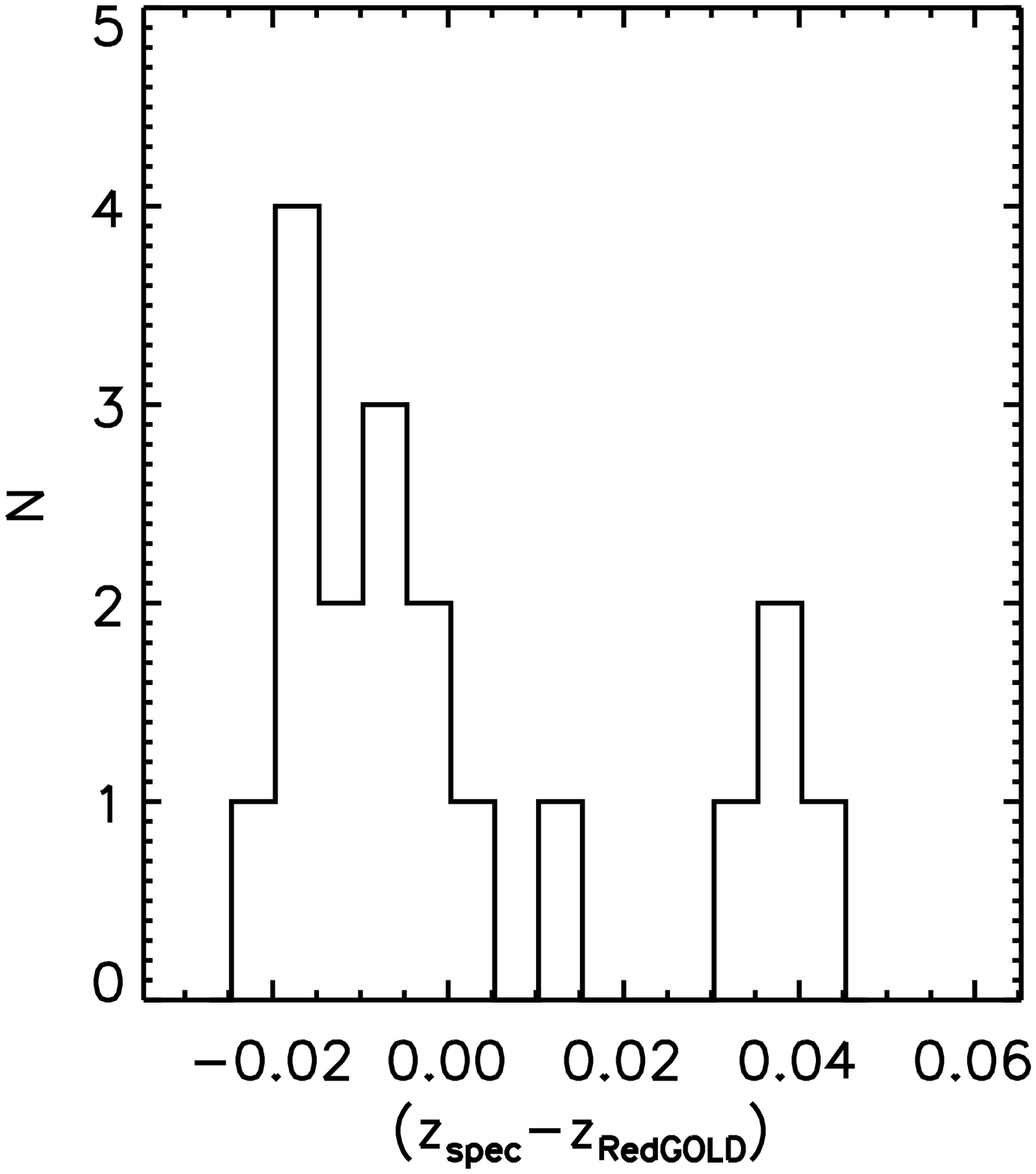}    
        \includegraphics[width=.5\textwidth]{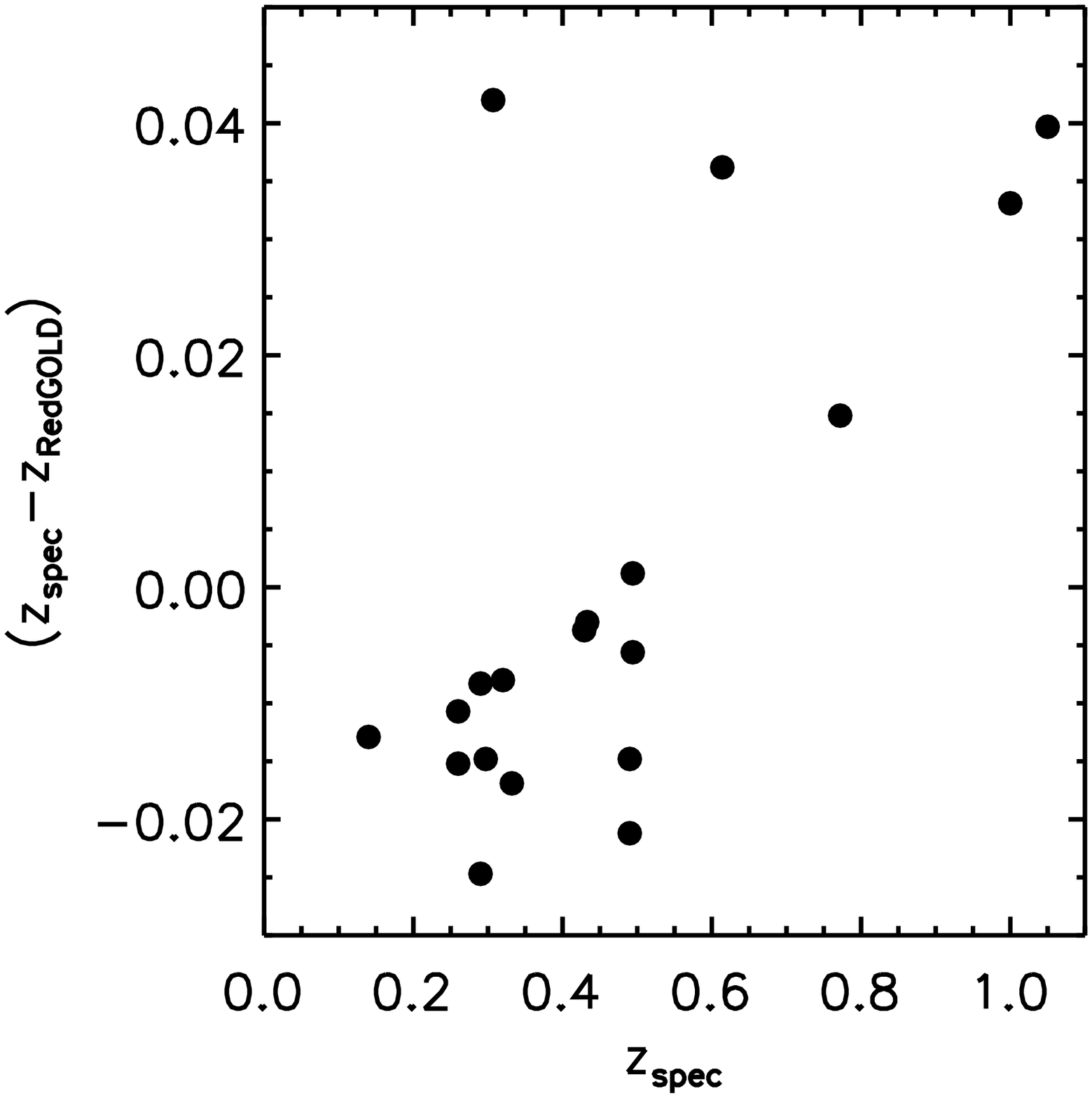}    
      \end{tabular}
        \caption{Left panel: the $(z_{spec}-z_{RedGOLD})$ distribution for the 19 spectroscopically confirmed clusters recovered by {\it RedGOLD}. The redshift difference is less than 0.05 for all detections up to $z\sim1$.  Right panel: $(z_{spec}-z_{RedGOLD})$ as a function of the spectroscopic redshift.}
    \label{fig:histozspec}
  \end{figure*}

The comparison of our detection algorithm with these known X--ray detections on the CFHT-LS W1 confirms that  the adopted cluster centre definition is efficient: in fact, the mean separation between the optical and the X--ray centre is $17.2''\pm 11.2''$ for all recovered confirmed clusters.

Up to redshift $z\sim1$, we accurately recover the cluster redshift. In fact,
the discrepancy between our cluster photometric redshifts and the corresponding spectroscopic measurement  is less than 0.05, as shown is Fig.\ref{fig:histozspec}, where the median $\delta z \sim 0.004$. The right panel of Fig.\ref{fig:histozspec} shows that the redshift difference $(z_{spec}-z_{RedGOLD})$ is larger at higher redshift (i.e. $z\ge0.5$), with four out of six objects with $|z_{spec}-z_{RedGOLD}|>0.02$. This is expected since the photometric redshift accuracy is lower at fainter magnitudes and increasing redshifts. 
However, this effect is negligible, being the redshift difference $|z_{spec}-z_{RedGOLD}|$  very low for all the spectroscopic confirmed clusters recovered by {\it RedGOLD}.
This result confirms that the BC03 model colours (from which we derive $ z_{RedGOLD}$) accurately reproduce galaxy colours in the redshift range that we considered.

From \citet{Berge2008}, we have a mass estimate based on weak lensing measurements for four clusters detected by {\itshape RedGOLD} in the XMM-LSS area: we recover all the four clusters in the CFHT-LS W1 field.
We show these values in Table \ref{tab:mass}.

\begin{table}
\begin{center}
\begin{tabular}{c c c c c }
\hline
XLSSC & RA & Dec & z & $M_{200}$ (WL) \\ 
 & $(^\circ)$ & $(^\circ)$ & & \footnotesize $(10^{13} {\rm h^{-1} M_{\odot}})$ \\
\hline
013 & 36.8497 & -4.5481 & 0.307 & $8.2^{+2.5}_{-1.9}$ \\ 
053 & 36.1229 & -4.8341 & 0.50 & $10.3^{+3.0}_{-2.6}$ \\ 
041 & 36.3723  & -4.2604 & 0.14 & $4.9^{+1.6}_{-1.2}$ \\ 
044 & 36.1389 & -4.2384 & 0.26 & $7.2^{+2.3}_{-1.7} $ \\ 
\hline
\end{tabular}
\caption{Mass estimates for 4 clusters in the XMM-LSS area from \citet{Berge2008}. We recover all 4 clusters.}
\label{tab:mass}
\end{center}
\end{table}

\subsection{Comparison with optically selected cluster catalogues}

Three optically detected cluster catalogues are publicly available in the CFHT-LS W1 field: (1) the redMaPPer catalogue from \citet{Rykoff2014}, obtained using SDSS observations; (2) the \citet{Milkeraitis2010} and the (3) \citet{Durret2011} catalogues, both obtained using CFHT-LS W1 observations and methods using photometric redshift catalogues.

\subsubsection{Comparison with redMaPPer}\label{sec:compRedMapper}

\begin{figure}
    \begin{center}
        \includegraphics[width=\columnwidth]{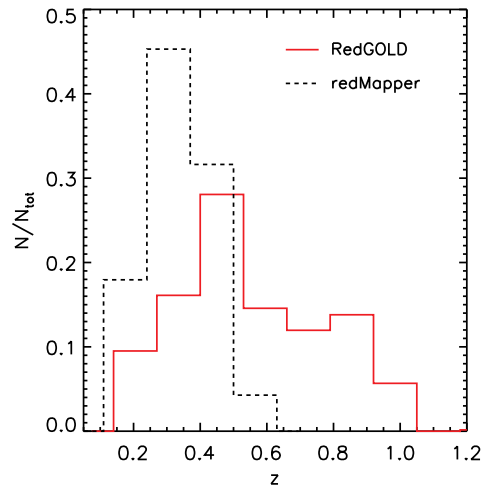}    
      \end{center}
    \caption{Redshift distribution of CFHT-LS W1 cluster detections in the $\sim 60\ deg^2$ (red solid line). The black dashed line represents the redMaPPer detections in the same region. Each histogram is normalised to the total number of detections.}
    \label{fig:redsDistrCFHTLSredMapper}
  \end{figure}

The first optically detected cluster catalogue to which we compare the {\it RedGOLD} 
cluster candidates is the redMaPPer catalogue  \protect\citep{Rykoff2014}, obtained using observations from the SDSS.
In Fig.~\ref{fig:redsDistrCFHTLSredMapper}, we show the redshift distribution of our cluster candidates: the red solid line represents 
the {\it RedGOLD} detections in the CFHT-LS W1 field  while the dashed black line shows the redshift distribution of the redMaPPer catalogue in the same area. Both histograms are normalised to the total number of detections found by the corresponding algorithm. As expected,  we detect cluster candidates at higher redshift than redMaPPer since the CFHTLenS data are deeper than the SDSS.

 To match the {\it RedGOLD} cluster candidates with the redMaPPer catalogue, we adopt the same matching algorithm described for the X--ray detected catalogue, with
 a maximum projected distance between the centres corresponding to $R_{200}+\sigma_{R200}$ and a maximum
redshift difference of  $\Delta z=|z_{redMaPPer}-z_{RedGOLD}|\le3\times \sigma_{photoz}=3\times 0.03\times(1+z)$, where $z_{redMaPPer}$ is the cluster redshift in the redMaPPer catalogue.

There are 116 redMaPPer cluster detections in our field,
115 detected with  {\itshape RedGOLD} (i.e. the $99\%$), when not applying any lower limit on the  radial galaxy distribution, $\lambda$ and $\sigma_{det}$. 
The only redMaPPer cluster that we do not detect has a sparse structure and has redshift $z=0.48$.
We discard seven additional redMaPPer detections when considering the optimal lower limits imposed on the radial galaxy distribution (two detections) and richness and $\sigma_{det}$ (five detections). With this final selection, we obtain 108 {\itshape RedGOLD} detections out of the 116 clusters detected with redMaPPer (i.e. $\sim 93\%$). 

All the redMaPPer detections in the area spanned by the \citet{Gozaliasl2014} catalogue, six clusters, have an X--ray counterpart. {\itshape RedGOLD} considers as detections only five of these six clusters. The unrecovered redMaPPer detection with an X--ray counterpart is at $z\sim0.5$ and has $M_{200}\sim8.5\times10^{13}\ {\rm M_{\odot}}$, i.e. it is in the mass range in which we are $\sim70\%$ complete (see section~\ref{sec:compl} and \ref{sec:compGoz}). 

\begin{figure}
    \begin{center}
        \includegraphics[width=\columnwidth]{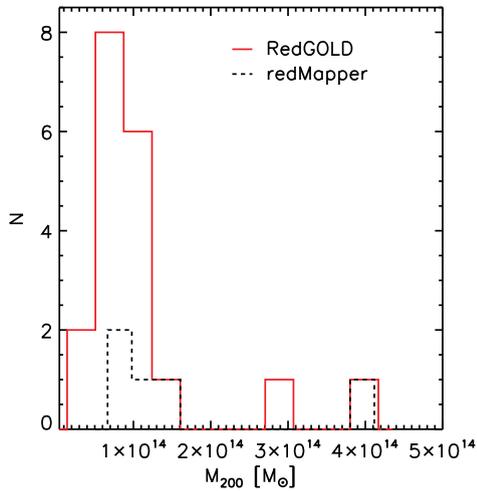}    
      \end{center}
    \caption{Mass distribution of the cluster detected by {\itshape RedGOLD} (red solid line) and  redMaPPer (black dashed line) with an X--ray counterpart in the \protect\citet{Gozaliasl2014} catalogue. The mass measurements are from \protect\citet{Gozaliasl2014}.}
    \label{fig:HistoMass}
  \end{figure}

 Fig. \ref{fig:HistoMass} shows the mass distribution of the clusters recovered by RedGOLD in red and those recovered by redMaPPer in black: our catalogue reaches lower cluster mass values with respect to the redMaPPer detections, as expected since the CFHTLenS is deeper than the SDSS, and the redMapper catalogue is cut at a given richness.
For this reason, our {\itshape RedGOLD} catalogue includes $\sim200$ 
detections up to $z=0.5$, unrecovered by  redMaPPer using the SDSS.

\begin{figure}
    \begin{center}
       \includegraphics[width=\columnwidth]{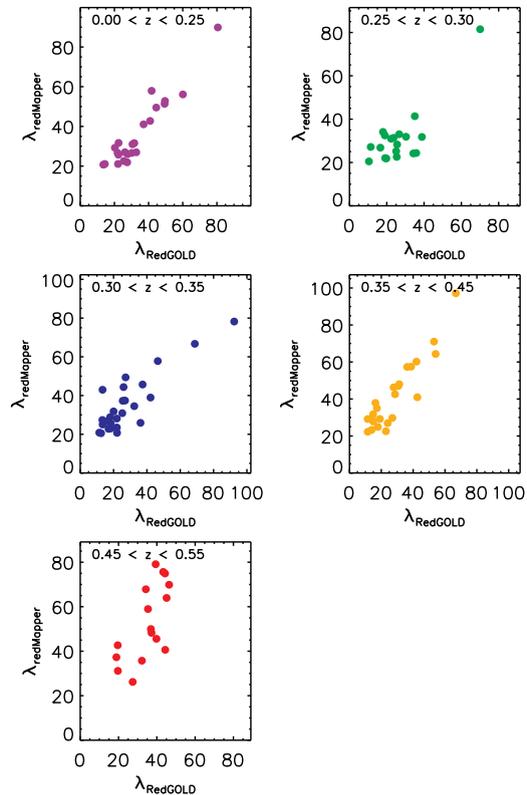}   
      \end{center}
    \caption{Comparison of the {\it RedGOLD} and redMaPPer richness, ($\lambda_{redMaPPer}$ vs $\lambda_{RedGOLD}$), in different redshift bins as indicated in each panel. }
    \label{fig:CompRich}
  \end{figure}

\begin{figure}
    \begin{center}
        \includegraphics[width=\columnwidth]{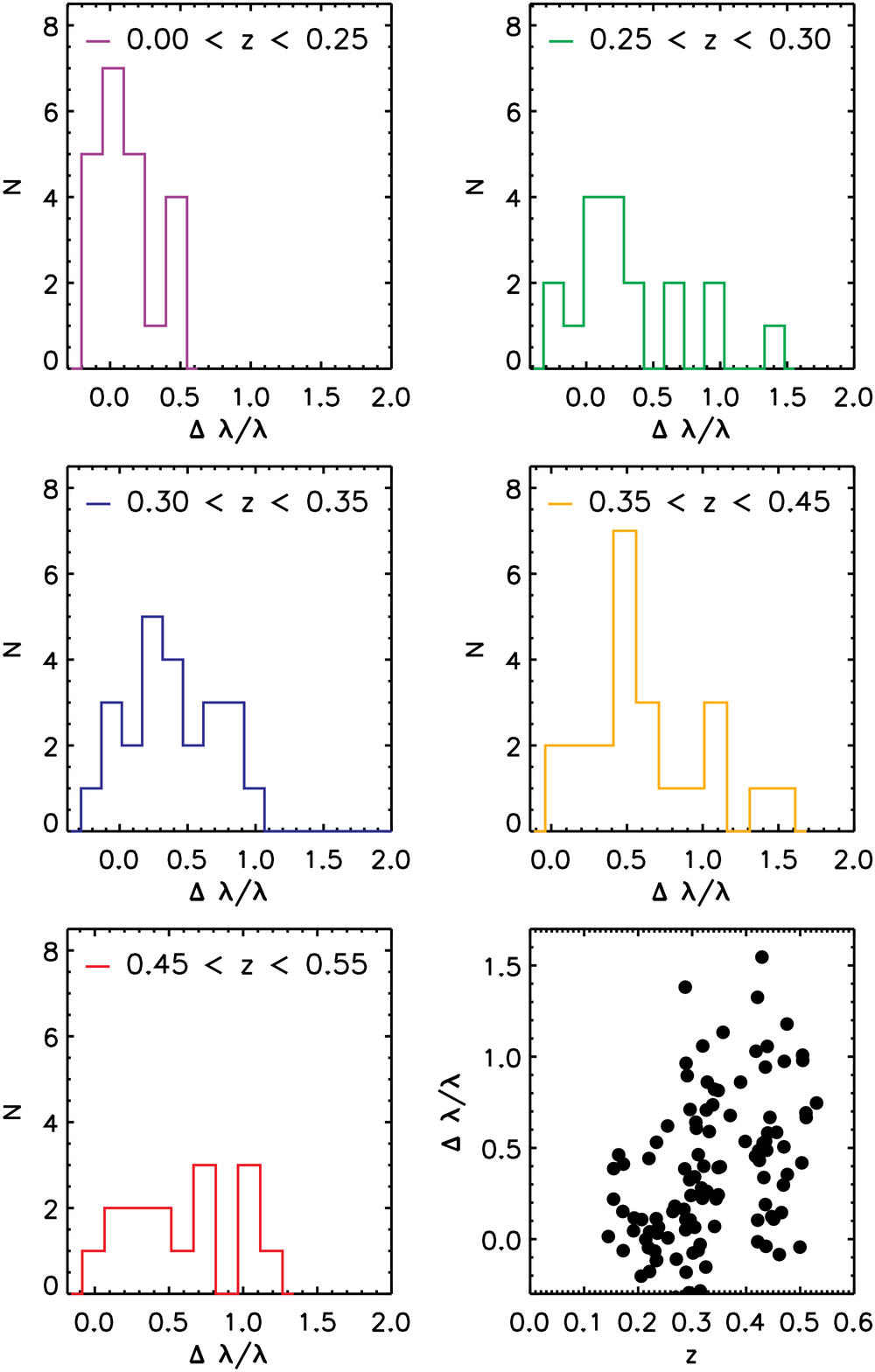} 
      \end{center}
    \caption{Histogram of the $\left(\frac{\lambda_{redMaPPer}-\lambda_{RedGOLD}}{\lambda_{RedGOLD}}\right)$, in different redshift bins as indicated in each panel. The bottom right panel shows the $\left( \frac{\lambda_{redMaPPer}-\lambda_{RedGOLD}}{\lambda_{RedGOLD}}\right )$ distribution as a function of the redshift.}
    \label{fig:CompRichHisto}
  \end{figure}

 In Fig. \ref{fig:CompRich} and Fig.~\ref{fig:CompRichHisto}, we compare the richness estimates obtained by redMaPPer and {\itshape RedGOLD} for the 108 common detections.  We show  the 
$\lambda_{RedGOLD}$ vs $\lambda_{redMaPPer}$ and the histogram of the difference between our richness definition and the richness adopted in \citet{Rykoff2014}, ($\lambda_{redMaPPer}-\lambda_{RedGOLD}$)/($\lambda_{RedGOLD}$), in different redshift bins, respectively.

Different colours show the observed difference in different redshift bins,  as indicated in each panel.
The  redMaPPer richness is systematically higher than the {\itshape RedGOLD} richness as defined in this paper. In the bottom right panel in Fig.~\ref{fig:CompRichHisto}, we plot the ($\lambda_{redMaPPer}-\lambda_{RedGOLD})/\lambda_{RedGOLD}$ as a function of redshift: the difference between the two richness estimates in the {\it RedGOLD} and redMaPPer catalogue is larger at higher redshift.
 In Fig.~\ref{fig:CompRichHisto}, there is an apparent lack of clusters at z=0.35. This depends on the lack of galaxies at $z\sim0.35$in the galaxy photometric redshift distribution. We check the galaxy photometric redshifts of the objects with a spectroscopic redshifts $0.3\lesssim z \lesssim 0.4$ and they 
are fully consistent with the spectroscopic measurements. Therefore,
we conclude that the apparent lack of clusters visible in Fig.~\ref{fig:CompRichHisto} is due the cosmic variance.

In Table~\ref{Tab:CompRich}, we present the median value of this richness difference as a function of redshift.  The median difference is small at low redshift ($\sim 5-15\%$) at $z<0.3$, but it increases up to $\sim 60\%$ at higher redshifts (with single values reaching the $\sim200\%$). At these redshifts, we keep a simple approach counting galaxies up to the depth reached by the CFHTLenS, while the redMaPPer richness estimate includes an extrapolation of the SDSS depth (which is lower than CFHTLenS) to our same limit in ${\rm L^*}$.  It would be worth to investigate the observed difference richness in a future work, considering a larger cluster sample.

\begin{table}
\begin{center}
\begin{tabular}{c c}
\hline
 redshift & median($\Delta \lambda/\lambda_{RedGOLD}$) \\
\hline
$z\le 0.25$ & 0.05 \\ 
$0.25 < z \le 0.30$ &  0.16\\ 
$0.30 < z \le 0.35$ &  0.39\\ 
$0.35 < z \le 0.45$ &  0.54\\ 
$0.45 < z \le 0.55$ &  0.59 \\ 
\hline
\end{tabular}
\caption{Median value of $(\lambda_{redMaPPer}-\lambda_{RedGOLD})/\lambda_{RedGOLD}$ in different redshift bins}
\label{Tab:CompRich}
\end{center}
\end{table}

\subsubsection{Comparison with other catalogues obtained with CFHT-LS W1 observations}

\begin{figure}
    \begin{center}
        \includegraphics[width=\columnwidth]{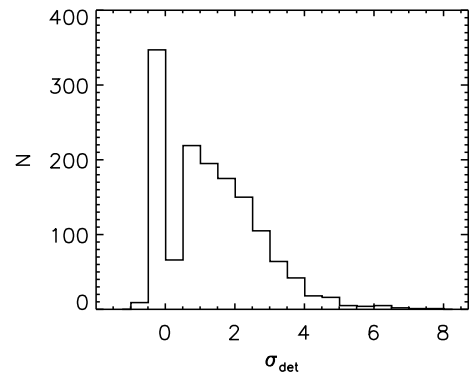}    
      \end{center}
    \caption{Histogram of the {\itshape RedGOLD}  sigma detection level $\sigma_{det}$ for the unrecovered detections in the catalogue provided by \protect\citet{Milkeraitis2010} with $\sigma_{Milkeraitis}\ge 5$. }
    \label{fig:HistoSigDetMil}
  \end{figure}

There are two public optically selected cluster candidate catalogues, obtained using the same CFHT-LS W1 observations as the CFHTLenS catalogue, the \citet{Milkeraitis2010}  and the  \protect\citet{Durret2011}  catalogues.

\citet{Milkeraitis2010} developed the {\itshape 3D-Matched-Filter} technique (3D-MF)  to detect galaxy clusters, and applied it to 
 the four wide fields of the CFHT-LS. Their detection algorithm is based on the matched filter technique, assuming  a cluster radial profile and luminosity function.  They used photometric redshifts to reduce contamination due to projection effects. 
 
 To compare our detections with the \citet{Milkeraitis2010} catalogue, we cut their catalogue  to $\sigma_{Milkeraitis}\ge 5$,  corresponding to $\sim1.6\times 10^{13}\ {\rm M_{\odot}}$ \citep{Ford2015}, with an expected false detection rate  $> 30\%$ \citep{Milkeraitis2010}. 
 We match the \citet{Milkeraitis2010} catalogue with the {\it RedGOLD} cluster candidates adopting less conservative constraints, with a maximum projected distance between the centres of 2 Mpc and a cluster redshift difference $|z_{Milkeraitis}-z_{RedGOLD}|\le 0.2$\footnote {with $z_{Milkeraitis}$ being the cluster redshift in the \citet{Milkeraitis2010} catalogue}, since the cluster redshift estimates in the \citet{Milkeraitis2010} catalogue are not refined and have a bin of 0.1.

In the CFTH-LS W1 subfield covered by this work, \citet{Milkeraitis2010}  detected 2871 cluster candidates with $\sigma_{Milkeraitis}\ge 5$.
 Of those, {\itshape RedGOLD} detects 1753 objects ($61\%$), when not applying any lower limit on the radial galaxy distribution, $\lambda$ and $\sigma_{det}$. 
When considering the optimal lower limits imposed on the radial galaxy distribution, richness and $\sigma_{det}$, we discard 1158 objects, and obtain  595 {\itshape RedGOLD} detections (i.e., the $21\%$ of the Milkeraitis' detections). These numbers are expected since {we find $\sim 11$ detections per $\rm deg^2$
while \citet{Milkeraitis2010} found more than $ 45$ detections} per $\rm deg^2$ at $\sigma_{Milkeraitis} \ge 5$. 

To understand which kind of objects {\itshape RedGOLD} does not detect or discards, we estimate our detection level at the positions of the centres of the
unrecovered detections of the Milkearitis' catalogue. Fig. \ref{fig:HistoSigDetMil} shows the distributions of our estimated 
$\sigma_{det}$, corresponding to the unrecovered candidates in the Milkeraitis' catalogue. 

 We find that only $\sim 3\%$ of the unrecovered Milkeraitis'  detections have a $\sigma_{det}\ge 4$ at $z\le0.6$ and $\sigma_{det}\ge 4.5$ at $z>0.6$: this implies that we do not select most of their detections because of their low $\sigma_{det}$. In fact, in this low $\sigma_{det}$ range, {we expect a lower purity that we do not accept (see previous section)}. The remaining $\sim 3\%$  with a $\sigma_{det}$ in our selection range, are discarded because of their low $\lambda$. The median significance of the Milkeraitis'  discarded detections is $\sigma_{Milkeraitis}  \sim 5$, which approximately corresponds to  $M\sim 1.6\times 10^{13}\ {\rm M_{\odot}}$ \citep{Ford2015}.

It is interesting that, when we consider higher $\sigma_{Milkeraitis}$ detections, at $\sigma_{Milkeraitis}  \ge 10$, which corresponds to $M\gtrsim10^{14}\ {\rm M_{\odot}}$ \citep{Ford2015}, {\itshape RedGOLD} recovers the $\sim 78\% (95\%)$ of the objects  with (without) the imposed criteria on the {\it RedGOLD} parameters. At  $\sigma_{Milkeraitis}  \ge 15$, {\itshape RedGOLD} recovers the $\sim 95\% (100\%)$ of the objects, at higher $\sigma_{Milkeraitis}>17$, we recover the same 13 objects with (without) the imposed criteria on the {\it RedGOLD} parameters. This means that for the most massive detections, we recover similar cluster candidates.

On the other hand, {\itshape RedGOLD} detects 652 cluster candidates and approximatively $75\%$ of those are also selected in the Milkeraitis' catalogue when considering all their detections with $\sigma_{Milkeraitis}\ge 5$. When considering all objects in the Milkeraitis' catalogue (i.e. $\sigma_{Milkeraitis} \ge 3.5$), we find $\sim 85\%$ of the {\itshape RedGOLD} detections.

 \begin{figure}
    \begin{center}
        \includegraphics[width=\columnwidth]{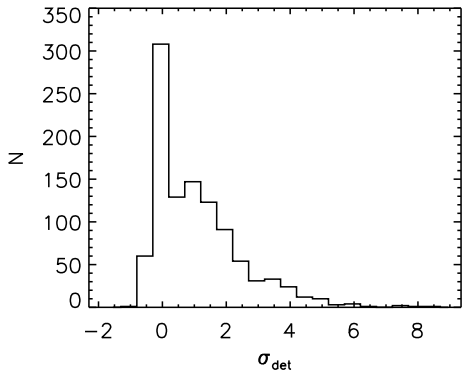}    
      \end{center}
    \caption{Histogram of the {\itshape RedGOLD} sigma detection level $\sigma_{det}$ for the unrecovered detections in the catalogue provided by \protect\citet{Durret2011} with $S/N\ge 3$ within  the redshift range $0.375<z<1.05$. }
    \label{fig:HistoSigDetDur}
  \end{figure}

\begin{figure*}
\centering
\begin{tabular}{@{}cccc@{}}
        \includegraphics[width=.5\textwidth]{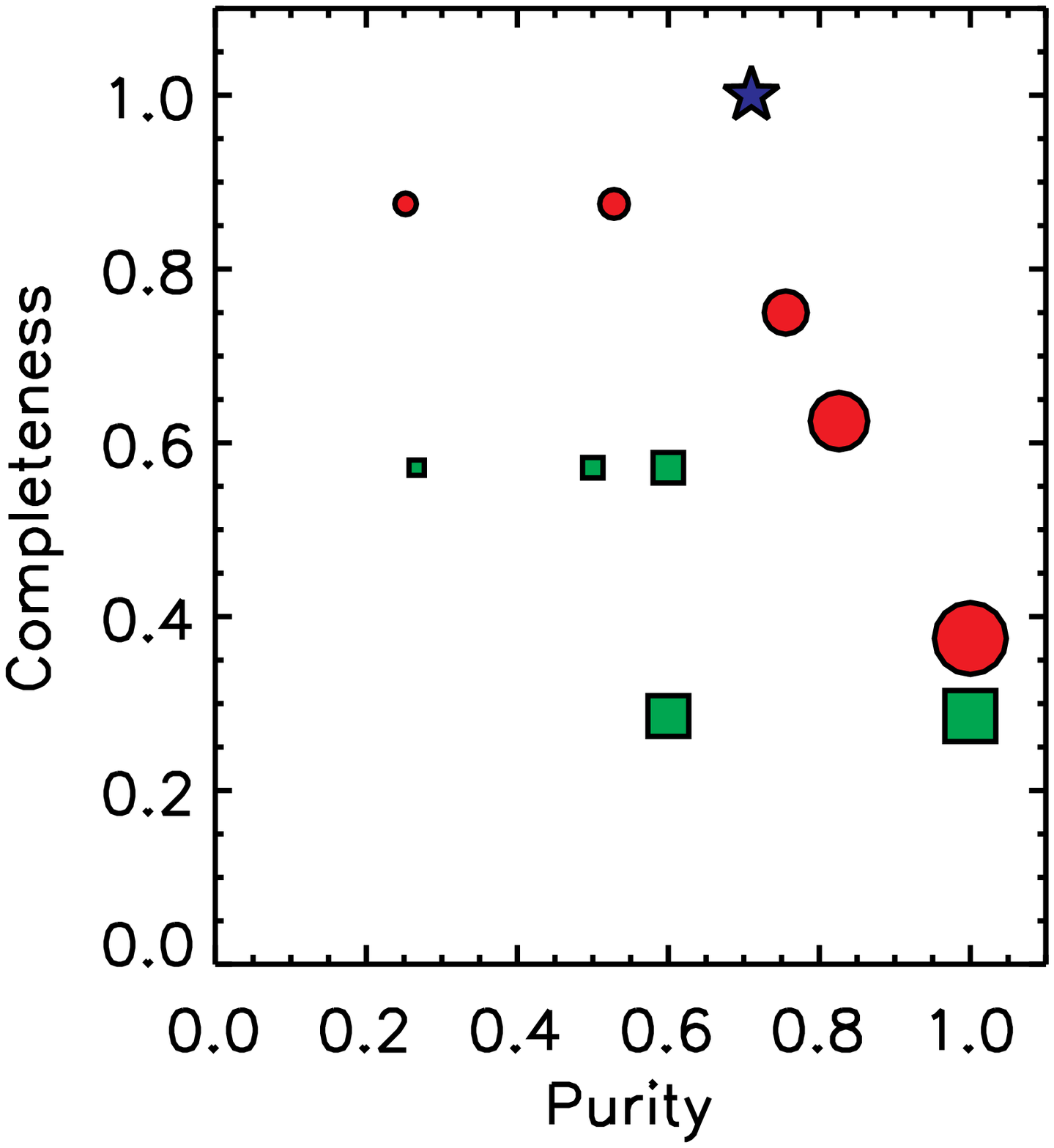}    
        \includegraphics[width=.5\textwidth]{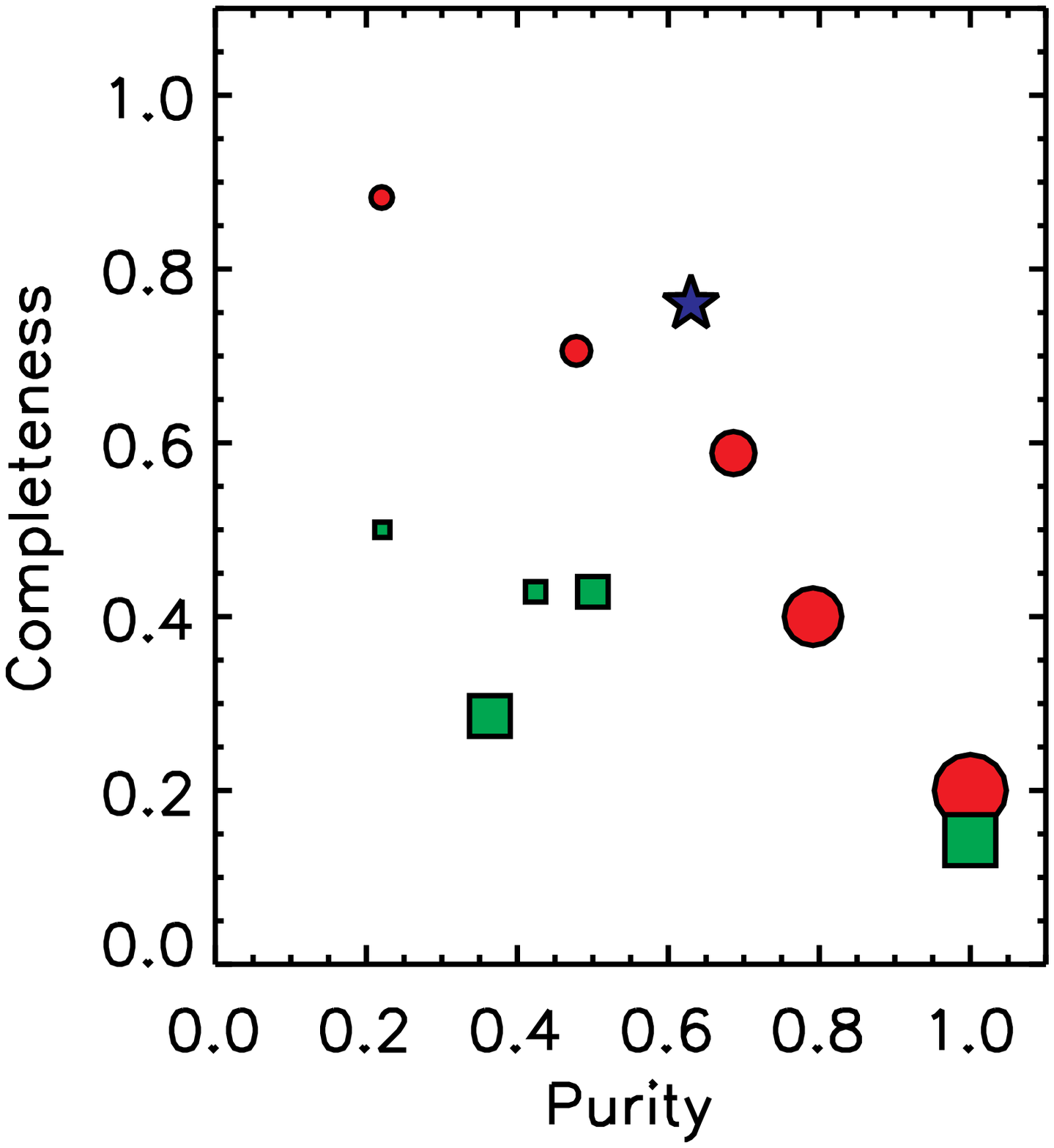}    
      \end{tabular}
        \caption{Completeness as a function of the purity for the \protect\citet{Milkeraitis2010} (red circles) and  \protect\citet{Durret2011} (green squares) catalogues, respectively,  estimated using the X--ray detected catalogue from \citet{Gozaliasl2014} sample. The size of the symbols shows different  thresholds of their detection level: from the smaller to the larger  $\sigma_{Milkeraitis}=3, 5, 6, 8, 10$ and $\sigma_{Durret}=2, 4, 5, 6, 7$. The left and right panels show results for $z <0.6$ and in the whole redshift range, respectively. The blue star represents the value of completeness and purity reached by {\itshape RedGOLD} with the optimised values of $\sigma_{det}$ and $\lambda$.}
 \label{fig:ComplPur_lam_var_Milk}
  \end{figure*}

\begin{table*}
\begin{center}
\resizebox{!}{1.5 cm}{
\begin{tabular}{c c c }
\hline
 & \citet{Milkeraitis2010} & \citet{Durret2011} \\
\hline
$\sigma_{catalogue}<\sigma_{limit}$ & 68\% (74\%) & 33\% (42\%) \\
$\sigma_{catalogue}\ge \sigma_{limit}$ & 32\% (26\%)& 67\% (58\%)\\
$N_{common}/ N_{tot}$ & 595(1753) / 2871 & 250 (732) / 1293 \\
\%$N_{common}/ N_{tot} $ & 21\% (61\%) & 19\% (57\%)\\
$N_{det} $ & 652 (3015) & 475 (2440) \\
$N_{common}/N_{det}$ & 91\% (58\%)& 53\% (30\%) \\
\hline
\end{tabular}
}

\caption{ Percentage of the matched cluster candidates from \protect\citet{Milkeraitis2010} and \protect\citet{Durret2011} obtained considering  our detection limit $\sigma_{det} \ge 4$ at $z\le0.6$, and  $\sigma_{det} \ge 4.5$ at $z>0.6$. For the comparison with  \protect\citet{Milkeraitis2010},  we consider $\sigma_{Milkeraitis}\ge5$, which correspond to $M>1.5\times 10^{13}\ {\rm M_{\odot}}$. The first and second row show the percentage of matched detections with $\sigma_{Milkeraitis}< 10$ and $\sigma_{Milkeraitis}\ge 10$, which correspond to $M=10^{14}\ {\rm M_{\odot}}$ \protect\citep{Ford2015}. 
For the comparison with \protect\citet{Durret2011},  we split their sample considering $\sigma_{Durret}<4$ and $\sigma_{Durret}\ge4$.
$N_{common}$ represents the number of the common detections (our minimum detection has $\sigma_{det}\ge3$) while $N_{tot}$ is the total number of cluster candidates in their optical cluster catalogue. $N_{det}$ is the number of the cluster candidates detected by {\itshape RedGOLD}. The values in parenthesis refer to the comparison without applying any lower limit on the radial galaxy distribution, $\lambda$ and $\sigma_{det}$.}
\label{tab:GrMiDumatch}
\end{center}
\end{table*}


\citet{Durret2011}  built an optical detected cluster catalogue, using a detection technique based on the galaxy density maps \citep{Adami2010}: they used photometric redshifts and detected overdensities in redshift slices, over a given threshold using the tool SExtractor \citep{Bertin1996}. For each detection, they provide the cluster candidate photometric redshift.

We match the \citet{Durret2011} catalogue with the {\it RedGOLD} cluster candidates adopting the same matching algorithm used for the \citet{Milkeraitis2010} cluster catalogue.

When comparing our detections to their catalogue, we only consider their most reliable detections, i.e. those  in the redshift range $0.375<z<1.05$ and with a signal--to--noise ratio $S/N\ge 3$ \citep{Durret2011}. Those are 1293 objects and {\itshape RedGOLD} detects  the $\sim 19\%$ (57\%) of the objects  with (without) the imposed criteria on the {\it RedGOLD} parameters.

As above, we estimate our sigma detection level $\sigma_{det}$ at the position of the unmatched Durret's candidates and we show their distribution in Fig. \ref{fig:HistoSigDetDur}. Also in this case, most of the missing detections have a low detection level, with only $3\%$ of the unrecovered Durret candidates having $\sigma_{det}\ge 4$ at $z\le0.6$ and $\sigma_{det}\ge 4.5$ at $z>0.6$: this implies that  they are mostly lower $\sigma_{det}$ (i.e. less massive) detections.

 If we consider the {\it RedGOLD} cluster candidate catalogue in the redshift range $0.35<z<1.1$ to 
match  the Durret's catalogue to our catalogue in the same redshift interval, we  find 475 (2440) with (without) imposing our constraints on the {\it RedGOLD} parameters, but only $\sim 34\%$ ($\sim19 \%$) are detected also by \citet{Durret2011}. The mean richness and detection significance of the {\it RedGOLD} cluster candidates not detected in the Durret's catalogue are $<\lambda>\simeq17\ (8)$ and $<\sigma_{det}>\simeq6\ (4)$ when considering our cluster sample with (without) lower limits on the {\it RedGOLD} parameters.

From this comparison, we conclude that most of the \citet{Durret2011} cluster candidates are objects less massive than ours, and that their algorithm does not find most of our massive candidates.

We summarise our results on the comparison with the other optical detected cluster catalogues obtained with CFHT-LS W1 observations in Table~\ref{tab:GrMiDumatch} and Table~\ref{tab:GrMiDu}.  

Table~\ref{tab:GrMiDumatch} shows the distribution of their detection significances  for the matched cluster candidates when applying the optimal values for $\lambda$ and $\sigma_{det}$ for the {\it RedGOLD} detections and considering the cluster candidates from the 
\citet{Milkeraitis2010} and \citet{Durret2011} catalogues, with $\sigma_{Milkeraitis}\ge 5$ and $\sigma_{Durret}\ge 3$, respectively. We show this distribution splitting  the detection significances of  the candidates recovered by {\it RedGOLD}
with respect to $\sigma_{limit}=10$ and $\sigma_{limit}=4$, for the \citet{Milkeraitis2010} and \citet{Durret2011} catalogue, respectively.
The corresponding values without imposing any lower limit on the {\it RedGOLD} parameters are shown in parenthesis. In Table ~\ref{tab:GrMiDumatch}, we also show the fraction $N_{common}/N_{tot}$, where $N_{common}$ is the number of the common detections and $N_{tot}$ is the total number of the cluster candidates in the two different cluster catalogues and the $N_{common}/N_{det}$ ratio, where $N_{det}$ is the number of the {\it RedGOLD} detections. 
 
Table~\ref{tab:GrMiDu} shows the detection significance $\sigma_{det}$ for the unmatched cluster candidates from the two other catalogues that we analysed, 
the \citet{Milkeraitis2010} and \citet{Durret2011} catalogues, when we run {\itshape RedGOLD} on their detection centres. Depending on the algorithm, we find that $\sim 70-80\%$ of the unmatched candidates have {\itshape RedGOLD} detections at $<2\sigma$, and only $\sim 3\%$  have {\itshape RedGOLD} detections at $>4.5\sigma$.

As already described for the {\it RedGOLD} cluster candidates, we estimate the completeness and purity  with respect 
to the \citet{Gozaliasl2014} catalogue  for the \citet{Milkeraitis2010} and the \citet{Durret2011} detections. In Fig.~\ref{fig:ComplPur_lam_var_Milk}, we show the completeness as a function of the purity for different detection levels,  up to $z\sim0.6$ (left panel) and in the whole redshift range (right panel). 
Red circles and green squares refer to the \citet{Milkeraitis2010} and \citet{Durret2011} catalogues, respectively.  The size of the symbols shows different  thresholds of their detection level: from the smaller to the larger  $\sigma_{Milkeraitis}=3, 5, 6, 8, 10$ and $\sigma_{Durret}=2, 4, 5, 6, 7$.

As expected, the completeness decreases with the increasing detection level thresholds, and reaches $\sim90\%$ at $\sigma_{Milkeraitis}\ge 3$ and $\sim 60\%$ at $\sigma_{Durret}\ge 2$ for the \citet{Milkeraitis2010} and \citet{Durret2011} catalogue, respectively. 
On the other hand, the purity increases with the detection significance: the best compromise between completeness and purity is found for $\sigma_{Milkeraitis}\ge 6$ and for $\sigma_{Durret}\ge 5$.

With this cut, the \citet{Milkeraitis2010} catalogue reaches a completeness of $\sim75\%$ ($\sim60\%$)
at $z\le0.6$ ($z\le 1.1$) and a purity of $\sim75\%$ ($\sim70\%$) at $z\le0.6$ ($z\le 1.1$). 
Similarly, the \citet{Durret2011} catalogue reaches a completeness of $\sim60\%$ ($\sim40\%$)
at $z\le0.6$ ($z\le 1.1$) and a purity of $\sim60\%$ ($\sim50\%$) at $z\le0.6$ ($z\le 1.1$). 
 The blue star represents the value of completeness and purity reached by RedGOLD with the optimised values of $\sigma_{det}$ and $\lambda$.

This comparison shows that the {\it RedGOLD} catalogue reaches a better compromise between completeness and purity at both low and high redshifts with respect to the \citet{Milkeraitis2010} and \citet{Durret2011} catalogues, being more complete and purer when using our thresholds on $\lambda$ and $\sigma_{det}$.

\begin{table}
\begin{center}
\resizebox{!}{1.3 cm}{
\begin{tabular}{c c c }
\hline
RedGOLD & \citet{Milkeraitis2010} & \citet{Durret2011} \\
\hline
$\sigma_{det}<1$ & 45\% & 57\% \\
$\sigma_{det}\ge1$ & 55\% & 43\% \\
$\sigma_{det}\ge2$ & 29\% & 21\% \\
$\sigma_{det}\ge3$ & 11\% & 10\% \\
$\sigma_{det}\ge4$ & 4\% &4\% \\
$\sigma_{det}\ge4.5$ & 2\% & 3\% \\
$N_{tot}$ & 1425 & 1036\\
\hline
\end{tabular}
}
\caption{Percentage of the unmatched cluster candidates for each detection limit when we run our algorithm on the cluster candidate centres from \protect\citet{Milkeraitis2010} and \protect\citet{Durret2011} with  $\sigma_{det}<1$, $\sigma_{det} \ge 1$, $\sigma_{det} \ge 2$, $\sigma_{det} \ge 3$, $\sigma_{det} \ge 4$ and $\sigma_{det} \ge 4.5$. $N_{tot}$ represents the total number of the unmatched detections for each optical cluster catalogue. } 
\label{tab:GrMiDu}
\end{center}
\end{table}

\section{Conclusions} \label{sec:conclusions}
 We present our galaxy cluster detection algorithm {\itshape RedGOLD} and apply it to $\sim 60$~deg$^2$ of the optical survey CFHT-LS W1 to detect clusters up to $z\sim1$ using the CFHTLenS data reduction.  {\itshape RedGOLD} is based on a revised red--sequence overdensity search technique. To properly detect overdensities of passive red-sequence galaxies, we use colour-colour diagrams and color cuts that correspond to 
the $(U-B)$ and $(B-V)$ rest--frame colours of passive ETGs. This permits us to discard blue star-forming galaxies and dusty star-forming galaxies with the same  $(U-B)$  rest--frame colour as passive galaxies at the same redshifts. Photometric redshifts  improve our selection on the red--sequence, and the spectral classification from the SED fitting identifies ETGs. We also impose a constraint on the cluster profile, and {\itshape RedGOLD} only retains detections with a radial distribution in agreement with the NFW profile.

{\itshape RedGOLD} detections are characterised by their significance $\sigma_{det}$. The algorithm also provides the candidate richness $\lambda$ as a proxy of the cluster mass. We adopt the modification of the richness definition from \citet{Rykoff2014} for the redMaPPer algorithm applied to the SDSS, and adapt it to the CFHTLenS depth.  
We show that our richness $\lambda$ is very similar to the richness from \citet{Rykoff2014} up to $z\sim0.3$. At higher redshift, the redMaPPer richness is systematically higher, up to a median difference of $\sim 60\%$. Because the CFHTLenS is deeper than the SDSS, we believe that this difference is partially due to the fact that  we are counting galaxies down to the CFHTLenS depth, while in redMaPPer the richness estimate is extrapolated to a larger depth than the SDSS, but we will investigate this observed difference in a future work by analysing a larger cluster sample.

The detection significance $\sigma_{det}$ and the cluster richness $\lambda$ are the two key parameters for the completeness and purity of the {\it RedGOLD} cluster catalogues.  We calibrate the optimal values of these two parameters using both simulations and X--ray observations from \citet{Gozaliasl2014}.  We apply {\it RedGOLD} to the Millennium Simulations, using the lightcones built by \citet{Henriques2012} based on the \citet{Guo2011} model.  We find that the red--sequence of clusters in their lightcones is not accurately reproduced, with a lack of ETGs and bluer colours than those predicted by the BC03 models (which accurately reproduce the observed colours as a function of redshift). We modify the simulations to correct their biases in colour and ETG fractions. 

From both our calibration on simulations and observations, we obtain the values of $\sigma_{det}$ and $\lambda$ that optimise completeness and purity at the same time: our final cluster catalogue in the CFHT-LS W1 includes candidates with $\lambda\ge10$ and $\sigma_{det}\ge 4$ at $z\le 0.6$, and $\sigma_{det}\ge 4.5$ at $z> 0.6$. For cluster mass $M_{200} > 10^{14}\ {\rm M_{\odot}}$, {\it RedGOLD} is $\sim80\%$ pure up to $z\sim 1.1$. In this mass range, for $z\lesssim0.6$ ($0.6<z<1.1$), the optimal values of $\lambda\ge10$ and $\sigma_{det}\ge4$ ($\lambda \ge10$ and $\sigma_{det}\ge4.5$) give us a completeness of  $\sim100\%$ ($\sim70\%$). 

In the CFHT-LS W1 area analysed in this work, and using the parameter range above, we find $\sim 11$ detections per $\rm deg^2$ up to $z\sim 1.1$. Approximatively $58\%$  of our detections have at least one galaxy with a confirmed spectroscopic redshift from public catalogues available in the area that is within the uncertainty of the cluster photometric redshift. The comparison of our detections with available X-ray detected cluster catalogs confirms our estimated completeness.

Our centring algorithm and our determination of the cluster photometric redshift are very precise: we find that the median separation between the peak of the X--ray emission and our cluster centres is $17.2''\pm11.2''$, and the redshift difference with spectroscopy is less than 0.05 up to $z \sim 1$.

Comparing our catalogue with the redMaPPer detections from \citep{Rykoff2014}, we recover $\sim99\%$ of their detections with no limits on $\lambda$ and $\sigma_{det}$. When applying  the limits on the {\it RedGOLD} parameters, we discard 7 small systems, recovering $\sim93\%$ of the detections in the redMaPPer catalogue.
When comparing with redMaPPer detections which are also in the \citet{Gozaliasl2014} X--ray group catalogue, we find that {\itshape RedGOLD} recovers all the  redMaPPer detections but one  with $M_{200}=8.5\times 10^{13}\ {\rm M_{\odot}}$ at $z\sim0.5$. Our cluster catalogue reaches lower cluster masses with respect to the redMaPPer detections. We believe that this is because  the CFHTLenS is deeper than the SDSS and because the redMaPPer catalogue was built using a different limit in the cluster candidate richness.

We study the $T_X-\lambda$ relation for the {\itshape RedGOLD}  detections using the \citet{Gozaliasl2014}  and \citet{Mehrtens2012} X--ray catalogues.  Up to $z=0.6$ and  using  the \citet{Gozaliasl2014} catalogue, for the relation $\ln(T_X)=A+\alpha\ln(\lambda/\lambda_{pivot})$ we obtain  $A=0.34\pm0.17$, $\alpha=0.81\pm 0.20$ and a scatter of $\sigma=0.28\pm 0.04$, corresponding to a scatter in mass at fixed richness of $\sigma_{M|\lambda}=0.39\pm0.07$.  Using the \citet{Mehrtens2012} catalogue, we obtain $A=1.41\pm0.32$, $\alpha=1.55\pm 0.75$ and a scatter $\sigma=0.23\pm 0.08$, corresponding to $\sigma_{M|\lambda}=0.30\pm0.13$. 
Our results are consistent with \citet{Rozo2014b}, when using both the \citet{Gozaliasl2014} and the \citet{Mehrtens2012} catalogue, even if  we find a slightly higher scatter at fixed richness for the  \citet{Gozaliasl2014} catalogue. This result is very promising because the {\it RedGOLD} catalogue reaches a lower mass threshold. If we apply richness cuts corresponding to $M_{200}\sim7\times 10^{13}\ M_{\odot}$ and $M_{200}\sim \times 10^{14}\ M_{\odot}$, we obtain 
smaller values of the scatter in mass at fixed richness. However, with these higher richness thresholds we only have a small number of points, and we need to extend this analysis to a larger cluster sample.

We compare our {\itshape RedGOLD} cluster catalogue to two optical cluster catalogues publicly available in the same area, the
\citet{Milkeraitis2010} and the \citet{Durret2011} catalogues. 
For cluster masses  $M_{200}\gtrsim10^{14}\ {\rm M_{\odot}}$, {\itshape RedGOLD} recovers  $\sim 80\%$ of the Milkeraitis' detections, and discards a significant fraction of small groups detected in the two catalogues. When we estimate the completeness and purity of these two algorithms, we obtain optimised values that are lower than those of {\itshape RedGOLD} at all redshifts. We find that the best compromise between completeness and purity is found for $\sigma_{Milkeraitis}\ge 6$ and for $\sigma_{Durret}\ge 5$. 
With this cut, the \citet{Milkeraitis2010} catalogue reaches a completeness of $\sim75\%$ ($\sim60\%$)
at $z\le0.6$ ($z\le 1.1$) and a purity of $\sim75\%$ ($\sim70\%$) at $z\le0.6$ ($z\le 1.1$). 
Similarly, the \citet{Durret2011} catalogue reaches a completeness of $\sim60\%$ ($\sim40\%$)
at $z\le0.6$ ($z\le 1.1$) and a purity of $\sim60\%$ ($\sim50\%$) at $z\le0.6$ ($z\le 1.1$).  Comparing these three catalogues at their optimal values of completeness and purity, {we find that \itshape RedGOLD}  is both more complete and purer than \citet{Milkeraitis2010} and \citet{Durret2011}.

Our results show that our cluster detection algorithm {\itshape RedGOLD} is able to effectively detect galaxy clusters with mass $M\gtrsim10^{14}\ {\rm M_{\odot}}$, with a purity of $\sim 80\%$ at $z\lesssim 1.1$, and a completeness of  $\sim100\%$  at $z\le 0.6$,  and  $\sim70\%$ up to $z\sim1$, at the CFHTLenS depth. 

\section*{Acknowledgments}
The Millennium Simulation databases used in this paper and the web application providing online access to them were constructed as part of the activities of the German Astrophysical Virtual Observatory (GAVO). We warmly thank our referee for his/her constructive comments that improved this paper. We thank Eduardo Rozo for insightful discussions on the method and the comparison with the redmaPPer algorithm. We thank James G. Bartlett for the interesting discussions and for carefully editing the abstract and the conclusions.
The French authors (R.L., S.M., A.R.) acknowledge the support of the French Agence Nationale
de la Recherche (ANR) under the reference ANR10-
BLANC-0506-01-Projet VIRAGE (PI: S.Mei). S.M. acknowledges financial support from the Institut Universitaire de France (IUF), of which she is senior member.
H.H. is supported by the DFG Emmy Noether grant Hi 1495/2-1. We thank the Observatory of Paris for hosting T.E. under its visitor program.

\bibliographystyle{mn2e_Daly}
\bibliography{myRefMajor}

\begin{thebibliography}{122}
\expandafter\ifx\csname natexlab\endcsname\relax\def\natexlab#1{#1}\fi

\bibitem[{{Adami} {et~al}\mbox{.}(2010){Adami}, {Durret}, {Benoist}, {Coupon},
  {Mazure}, {Meneux}, {Ilbert}, {Blaizot}, {Arnouts}, {Cappi}, {Garilli},
  {Guennou}, {Lebrun}, {Lef{\`e}vre}, {Maurogordato}, {McCracken}, {Mellier},
  {Slezak}, {Tresse}, \& {Ulmer}}]{Adami2010}
{Adami} C. {et~al.}, 2010, \aap, 509, A81

\bibitem[{{Andreon} \& {Hurn}(2010)}]{Andreon2010}
{Andreon} S., {Hurn} M.~A., 2010, \mnras, 404, 1922

\bibitem[{{Andreon} {et~al}\mbox{.}(2005){Andreon}, {Valtchanov}, {Jones},
  {Altieri}, {Bremer}, {Willis}, {Pierre}, \& {Quintana}}]{Andreon2005}
{Andreon} S., {Valtchanov} I., {Jones} L.~R., {Altieri} B., {Bremer} M.,
  {Willis} J., {Pierre} M., {Quintana} H., 2005, \mnras, 359, 1250

\bibitem[{{Andreon} {et~al}\mbox{.}(2004){Andreon}, {Willis}, {Quintana},
  {Valtchanov}, {Pierre}, \& {Pacaud}}]{Andreon2004}
{Andreon} S., {Willis} J., {Quintana} H., {Valtchanov} I., {Pierre} M.,
  {Pacaud} F., 2004, \mnras, 353, 353

\bibitem[{{Arnouts} {et~al}\mbox{.}(1999){Arnouts}, {Cristiani}, {Moscardini},
  {Matarrese}, {Lucchin}, {Fontana}, \& {Giallongo}}]{Arnouts1999}
{Arnouts} S., {Cristiani} S., {Moscardini} L., {Matarrese} S., {Lucchin} F.,
  {Fontana} A., {Giallongo} E., 1999, \mnras, 310, 540

\bibitem[{{Arnouts} {et~al}\mbox{.}(2002){Arnouts}, {Moscardini}, {Vanzella},
  {Colombi}, {Cristiani}, {Fontana}, {Giallongo}, {Matarrese}, \&
  {Saracco}}]{Arnouts2002}
{Arnouts} S. {et~al.}, 2002, \mnras, 329, 355

\bibitem[{{Ascaso} {et~al}\mbox{.}(2015){Ascaso}, {Mei}, \&
  {Ben{\'{\i}}tez}}]{Ascaso2015}
{Ascaso} B., {Mei} S., {Ben{\'{\i}}tez} N., 2015, ArXiv e-prints

\bibitem[{{Ascaso} {et~al}\mbox{.}(2012){Ascaso}, {Wittman}, \&
  {Ben{\'{\i}}tez}}]{Ascaso2012}
{Ascaso} B., {Wittman} D., {Ben{\'{\i}}tez} N., 2012, \mnras, 420, 1167

\bibitem[{{Bartelmann}(1996)}]{Bartelmann1996}
{Bartelmann} M., 1996, \aap, 313, 697

\bibitem[{{Becker} \& {Kravtsov}(2011)}]{Becker2011}
{Becker} M.~R., {Kravtsov} A.~V., 2011, \apj, 740, 25

\bibitem[{{Ben{\'{\i}}tez}(2000)}]{Benitez2000}
{Ben{\'{\i}}tez} N., 2000, \apj, 536, 571

\bibitem[{{Ben{\'{\i}}tez} {et~al}\mbox{.}(2004){Ben{\'{\i}}tez}, {Ford},
  {Bouwens}, {Menanteau}, {Blakeslee}, {Gronwall}, {Illingworth}, {Meurer},
  {Broadhurst}, {Clampin}, {Franx}, {Hartig}, {Magee}, {Sirianni}, {Ardila},
  {Bartko}, {Brown}, {Burrows}, {Cheng}, {Cross}, {Feldman}, {Golimowski},
  {Infante}, {Kimble}, {Krist}, {Lesser}, {Levay}, {Martel}, {Miley},
  {Postman}, {Rosati}, {Sparks}, {Tran}, {Tsvetanov}, {White}, \&
  {Zheng}}]{Benitez2004}
{Ben{\'{\i}}tez} N. {et~al.}, 2004, \apjs, 150, 1

\bibitem[{{Benjamin} {et~al}\mbox{.}(2013){Benjamin}, {Van Waerbeke},
  {Heymans}, {Kilbinger}, {Erben}, {Hildebrandt}, {Hoekstra}, {Kitching},
  {Mellier}, {Miller}, {Rowe}, {Schrabback}, {Simpson}, {Coupon}, {Fu},
  {Harnois-D{\'e}raps}, {Hudson}, {Kuijken}, {Semboloni}, {Vafaei}, \&
  {Velander}}]{Benjamin2013}
{Benjamin} J. {et~al.}, 2013, \mnras, 431, 1547

\bibitem[{{Berg{\'e}} {et~al}\mbox{.}(2008){Berg{\'e}}, {Pacaud},
  {R{\'e}fr{\'e}gier}, {Massey}, {Pierre}, {Amara}, {Birkinshaw},
  {Paulin-Henriksson}, {Smith}, \& {Willis}}]{Berge2008}
{Berg{\'e}} J. {et~al.}, 2008, \mnras, 385, 695

\bibitem[{{Bertin} \& {Arnouts}(1996)}]{Bertin1996}
{Bertin} E., {Arnouts} S., 1996, \aaps, 117, 393

\bibitem[{{Bessell}(1990)}]{Bessell1990}
{Bessell} M.~S., 1990, \pasp, 102, 1181

\bibitem[{{Boulade} {et~al}\mbox{.}(2003){Boulade}, {Charlot}, {Abbon}, {Aune},
  {Borgeaud}, {Carton}, {Carty}, {Da Costa}, {Deschamps}, {Desforge},
  {Eppell{\'e}}, {Gallais}, {Gosset}, {Granelli}, {Gros}, {de Kat}, {Loiseau},
  {Ritou}, {Rouss{\'e}}, {Starzynski}, {Vignal}, \& {Vigroux}}]{Boulade2003}
{Boulade} O. {et~al.}, 2003, in Society of Photo-Optical Instrumentation
  Engineers (SPIE) Conference Series, Vol. 4841, Instrument Design and
  Performance for Optical/Infrared Ground-based Telescopes, {Iye} M.,
  {Moorwood} A.~F.~M., eds., pp. 72--81

\bibitem[{{Bower} {et~al}\mbox{.}(1992){Bower}, {Lucey}, \&
  {Ellis}}]{Bower1992}
{Bower} R.~G., {Lucey} J.~R., {Ellis} R.~S., 1992, \mnras, 254, 601

\bibitem[{{Brodwin} {et~al}\mbox{.}(2013){Brodwin}, {Stanford}, {Gonzalez},
  {Zeimann}, {Snyder}, {Mancone}, {Pope}, {Eisenhardt}, {Stern}, {Alberts},
  {Ashby}, {Brown}, {Chary}, {Dey}, {Galametz}, {Gettings}, {Jannuzi},
  {Miller}, {Moustakas}, \& {Moustakas}}]{Brodwin2013}
{Brodwin} M. {et~al.}, 2013, \apj, 779, 138

\bibitem[{{Bruzual} \& {Charlot}(2003)}]{Bruzual2003}
{Bruzual} G., {Charlot} S., 2003, \mnras, 344, 1000

\bibitem[{{Cameron}(2011)}]{Cameron2011}
{Cameron} E., 2011, \pasa, 28, 128

\bibitem[{{Capak} {et~al}\mbox{.}(2004){Capak}, {Cowie}, {Hu}, {Barger},
  {Dickinson}, {Fernandez}, {Giavalisco}, {Komiyama}, {Kretchmer}, {McNally},
  {Miyazaki}, {Okamura}, \& {Stern}}]{Capak2004}
{Capak} P. {et~al.}, 2004, \aj, 127, 180

\bibitem[{{Chiang} {et~al}\mbox{.}(2013){Chiang}, {Overzier}, \&
  {Gebhardt}}]{ChiangOverzier2013}
{Chiang} Y.-K., {Overzier} R., {Gebhardt} K., 2013, \apj, 779, 127

\bibitem[{{Coe} {et~al}\mbox{.}(2006){Coe}, {Ben{\'{\i}}tez}, {S{\'a}nchez},
  {Jee}, {Bouwens}, \& {Ford}}]{Coe2006}
{Coe} D., {Ben{\'{\i}}tez} N., {S{\'a}nchez} S.~F., {Jee} M., {Bouwens} R.,
  {Ford} H., 2006, \aj, 132, 926

\bibitem[{{Cohn} {et~al}\mbox{.}(2007){Cohn}, {Evrard}, {White}, {Croton}, \&
  {Ellingson}}]{Cohn2007}
{Cohn} J.~D., {Evrard} A.~E., {White} M., {Croton} D., {Ellingson} E., 2007,
  \mnras, 382, 1738

\bibitem[{{Coleman} {et~al}\mbox{.}(1980){Coleman}, {Wu}, \&
  {Weedman}}]{Coleman1980}
{Coleman} G.~D., {Wu} C.-C., {Weedman} D.~W., 1980, \apjs, 43, 393

\bibitem[{{Collister} \& {Lahav}(2005)}]{Collister2005}
{Collister} A.~A., {Lahav} O., 2005, \mnras, 361, 415

\bibitem[{{Croton} {et~al}\mbox{.}(2006){Croton}, {Springel}, {White}, {De
  Lucia}, {Frenk}, {Gao}, {Jenkins}, {Kauffmann}, {Navarro}, \&
  {Yoshida}}]{Croton2006}
{Croton} D.~J. {et~al.}, 2006, \mnras, 365, 11

\bibitem[{{Desai} {et~al}\mbox{.}(2007){Desai}, {Dalcanton},
  {Arag{\'o}n-Salamanca}, {Jablonka}, {Poggianti}, {Gogarten}, {Simard},
  {Milvang-Jensen}, {Rudnick}, {Zaritsky}, {Clowe}, {Halliday}, {Pell{\'o}},
  {Saglia}, \& {White}}]{Desai2007}
{Desai} V. {et~al.}, 2007, \apj, 660, 1151

\bibitem[{{Dressler}(1980)}]{Dressler1980}
{Dressler} A., 1980, \apj, 236, 351

\bibitem[{{Duffy} {et~al}\mbox{.}(2008){Duffy}, {Schaye}, {Kay}, \& {Dalla
  Vecchia}}]{Duffy2008}
{Duffy} A.~R., {Schaye} J., {Kay} S.~T., {Dalla Vecchia} C., 2008, \mnras, 390,
  L64

\bibitem[{{Durret} {et~al}\mbox{.}(2011){Durret}, {Adami}, {Cappi},
  {Maurogordato}, {M{\'a}rquez}, {Ilbert}, {Coupon}, {Arnouts}, {Benoist},
  {Blaizot}, {Edorh}, {Garilli}, {Guennou}, {Le Brun}, {Le F{\`e}vre},
  {Mazure}, {McCracken}, {Mellier}, {Mezrag}, {Slezak}, {Tresse}, \&
  {Ulmer}}]{Durret2011}
{Durret} F. {et~al.}, 2011, \aap, 535, A65

\bibitem[{{Dutton} \& {Macci{\`o}}(2014)}]{Dutton2014}
{Dutton} A.~A., {Macci{\`o}} A.~V., 2014, \mnras, 441, 3359

\bibitem[{{Eisenhardt} {et~al}\mbox{.}(2008){Eisenhardt}, {Brodwin},
  {Gonzalez}, {Stanford}, {Stern}, {Barmby}, {Brown}, {Dawson}, {Dey}, {Doi},
  {Galametz}, {Jannuzi}, {Kochanek}, {Meyers}, {Morokuma}, \&
  {Moustakas}}]{Eisenhardt2008}
{Eisenhardt} P.~R.~M. {et~al.}, 2008, \apj, 684, 905

\bibitem[{{Erben} {et~al}\mbox{.}(2013){Erben}, {Hildebrandt}, {Miller}, {van
  Waerbeke}, {Heymans}, {Hoekstra}, {Kitching}, {Mellier}, {Benjamin}, {Blake},
  {Bonnett}, {Cordes}, {Coupon}, {Fu}, {Gavazzi}, {Gillis}, {Grocutt}, {Gwyn},
  {Holhjem}, {Hudson}, {Kilbinger}, {Kuijken}, {Milkeraitis}, {Rowe},
  {Schrabback}, {Semboloni}, {Simon}, {Smit}, {Toader}, {Vafaei}, {van Uitert},
  \& {Velander}}]{Erben2013}
{Erben} T. {et~al.}, 2013, \mnras, 433, 2545

\bibitem[{{Evrard} {et~al}\mbox{.}(2008){Evrard}, {Bialek}, {Busha}, {White},
  {Habib}, {Heitmann}, {Warren}, {Rasia}, {Tormen}, {Moscardini}, {Power},
  {Jenkins}, {Gao}, {Frenk}, {Springel}, {White}, \& {Diemand}}]{Evrard2008}
{Evrard} A.~E. {et~al.}, 2008, \apj, 672, 122

\bibitem[{{Ferrarese} {et~al}\mbox{.}(2012){Ferrarese}, {C{\^o}t{\'e}},
  {Cuillandre}, {Gwyn}, {Peng}, {MacArthur}, {Duc}, {Boselli}, {Mei}, {Erben},
  {McConnachie}, {Durrell}, {Mihos}, {Jord{\'a}n}, {Lan{\c c}on}, {Puzia},
  {Emsellem}, {Balogh}, {Blakeslee}, {van Waerbeke}, {Gavazzi}, {Vollmer},
  {Kavelaars}, {Woods}, {Ball}, {Boissier}, {Courteau}, {Ferriere}, {Gavazzi},
  {Hildebrandt}, {Hudelot}, {Huertas-Company}, {Liu}, {McLaughlin}, {Mellier},
  {Milkeraitis}, {Schade}, {Balkowski}, {Bournaud}, {Carlberg}, {Chapman},
  {Hoekstra}, {Peng}, {Sawicki}, {Simard}, {Taylor}, {Tully}, {van Driel},
  {Wilson}, {Burdullis}, {Mahoney}, \& {Manset}}]{Ferrarese2012}
{Ferrarese} L. {et~al.}, 2012, \apjs, 200, 4

\bibitem[{{Finoguenov} {et~al}\mbox{.}(2003){Finoguenov}, {Borgani},
  {Tornatore}, \& {B{\"o}hringer}}]{Finoguenov2003}
{Finoguenov} A., {Borgani} S., {Tornatore} L., {B{\"o}hringer} H., 2003, \aap,
  398, L35

\bibitem[{{Finoguenov} {et~al}\mbox{.}(2009){Finoguenov}, {Connelly}, {Parker},
  {Wilman}, {Mulchaey}, {Saglia}, {Balogh}, {Bower}, \&
  {McGee}}]{Finoguenov2009}
{Finoguenov} A. {et~al.}, 2009, \apj, 704, 564

\bibitem[{{Ford} {et~al}\mbox{.}(2015){Ford}, {Van Waerbeke}, {Milkeraitis},
  {Laigle}, {Hildebrandt}, {Erben}, {Heymans}, {Hoekstra}, {Kitching},
  {Mellier}, {Miller}, {Choi}, {Coupon}, {Fu}, {Hudson}, {Kuijken},
  {Robertson}, {Rowe}, {Schrabback}, \& {Velander}}]{Ford2015}
{Ford} J. {et~al.}, 2015, \mnras, 447, 1304

\bibitem[{{Gehrels}(1986)}]{Gehrels1986}
{Gehrels} N., 1986, \apj, 303, 336

\bibitem[{{George} {et~al}\mbox{.}(2012){George}, {Leauthaud}, {Bundy},
  {Finoguenov}, {Ma}, {Rykoff}, {Tinker}, {Wechsler}, {Massey}, \&
  {Mei}}]{George2012}
{George} M.~R. {et~al.}, 2012, \apj, 757, 2

\bibitem[{{George} {et~al}\mbox{.}(2011){George}, {Leauthaud}, {Bundy},
  {Finoguenov}, {Tinker}, {Lin}, {Mei}, {Kneib}, {Aussel}, {Behroozi}, {Busha},
  {Capak}, {Coccato}, {Covone}, {Faure}, {Fiorenza}, {Ilbert}, {Le Floc'h},
  {Koekemoer}, {Tanaka}, {Wechsler}, \& {Wolk}}]{George2011}
{George} M.~R. {et~al.}, 2011, \apj, 742, 125

\bibitem[{{Gillis} {et~al}\mbox{.}(2013){Gillis}, {Hudson}, {Erben}, {Heymans},
  {Hildebrandt}, {Hoekstra}, {Kitching}, {Mellier}, {Miller}, {van Waerbeke},
  {Bonnett}, {Coupon}, {Fu}, {Hilbert}, {Rowe}, {Schrabback}, {Semboloni}, {van
  Uitert}, \& {Velander}}]{Gillis2013}
{Gillis} B.~R. {et~al.}, 2013, \mnras, 431, 1439

\bibitem[{{Gladders} \& {Yee}(2000)}]{Gladders2000}
{Gladders} M.~D., {Yee} H.~K.~C., 2000, \aj, 120, 2148

\bibitem[{{Gozaliasl} {et~al}\mbox{.}(2014){Gozaliasl}, {Finoguenov},
  {Khosroshahi}, {Mirkazemi}, {Salvato}, {Jassur}, {Erfanianfar}, {Popesso},
  {Tanaka}, {Lerchster}, {Kneib}, {McCracken}, {Mellier}, {Egami}, {Pereira},
  {Brimioulle}, {Erben}, \& {Seitz}}]{Gozaliasl2014}
{Gozaliasl} G. {et~al.}, 2014, \aap, 566, A140

\bibitem[{{Grove} {et~al}\mbox{.}(2009){Grove}, {Benoist}, \&
  {Martel}}]{Grove2009}
{Grove} L.~F., {Benoist} C., {Martel} F., 2009, \aap, 494, 845

\bibitem[{{Guo} {et~al}\mbox{.}(2011){Guo}, {White}, {Boylan-Kolchin}, {De
  Lucia}, {Kauffmann}, {Lemson}, {Li}, {Springel}, \& {Weinmann}}]{Guo2011}
{Guo} Q. {et~al.}, 2011, \mnras, 413, 101

\bibitem[{{Guzzo} {et~al}\mbox{.}(2014){Guzzo}, {Scodeggio}, {Garilli},
  {Granett}, {Fritz}, {Abbas}, {Adami}, {Arnouts}, {Bel}, {Bolzonella},
  {Bottini}, {Branchini}, {Cappi}, {Coupon}, {Cucciati}, {Davidzon}, {De
  Lucia}, {de la Torre}, {Franzetti}, {Fumana}, {Hudelot}, {Ilbert}, {Iovino},
  {Krywult}, {Le Brun}, {Le F{\`e}vre}, {Maccagni}, {Ma{\l}ek}, {Marulli},
  {McCracken}, {Paioro}, {Peacock}, {Polletta}, {Pollo}, {Schlagenhaufer},
  {Tasca}, {Tojeiro}, {Vergani}, {Zamorani}, {Zanichelli}, {Burden}, {Di
  Porto}, {Marchetti}, {Marinoni}, {Mellier}, {Moscardini}, {Nichol},
  {Percival}, {Phleps}, \& {Wolk}}]{Guzzo2014}
{Guzzo} L. {et~al.}, 2014, \aap, 566, A108

\bibitem[{{Gwyn}(2012)}]{Gwyn2012}
{Gwyn} S.~D.~J., 2012, \apj, 143, 38

\bibitem[{{Henriques} {et~al}\mbox{.}(2012){Henriques}, {White}, {Lemson},
  {Thomas}, {Guo}, {Marleau}, \& {Overzier}}]{Henriques2012}
{Henriques} B.~M.~B., {White} S.~D.~M., {Lemson} G., {Thomas} P.~A., {Guo} Q.,
  {Marleau} G.-D., {Overzier} R.~A., 2012, \mnras, 421, 2904

\bibitem[{{Heymans} {et~al}\mbox{.}(2012){Heymans}, {Van Waerbeke}, {Miller},
  {Erben}, {Hildebrandt}, {Hoekstra}, {Kitching}, {Mellier}, {Simon},
  {Bonnett}, {Coupon}, {Fu}, {Harnois D{\'e}raps}, {Hudson}, {Kilbinger},
  {Kuijken}, {Rowe}, {Schrabback}, {Semboloni}, {van Uitert}, {Vafaei}, \&
  {Velander}}]{Heymans2012}
{Heymans} C. {et~al.}, 2012, \mnras, 427, 146

\bibitem[{{High} {et~al}\mbox{.}(2010){High}, {Stalder}, {Song}, {Ade}, {Aird},
  {Allam}, {Armstrong}, {Barkhouse}, {Benson}, {Bertin}, {Bhattacharya},
  {Bleem}, {Brodwin}, {Buckley-Geer}, {Carlstrom}, {Challis}, {Chang},
  {Crawford}, {Crites}, {de Haan}, {Desai}, {Dobbs}, {Dudley}, {Foley},
  {George}, {Gladders}, {Halverson}, {Hamuy}, {Hansen}, {Holder}, {Holzapfel},
  {Hrubes}, {Joy}, {Keisler}, {Lee}, {Leitch}, {Lin}, {Lin}, {Loehr}, {Lueker},
  {Marrone}, {McMahon}, {Mehl}, {Meyer}, {Mohr}, {Montroy}, {Morell}, {Ngeow},
  {Padin}, {Plagge}, {Pryke}, {Reichardt}, {Rest}, {Ruel}, {Ruhl}, {Schaffer},
  {Shaw}, {Shirokoff}, {Smith}, {Spieler}, {Staniszewski}, {Stark}, {Stubbs},
  {Tucker}, {Vanderlinde}, {Vieira}, {Williamson}, {Wood-Vasey}, {Yang},
  {Zahn}, \& {Zenteno}}]{High2010}
{High} F.~W. {et~al.}, 2010, \apj, 723, 1736

\bibitem[{{Hilbert} \& {White}(2010)}]{Hilbert2010}
{Hilbert} S., {White} S.~D.~M., 2010, \mnras, 404, 486

\bibitem[{{Hildebrandt} {et~al}\mbox{.}(2012){Hildebrandt}, {Erben}, {Kuijken},
  {van Waerbeke}, {Heymans}, {Coupon}, {Benjamin}, {Bonnett}, {Fu}, {Hoekstra},
  {Kitching}, {Mellier}, {Miller}, {Velander}, {Hudson}, {Rowe}, {Schrabback},
  {Semboloni}, \& {Ben{\'{\i}}tez}}]{Hildebrandt2012}
{Hildebrandt} H. {et~al.}, 2012, \mnras, 421, 2355

\bibitem[{{Holland} {et~al}\mbox{.}(2015){Holland}, {B{\"o}hringer}, {Chon}, \&
  {Pierini}}]{Holland2015}
{Holland} J.~G., {B{\"o}hringer} H., {Chon} G., {Pierini} D., 2015, \mnras,
  448, 2644

\bibitem[{{Ilbert} {et~al}\mbox{.}(2006){Ilbert}, {Arnouts}, {McCracken},
  {Bolzonella}, {Bertin}, {Le F{\`e}vre}, {Mellier}, {Zamorani}, {Pell{\`o}},
  {Iovino}, {Tresse}, {Le Brun}, {Bottini}, {Garilli}, {Maccagni}, {Picat},
  {Scaramella}, {Scodeggio}, {Vettolani}, {Zanichelli}, {Adami}, {Bardelli},
  {Cappi}, {Charlot}, {Ciliegi}, {Contini}, {Cucciati}, {Foucaud}, {Franzetti},
  {Gavignaud}, {Guzzo}, {Marano}, {Marinoni}, {Mazure}, {Meneux}, {Merighi},
  {Paltani}, {Pollo}, {Pozzetti}, {Radovich}, {Zucca}, {Bondi}, {Bongiorno},
  {Busarello}, {de La Torre}, {Gregorini}, {Lamareille}, {Mathez}, {Merluzzi},
  {Ripepi}, {Rizzo}, \& {Vergani}}]{Ilbert2006}
{Ilbert} O. {et~al.}, 2006, \aap, 457, 841

\bibitem[{{Johnston} {et~al}\mbox{.}(2007){Johnston}, {Sheldon}, {Wechsler},
  {Rozo}, {Koester}, {Frieman}, {McKay}, {Evrard}, {Becker}, \&
  {Annis}}]{Johnston2007}
{Johnston} D.~E. {et~al.}, 2007, ArXiv e-prints

\bibitem[{{Jones} {et~al}\mbox{.}(2003){Jones}, {Ponman}, {Horton}, {Babul},
  {Ebeling}, \& {Burke}}]{Jones2003}
{Jones} L.~R., {Ponman} T.~J., {Horton} A., {Babul} A., {Ebeling} H., {Burke}
  D.~J., 2003, \mnras, 343, 627

\bibitem[{{Kim} {et~al}\mbox{.}(2002){Kim}, {Kepner}, {Postman}, {Strauss},
  {Bahcall}, {Gunn}, {Lupton}, {Annis}, {Nichol}, {Castander}, {Brinkmann},
  {Brunner}, {Connolly}, {Csabai}, {Hindsley}, {Ivezi{\'c}}, {Vogeley}, \&
  {York}}]{Kim2002}
{Kim} R.~S.~J. {et~al.}, 2002, \aj, 123, 20

\bibitem[{{Kinney} {et~al}\mbox{.}(1996){Kinney}, {Calzetti}, {Bohlin},
  {McQuade}, {Storchi-Bergmann}, \& {Schmitt}}]{Kinney1996}
{Kinney} A.~L., {Calzetti} D., {Bohlin} R.~C., {McQuade} K., {Storchi-Bergmann}
  T., {Schmitt} H.~R., 1996, \apj, 467, 38

\bibitem[{{Kitzbichler} \& {White}(2007)}]{Kitzbichler2007}
{Kitzbichler} M.~G., {White} S.~D.~M., 2007, \mnras, 376, 2

\bibitem[{{Klypin} {et~al}\mbox{.}(2014){Klypin}, {Yepes}, {Gottlober},
  {Prada}, \& {Hess}}]{Klypin2014}
{Klypin} A., {Yepes} G., {Gottlober} S., {Prada} F., {Hess} S., 2014, ArXiv
  e-prints

\bibitem[{{Koester} {et~al}\mbox{.}(2007){Koester}, {McKay}, {Annis},
  {Wechsler}, {Evrard}, {Bleem}, {Becker}, {Johnston}, {Sheldon}, {Nichol},
  {Miller}, {Scranton}, {Bahcall}, {Barentine}, {Brewington}, {Brinkmann},
  {Harvanek}, {Kleinman}, {Krzesinski}, {Long}, {Nitta}, {Schneider},
  {Sneddin}, {Voges}, \& {York}}]{Koester2007}
{Koester} B.~P. {et~al.}, 2007, \apj, 660, 239

\bibitem[{{Larson} \& {Tinsley}(1978)}]{Larson1978}
{Larson} R.~B., {Tinsley} B.~M., 1978, \apj, 219, 46

\bibitem[{{Laureijs} {et~al}\mbox{.}(2011){Laureijs}, {Amiaux}, {Arduini},
  {Augu{\`e}res}, {Brinchmann}, {Cole}, {Cropper}, {Dabin}, {Duvet}, {Ealet},
  \& et~al.}]{laureijs2011}
{Laureijs} R. {et~al.}, 2011, ArXiv e-prints

\bibitem[{{Le F{\`e}vre} {et~al}\mbox{.}(2013){Le F{\`e}vre}, {Cassata},
  {Cucciati}, {Garilli}, {Ilbert}, {Le Brun}, {Maccagni}, {Moreau},
  {Scodeggio}, {Tresse}, {Zamorani}, {Adami}, {Arnouts}, {Bardelli},
  {Bolzonella}, {Bondi}, {Bongiorno}, {Bottini}, {Cappi}, {Charlot}, {Ciliegi},
  {Contini}, {de la Torre}, {Foucaud}, {Franzetti}, {Gavignaud}, {Guzzo},
  {Iovino}, {Lemaux}, {L{\'o}pez-Sanjuan}, {McCracken}, {Marano}, {Marinoni},
  {Mazure}, {Mellier}, {Merighi}, {Merluzzi}, {Paltani}, {Pell{\`o}}, {Pollo},
  {Pozzetti}, {Scaramella}, {Tasca}, {Vergani}, {Vettolani}, {Zanichelli}, \&
  {Zucca}}]{Lefevre2013}
{Le F{\`e}vre} O. {et~al.}, 2013, \aap, 559, A14

\bibitem[{{Le F{\`e}vre} {et~al}\mbox{.}(2005){Le F{\`e}vre}, {Vettolani},
  {Garilli}, {Tresse}, {Bottini}, {Le Brun}, {Maccagni}, {Picat}, {Scaramella},
  {Scodeggio}, {Zanichelli}, {Adami}, {Arnaboldi}, {Arnouts}, {Bardelli},
  {Bolzonella}, {Cappi}, {Charlot}, {Ciliegi}, {Contini}, {Foucaud},
  {Franzetti}, {Gavignaud}, {Guzzo}, {Ilbert}, {Iovino}, {McCracken}, {Marano},
  {Marinoni}, {Mathez}, {Mazure}, {Meneux}, {Merighi}, {Paltani}, {Pell{\`o}},
  {Pollo}, {Pozzetti}, {Radovich}, {Zamorani}, {Zucca}, {Bondi}, {Bongiorno},
  {Busarello}, {Lamareille}, {Mellier}, {Merluzzi}, {Ripepi}, \&
  {Rizzo}}]{Lefevre2005}
{Le F{\`e}vre} O. {et~al.}, 2005, \aap, 439, 845

\bibitem[{{Lin} {et~al}\mbox{.}(2004){Lin}, {Mohr}, \& {Stanford}}]{Lin2004}
{Lin} Y.-T., {Mohr} J.~J., {Stanford} S.~A., 2004, \apj, 610, 745

\bibitem[{{LSST Dark Energy Science Collaboration}(2012)}]{lsstsb2012}
{LSST Dark Energy Science Collaboration}, 2012, ArXiv e-prints

\bibitem[{{Ma{\'{\i}}z Apell{\'a}niz}(2006)}]{MaizApellaniz2006}
{Ma{\'{\i}}z Apell{\'a}niz} J., 2006, \aj, 131, 1184

\bibitem[{{Mantz} {et~al}\mbox{.}(2010){Mantz}, {Allen}, {Ebeling}, {Rapetti},
  \& {Drlica-Wagner}}]{Mantz2010}
{Mantz} A., {Allen} S.~W., {Ebeling} H., {Rapetti} D., {Drlica-Wagner} A.,
  2010, \mnras, 406, 1773

\bibitem[{{Mazure} {et~al}\mbox{.}(2007){Mazure}, {Adami}, {Pierre}, {Le
  F{\`e}vre}, {Arnouts}, {Duc}, {Ilbert}, {Lebrun}, {Meneux}, {Pacaud},
  {Surdej}, \& {Valtchanov}}]{Mazure2007}
{Mazure} A. {et~al.}, 2007, \aap, 467, 49

\bibitem[{{McGee} {et~al}\mbox{.}(2009){McGee}, {Balogh}, {Bower}, {Font}, \&
  {McCarthy}}]{McGee2009}
{McGee} S.~L., {Balogh} M.~L., {Bower} R.~G., {Font} A.~S., {McCarthy} I.~G.,
  2009, \mnras, 400, 937

\bibitem[{{Mead} {et~al}\mbox{.}(2010){Mead}, {King}, {Sijacki}, {Leonard},
  {Puchwein}, \& {McCarthy}}]{Mead2010}
{Mead} J.~M.~G., {King} L.~J., {Sijacki} D., {Leonard} A., {Puchwein} E.,
  {McCarthy} I.~G., 2010, \mnras, 406, 434

\bibitem[{{Mehrtens} {et~al}\mbox{.}(2012){Mehrtens}, {Romer}, {Hilton},
  {Lloyd-Davies}, {Miller}, {Stanford}, {Hosmer}, {Hoyle}, {Collins}, {Liddle},
  {Viana}, {Nichol}, {Stott}, {Dubois}, {Kay}, {Sahl{\'e}n}, {Young}, {Short},
  {Christodoulou}, {Watson}, {Davidson}, {Harrison}, {Baruah}, {Smith},
  {Burke}, {Mayers}, {Deadman}, {Rooney}, {Edmondson}, {West}, {Campbell},
  {Edge}, {Mann}, {Sabirli}, {Wake}, {Benoist}, {da Costa}, {Maia}, \&
  {Ogando}}]{Mehrtens2012}
{Mehrtens} N. {et~al.}, 2012, \mnras, 423, 1024

\bibitem[{{Mei} {et~al}\mbox{.}(2009){Mei}, {Holden}, {Blakeslee}, {Ford},
  {Franx}, {Homeier}, {Illingworth}, {Jee}, {Overzier}, {Postman}, {Rosati},
  {Van der Wel}, \& {Bartlett}}]{Mei2009}
{Mei} S. {et~al.}, 2009, \apj, 690, 42

\bibitem[{{Mei} {et~al}\mbox{.}(2015){Mei}, {Scarlata}, {Pentericci}, {Newman},
  {Weiner}, {Ashby}, {Castellano}, {Conselice}, {Finkelstein}, {Galametz},
  {Grogin}, {Koekemoer}, {Huertas-Company}, {Lani}, {Lucas}, {Papovich},
  {Rafelski}, \& {Teplitz}}]{Mei2015}
{Mei} S. {et~al.}, 2015, \apj, 804, 117

\bibitem[{{Mei} {et~al}\mbox{.}(2012){Mei}, {Stanford}, {Holden}, {Raichoor},
  {Postman}, {Nakata}, {Finoguenov}, {Ford}, {Illingworth}, {Kodama}, {Rosati},
  {Tanaka}, {Huertas-Company}, {Rettura}, {Shankar}, {Carrasco}, {Demarco},
  {Eisenhardt}, {Jee}, {Koyama}, \& {White}}]{Mei2012}
{Mei} S. {et~al.}, 2012, \apj, 754, 141

\bibitem[{{Milkeraitis} {et~al}\mbox{.}(2010){Milkeraitis}, {van Waerbeke},
  {Heymans}, {Hildebrandt}, {Dietrich}, \& {Erben}}]{Milkeraitis2010}
{Milkeraitis} M., {van Waerbeke} L., {Heymans} C., {Hildebrandt} H., {Dietrich}
  J.~P., {Erben} T., 2010, \mnras, 406, 673

\bibitem[{{Miller} {et~al}\mbox{.}(2013){Miller}, {Heymans}, {Kitching}, {van
  Waerbeke}, {Erben}, {Hildebrandt}, {Hoekstra}, {Mellier}, {Rowe}, {Coupon},
  {Dietrich}, {Fu}, {Harnois-D{\'e}raps}, {Hudson}, {Kilbinger}, {Kuijken},
  {Schrabback}, {Semboloni}, {Vafaei}, \& {Velander}}]{Miller2013}
{Miller} L. {et~al.}, 2013, \mnras, 429, 2858

\bibitem[{{Miyazaki} {et~al}\mbox{.}(2007){Miyazaki}, {Hamana}, {Ellis},
  {Kashikawa}, {Massey}, {Taylor}, \& {Refregier}}]{Miyazaki2007}
{Miyazaki} S., {Hamana} T., {Ellis} R.~S., {Kashikawa} N., {Massey} R.~J.,
  {Taylor} J., {Refregier} A., 2007, \apj, 669, 714

\bibitem[{{Navarro} {et~al}\mbox{.}(1996){Navarro}, {Frenk}, \&
  {White}}]{Navarro1996}
{Navarro} J.~F., {Frenk} C.~S., {White} S.~D.~M., 1996, \apj, 462, 563

\bibitem[{{Oke} \& {Gunn}(1983)}]{Oke1983}
{Oke} J.~B., {Gunn} J.~E., 1983, \apj, 266, 713

\bibitem[{{Olsen} {et~al}\mbox{.}(2007){Olsen}, {Benoist}, {Cappi},
  {Maurogordato}, {Mazure}, {Slezak}, {Adami}, {Ferrari}, \&
  {Martel}}]{Olsen2007}
{Olsen} L.~F. {et~al.}, 2007, \aap, 461, 81

\bibitem[{{Pacaud} {et~al}\mbox{.}(2007){Pacaud}, {Pierre}, {Adami}, {Altieri},
  {Andreon}, {Chiappetti}, {Detal}, {Duc}, {Galaz}, {Gueguen}, {Le F{\`e}vre},
  {Hertling}, {Libbrecht}, {Melin}, {Ponman}, {Quintana}, {Refregier},
  {Sprimont}, {Surdej}, {Valtchanov}, {Willis}, {Alloin}, {Birkinshaw},
  {Bremer}, {Garcet}, {Jean}, {Jones}, {Le F{\`e}vre}, {Maccagni}, {Mazure},
  {Proust}, {R{\"o}ttgering}, \& {Trinchieri}}]{Pacaud2007}
{Pacaud} F. {et~al.}, 2007, \mnras, 382, 1289

\bibitem[{{Pierre} {et~al}\mbox{.}(2006){Pierre}, {Pacaud}, {Duc}, {Willis},
  {Andreon}, {Valtchanov}, {Altieri}, {Galaz}, {Gueguen}, {Le F{\`e}vre},
  {F{\`e}vre}, {Ponman}, {Sprimont}, {Surdej}, {Adami}, {Alshino}, {Bremer},
  {Chiappetti}, {Detal}, {Garcet}, {Gosset}, {Jean}, {Maccagni}, {Marinoni},
  {Mazure}, {Quintana}, \& {Read}}]{Pierre2006}
{Pierre} M. {et~al.}, 2006, \mnras, 372, 591

\bibitem[{{Postman} {et~al}\mbox{.}(2005){Postman}, {Franx}, {Cross}, {Holden},
  {Ford}, {Illingworth}, {Goto}, {Demarco}, {Rosati}, {Blakeslee}, {Tran},
  {Ben{\'{\i}}tez}, {Clampin}, {Hartig}, {Homeier}, {Ardila}, {Bartko},
  {Bouwens}, {Bradley}, {Broadhurst}, {Brown}, {Burrows}, {Cheng}, {Feldman},
  {Golimowski}, {Gronwall}, {Infante}, {Kimble}, {Krist}, {Lesser}, {Martel},
  {Mei}, {Menanteau}, {Meurer}, {Miley}, {Motta}, {Sirianni}, {Sparks}, {Tran},
  {Tsvetanov}, {White}, \& {Zheng}}]{Postman2005}
{Postman} M. {et~al.}, 2005, \apj, 623, 721

\bibitem[{{Postman} {et~al}\mbox{.}(1996){Postman}, {Lubin}, {Gunn}, {Oke},
  {Hoessel}, {Schneider}, \& {Christensen}}]{Postman1996}
{Postman} M., {Lubin} L.~M., {Gunn} J.~E., {Oke} J.~B., {Hoessel} J.~G.,
  {Schneider} D.~P., {Christensen} J.~A., 1996, \aj, 111, 615

\bibitem[{{Povi{\'c}} {et~al}\mbox{.}(2013){Povi{\'c}}, {Huertas-Company},
  {Aguerri}, {M{\'a}rquez}, {Masegosa}, {Husillos}, {Molino},
  {Crist{\'o}bal-Hornillos}, {Perea}, {Ben{\'{\i}}tez}, {Olmo},
  {Fern{\'a}ndez-Soto}, {Jim{\'e}nez-Teja}, {Moles}, {Alfaro},
  {Aparicio-Villegas}, {Ascaso}, {Broadhurst}, {Cabrera-Ca{\~n}o}, {Castander},
  {Cepa}, {Fernandez Lorenzo}, {Cervi{\~n}o}, {Delgado}, {Infante},
  {L{\'o}pez-Sanjuan}, {Mart{\'{\i}}nez}, {Matute}, {Oteo},
  {P{\'e}rez-Garc{\'{\i}}a}, {Prada}, \& {Quintana}}]{Povic2013}
{Povi{\'c}} M. {et~al.}, 2013, \mnras, 435, 3444

\bibitem[{{Press} {et~al}\mbox{.}(1992){Press}, {Teukolsky}, {Vetterling}, \&
  {Flannery}}]{Press1992}
{Press} W.~H., {Teukolsky} S.~A., {Vetterling} W.~T., {Flannery} B.~P., 1992,
  {Numerical recipes in C. The art of scientific computing}

\bibitem[{{Prevot} {et~al}\mbox{.}(1984){Prevot}, {Lequeux}, {Prevot},
  {Maurice}, \& {Rocca-Volmerange}}]{Prevot1984}
{Prevot} M.~L., {Lequeux} J., {Prevot} L., {Maurice} E., {Rocca-Volmerange} B.,
  1984, \aap, 132, 389

\bibitem[{{Proctor} {et~al}\mbox{.}(2011){Proctor}, {de Oliveira}, {Dupke}, {de
  Oliveira}, {Cypriano}, {Miller}, \& {Rykoff}}]{Proctor2011}
{Proctor} R.~N., {de Oliveira} C.~M., {Dupke} R., {de Oliveira} R.~L.,
  {Cypriano} E.~S., {Miller} E.~D., {Rykoff} E., 2011, \mnras, 418, 2054

\bibitem[{{Raichoor} {et~al}\mbox{.}(2014){Raichoor}, {Mei}, {Erben},
  {Hildebrandt}, {Huertas-Company}, {Ilbert}, {Licitra}, {Ball}, {Boissier},
  {Boselli}, {Chen}, {C{\^o}t{\'e}}, {Cuillandre}, {Duc}, {Durrell},
  {Ferrarese}, {Guhathakurta}, {Gwyn}, {Kavelaars}, {Lan{\c c}on}, {Liu},
  {MacArthur}, {Muller}, {Mu{\~n}oz}, {Peng}, {Puzia}, {Sawicki}, {Toloba},
  {Van Waerbeke}, {Woods}, \& {Zhang}}]{Raichoor2014}
{Raichoor} A. {et~al.}, 2014, \apj, 797, 102

\bibitem[{{Rasia} {et~al}\mbox{.}(2006){Rasia}, {Ettori}, {Moscardini},
  {Mazzotta}, {Borgani}, {Dolag}, {Tormen}, {Cheng}, \& {Diaferio}}]{Rasia2006}
{Rasia} E. {et~al.}, 2006, \mnras, 369, 2013

\bibitem[{{Rozo} {et~al}\mbox{.}(2014){Rozo}, {Bartlett}, {Evrard}, \&
  {Rykoff}}]{Rozo2014c}
{Rozo} E., {Bartlett} J.~G., {Evrard} A.~E., {Rykoff} E.~S., 2014, \mnras, 438,
  78

\bibitem[{{Rozo} {et~al}\mbox{.}(2011){Rozo}, {Rykoff}, {Koester}, {Nord},
  {Wu}, {Evrard}, \& {Wechsler}}]{Rozo2011}
{Rozo} E., {Rykoff} E., {Koester} B., {Nord} B., {Wu} H.-Y., {Evrard} A.,
  {Wechsler} R., 2011, \apj, 740, 53

\bibitem[{{Rozo} \& {Rykoff}(2014)}]{Rozo2014b}
{Rozo} E., {Rykoff} E.~S., 2014, \apj, 783, 80

\bibitem[{{Rozo} {et~al}\mbox{.}(2009{\natexlab{a}}){Rozo}, {Rykoff}, {Evrard},
  {Becker}, {McKay}, {Wechsler}, {Koester}, {Hao}, {Hansen}, {Sheldon},
  {Johnston}, {Annis}, \& {Frieman}}]{Rozo2009a}
{Rozo} E. {et~al.}, 2009{\natexlab{a}}, \apj, 699, 768

\bibitem[{{Rozo} {et~al}\mbox{.}(2009{\natexlab{b}}){Rozo}, {Rykoff},
  {Koester}, {McKay}, {Hao}, {Evrard}, {Wechsler}, {Hansen}, {Sheldon},
  {Johnston}, {Becker}, {Annis}, {Bleem}, \& {Scranton}}]{Rozo2009b}
{Rozo} E. {et~al.}, 2009{\natexlab{b}}, \apj, 703, 601

\bibitem[{{Rykoff} {et~al}\mbox{.}(2012){Rykoff}, {Koester}, {Rozo}, {Annis},
  {Evrard}, {Hansen}, {Hao}, {Johnston}, {McKay}, \& {Wechsler}}]{Rykoff2012}
{Rykoff} E.~S. {et~al.}, 2012, \apj, 746, 178

\bibitem[{{Rykoff} {et~al}\mbox{.}(2014){Rykoff}, {Rozo}, {Busha}, {Cunha},
  {Finoguenov}, {Evrard}, {Hao}, {Koester}, {Leauthaud}, {Nord}, {Pierre},
  {Reddick}, {Sadibekova}, {Sheldon}, \& {Wechsler}}]{Rykoff2014}
{Rykoff} E.~S. {et~al.}, 2014, \apj, 785, 104

\bibitem[{{Salpeter}(1955)}]{Salpeter1955}
{Salpeter} E.~E., 1955, \apj, 121, 161

\bibitem[{{Schechter}(1976)}]{Schechter1976}
{Schechter} P., 1976, \apj, 203, 297

\bibitem[{{Shankar} {et~al}\mbox{.}(2013){Shankar}, {Marulli}, {Bernardi},
  {Mei}, {Meert}, \& {Vikram}}]{Shankar2013}
{Shankar} F., {Marulli} F., {Bernardi} M., {Mei} S., {Meert} A., {Vikram} V.,
  2013, \mnras, 428, 109

\bibitem[{{Shankar} {et~al}\mbox{.}(2014){Shankar}, {Mei}, {Huertas-Company},
  {Moreno}, {Fontanot}, {Monaco}, {Bernardi}, {Cattaneo}, {Sheth}, {Licitra},
  {Delaye}, \& {Raichoor}}]{Shankar2014}
{Shankar} F. {et~al.}, 2014, \mnras, 439, 3189

\bibitem[{{Sirianni} {et~al}\mbox{.}(2005){Sirianni}, {Jee}, {Ben{\'{\i}}tez},
  {Blakeslee}, {Martel}, {Meurer}, {Clampin}, {De Marchi}, {Ford}, {Gilliland},
  {Hartig}, {Illingworth}, {Mack}, \& {McCann}}]{Sirianni2005}
{Sirianni} M. {et~al.}, 2005, \pasp, 117, 1049

\bibitem[{{Smith} {et~al}\mbox{.}(2005){Smith}, {Treu}, {Ellis}, {Moran}, \&
  {Dressler}}]{Smith2005}
{Smith} G.~P., {Treu} T., {Ellis} R.~S., {Moran} S.~M., {Dressler} A., 2005,
  \apj, 620, 78

\bibitem[{{Snyder} {et~al}\mbox{.}(2012){Snyder}, {Brodwin}, {Mancone},
  {Zeimann}, {Stanford}, {Gonzalez}, {Stern}, {Eisenhardt}, {Brown}, {Dey},
  {Jannuzi}, \& {Perlmutter}}]{Snyder2012}
{Snyder} G.~F. {et~al.}, 2012, \apj, 756, 114

\bibitem[{{Soares-Santos} {et~al}\mbox{.}(2011){Soares-Santos}, {de Carvalho},
  {Annis}, {Gal}, {La Barbera}, {Lopes}, {Wechsler}, {Busha}, \&
  {Gerke}}]{Soares-Santos2011}
{Soares-Santos} M. {et~al.}, 2011, \apj, 727, 45

\bibitem[{{Springel} {et~al}\mbox{.}(2005){Springel}, {White}, {Jenkins},
  {Frenk}, {Yoshida}, {Gao}, {Navarro}, {Thacker}, {Croton}, {Helly},
  {Peacock}, {Cole}, {Thomas}, {Couchman}, {Evrard}, {Colberg}, \&
  {Pearce}}]{Springel2005}
{Springel} V. {et~al.}, 2005, \nat, 435, 629

\bibitem[{{Sunyaev} \& {Zeldovich}(1970)}]{Sunyaev1970}
{Sunyaev} R.~A., {Zeldovich} Y.~B., 1970, Comments on Astrophysics and Space
  Physics, 2, 66

\bibitem[{{Thanjavur} {et~al}\mbox{.}(2009){Thanjavur}, {Willis}, \&
  {Crampton}}]{Thanjavur2009}
{Thanjavur} K., {Willis} J., {Crampton} D., 2009, \apj, 706, 571

\bibitem[{{Treu} {et~al}\mbox{.}(2003){Treu}, {Ellis}, {Kneib}, {Dressler},
  {Smail}, {Czoske}, {Oemler}, \& {Natarajan}}]{Treu2003}
{Treu} T., {Ellis} R.~S., {Kneib} J.-P., {Dressler} A., {Smail} I., {Czoske}
  O., {Oemler} A., {Natarajan} P., 2003, \apj, 591, 53

\bibitem[{{Valtchanov} {et~al}\mbox{.}(2004){Valtchanov}, {Pierre}, {Willis},
  {Dos Santos}, {Jones}, {Andreon}, {Adami}, {Altieri}, {Bolzonella}, {Bremer},
  {Duc}, {Gosset}, {Jean}, \& {Surdej}}]{Valtchanov2004}
{Valtchanov} I. {et~al.}, 2004, \aap, 423, 75

\bibitem[{{Vikhlinin} {et~al}\mbox{.}(2009){Vikhlinin}, {Burenin}, {Ebeling},
  {Forman}, {Hornstrup}, {Jones}, {Kravtsov}, {Murray}, {Nagai}, {Quintana}, \&
  {Voevodkin}}]{Vikhlinin2009}
{Vikhlinin} A. {et~al.}, 2009, \apj, 692, 1033

\bibitem[{{Vikhlinin} {et~al}\mbox{.}(2006){Vikhlinin}, {Kravtsov}, {Forman},
  {Jones}, {Markevitch}, {Murray}, \& {Van Speybroeck}}]{Vikhlinin2006}
{Vikhlinin} A., {Kravtsov} A., {Forman} W., {Jones} C., {Markevitch} M.,
  {Murray} S.~S., {Van Speybroeck} L., 2006, \apj, 640, 691

\bibitem[{{Voit}(2005)}]{Voit2005}
{Voit} G.~M., 2005, Reviews of Modern Physics, 77, 207

\bibitem[{{von der Linden} {et~al}\mbox{.}(2014){von der Linden}, {Allen},
  {Applegate}, {Kelly}, {Allen}, {Ebeling}, {Burchat}, {Burke}, {Donovan},
  {Morris}, {Blandford}, {Erben}, \& {Mantz}}]{VonDerLinden2014}
{von der Linden} A. {et~al.}, 2014, \mnras, 439, 2

\bibitem[{{Weinberg} {et~al}\mbox{.}(2013){Weinberg}, {Mortonson},
  {Eisenstein}, {Hirata}, {Riess}, \& {Rozo}}]{Weinberg2013}
{Weinberg} D.~H., {Mortonson} M.~J., {Eisenstein} D.~J., {Hirata} C., {Riess}
  A.~G., {Rozo} E., 2013, \physrep, 530, 87

\bibitem[{{Wen} {et~al}\mbox{.}(2012){Wen}, {Han}, \& {Liu}}]{Wen2012}
{Wen} Z.~L., {Han} J.~L., {Liu} F.~S., 2012, \apjs, 199, 34

\bibitem[{{York} {et~al}\mbox{.}(2000){York}, {Adelman}, {Anderson},
  {Anderson}, {Annis}, {Bahcall}, {Bakken}, {Barkhouser}, {Bastian}, {Berman},
  {Boroski}, {Bracker}, {Briegel}, {Briggs}, {Brinkmann}, {Brunner}, {Burles},
  {Carey}, {Carr}, {Castander}, {Chen}, {Colestock}, {Connolly}, {Crocker},
  {Csabai}, {Czarapata}, {Davis}, {Doi}, {Dombeck}, {Eisenstein}, {Ellman},
  {Elms}, {Evans}, {Fan}, {Federwitz}, {Fiscelli}, {Friedman}, {Frieman},
  {Fukugita}, {Gillespie}, {Gunn}, {Gurbani}, {de Haas}, {Haldeman}, {Harris},
  {Hayes}, {Heckman}, {Hennessy}, {Hindsley}, {Holm}, {Holmgren}, {Huang},
  {Hull}, {Husby}, {Ichikawa}, {Ichikawa}, {Ivezi{\'c}}, {Kent}, {Kim},
  {Kinney}, {Klaene}, {Kleinman}, {Kleinman}, {Knapp}, {Korienek}, {Kron},
  {Kunszt}, {Lamb}, {Lee}, {Leger}, {Limmongkol}, {Lindenmeyer}, {Long},
  {Loomis}, {Loveday}, {Lucinio}, {Lupton}, {MacKinnon}, {Mannery}, {Mantsch},
  {Margon}, {McGehee}, {McKay}, {Meiksin}, {Merelli}, {Monet}, {Munn},
  {Narayanan}, {Nash}, {Neilsen}, {Neswold}, {Newberg}, {Nichol}, {Nicinski},
  {Nonino}, {Okada}, {Okamura}, {Ostriker}, {Owen}, {Pauls}, {Peoples},
  {Peterson}, {Petravick}, {Pier}, {Pope}, {Pordes}, {Prosapio},
  {Rechenmacher}, {Quinn}, {Richards}, {Richmond}, {Rivetta}, {Rockosi},
  {Ruthmansdorfer}, {Sandford}, {Schlegel}, {Schneider}, {Sekiguchi}, {Sergey},
  {Shimasaku}, {Siegmund}, {Smee}, {Smith}, {Snedden}, {Stone}, {Stoughton},
  {Strauss}, {Stubbs}, {SubbaRao}, {Szalay}, {Szapudi}, {Szokoly}, {Thakar},
  {Tremonti}, {Tucker}, {Uomoto}, {Vanden Berk}, {Vogeley}, {Waddell}, {Wang},
  {Watanabe}, {Weinberg}, {Yanny}, {Yasuda}, \& {SDSS
  Collaboration}}]{York2000}
{York} D.~G. {et~al.}, 2000, \aj, 120, 1579

\end{thebibliography}

\end{document}